\documentclass[10pt,journal,compsoc]{IEEEtran}
\ifCLASSOPTIONcompsoc
\usepackage[nocompress]{cite}
\else
\usepackage{cite}
\fi

\ifCLASSINFOpdf
\usepackage[pdftex]{graphicx}
\graphicspath{{figures/}{pictures/}{images/}{./}} 
\DeclareGraphicsExtensions{.pdf,.png,.jpg,.jpeg}
\else
\fi

\graphicspath{{figures/}{pictures/}{images/}{./}} 
\usepackage{microtype} 
\PassOptionsToPackage{warn}{textcomp}  
\usepackage{textcomp}  
\usepackage{mathptmx}  
\usepackage{times} 
\usepackage{cite}  
\usepackage{tabu}  
\usepackage{booktabs}  
\usepackage{lscape}
\usepackage{amssymb}
\usepackage{tabularx}
\usepackage[figuresleft]{rotating}
\usepackage{graphicx}
\usepackage{subfigure}
\usepackage{multirow}
\usepackage{colortbl}
\usepackage{enumitem}
\usepackage{enumerate}
\usepackage{threeparttable}
\usepackage{makecell}
\usepackage{stfloats}
\usepackage[table]{xcolor}
\usepackage[colorlinks,
linkcolor=red,
anchorcolor=blue,
citecolor=green
]{hyperref}
\definecolor{bluecrayola}{rgb}{0.12,0.46,1.0}

\definecolor{p}{RGB}{236,117,36}
\definecolor{d}{RGB}{112,48,160}
\definecolor{v}{RGB}{255,51,153}

\title{Towards Natural Language Interfaces for Data Visualization: A Survey}

\begin{document}
	
	\author{
		Leixian~Shen,
		Enya~Shen,
		Yuyu~Luo,
		Xiaocong~Yang,
		Xuming~Hu,\\
		Xiongshuai~Zhang,
		Zhiwei~Tai,
		and~Jianmin~Wang

		\IEEEcompsocitemizethanks{
			\IEEEcompsocthanksitem 
			All authors are from Tsinghua University, Beijing, China.
			E-mail: $\left\{slx20, luoyy18, yangxc18, hxm19, zxs21, tzw20\right\}$@mails.tsinghua.edu.cn, $\left\{shenenya, jimwang\right\}$@tsinghua.edu.cn.
			
		}
		\thanks{Manuscript received XX XX, 2021; revised XX XX, 2022.}}
	
	\markboth{\MakeUppercase{IEEE TRANSACTIONS ON VISUALIZATION AND COMPUTER GRAPHICS},~Vol.~XX, No.~X, XX~2022}%
	{Shell \MakeLowercase{\textit{\textit{et al.}}}: Bare Demo of IEEEtran.cls for Computer Society Journals}

	\IEEEtitleabstractindextext{
		\vspace{-0.5cm}
		\begin{abstract}
			\raggedright
			Utilizing Visualization-oriented Natural Language Interfaces (V-NLI) as a complementary input modality to direct manipulation for visual analytics can provide an engaging user experience. 
It enables users to focus on their tasks rather than having to worry about how to operate visualization tools on the interface.
			In the past two decades, leveraging advanced natural language processing technologies, numerous V-NLI systems have been developed in academic research and commercial software, especially in recent years. In this article, we conduct a comprehensive review of the existing V-NLIs. In order to classify each paper, we develop categorical dimensions based on a classic information visualization pipeline with the extension of a V-NLI layer. The following seven stages are used: query interpretation, data transformation, visual mapping, view transformation, human interaction, dialogue management, and presentation. Finally, we also shed light on several promising directions for future work in the V-NLI community.
			\vspace{-0.3cm}
		\end{abstract}
		
		\begin{IEEEkeywords}
			Data Visualization, Natural Language Interfaces, Survey.
			\vspace{-0.1cm}
	\end{IEEEkeywords}}
	\maketitle
	\IEEEdisplaynontitleabstractindextext
	\IEEEpeerreviewmaketitle

	\section{Introduction}\label{sec:intro}
	\vspace{-0.3cm}
	
\begin{figure}[t]
		\setlength{\abovecaptionskip}{0cm}
		\setlength{\belowcaptionskip}{-0.1cm}
		\centering
		\includegraphics[width=0.9\columnwidth]{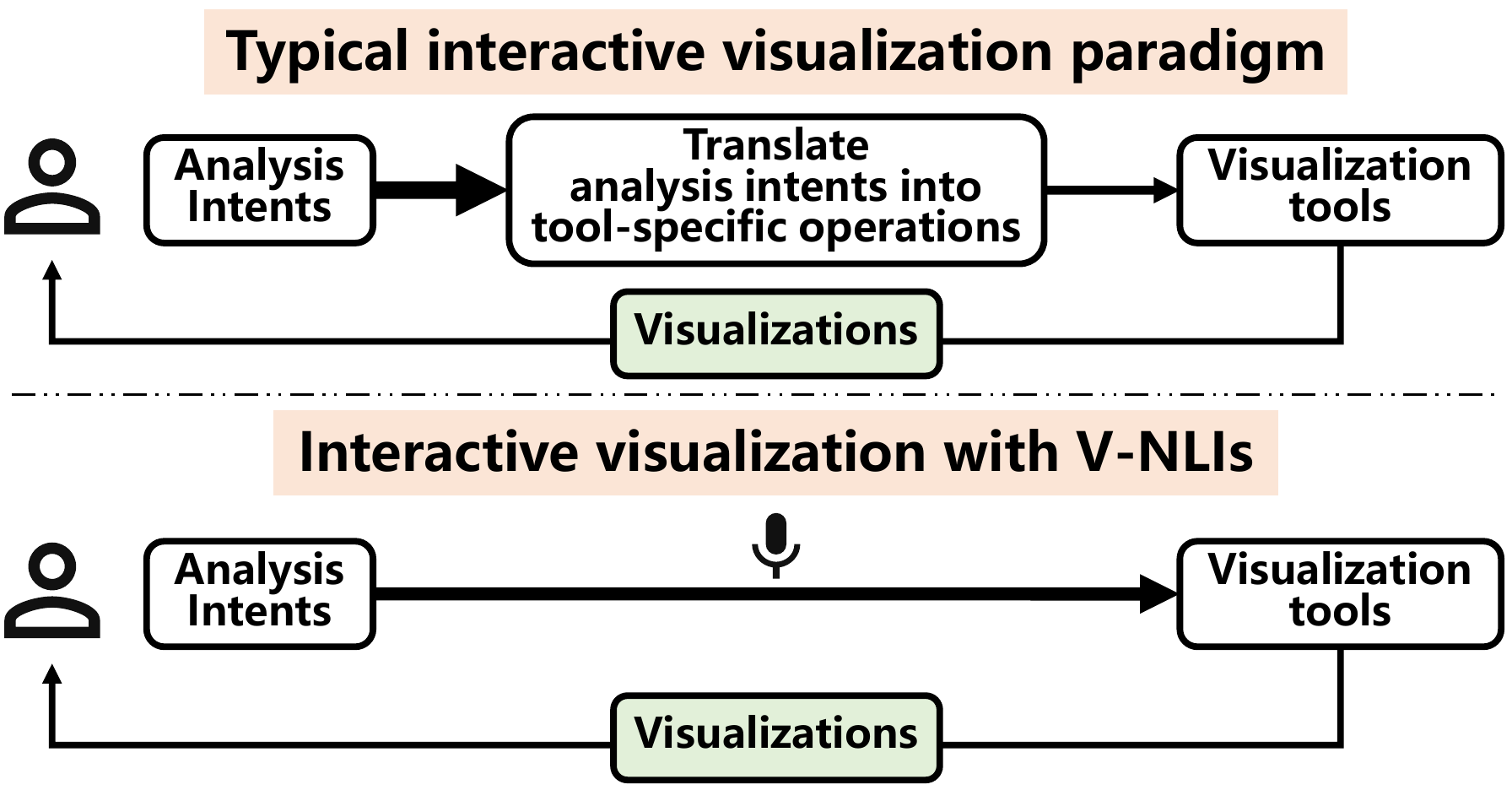}
		\caption{The traditional interaction paradigm requires users to translate their analysis intents into tool-specific operations\cite{Lee2012}. With the help of V-NLI, users can express their analysis intents in their terms.
		}
		\label{fig:pipeline}
		\vspace{-0.6cm}
	\end{figure}

	
	\IEEEPARstart{T}{he} use of interactive visualization is becoming increasingly popular in data analytics\cite{Battle2021}. As a common part of analytics suites, Windows, Icons, Menus, and Pointer (WIMP) interfaces have been widely employed to facilitate interactive visual analysis in current practices.
	However, this interaction paradigm presents a steep learning curve in visualization tools since it requires users to translate their analysis intents into tool-specific operations\cite{Lee2012}, as shown in the upper part of Figure \ref{fig:pipeline}.
	
	Over the years, the rapid development of Natural Language Processing (NLP) technology has provided great opportunities to explore a natural-language-based interaction paradigm for data visualization\cite{Belinkov2019,Young2018}.
	With the help of advanced NLP toolkits\cite{Honnibal2017,Bird2006,Manning2014,opennlp,googlenlp}, a surge of Visualization-oriented Natural Language Interfaces (V-NLI) emerged recently as a complementary input modality to traditional WIMP interaction.
	V-NLIs accept the user's natural language queries (e.g., ``visualize the distribution of budget as a histogram") as input and output appropriate visualizations (e.g., correspondingly, a histogram with the \textit{production budget}, a data attribute of cars dataset).
	The emergence of V-NLI can greatly enhance the usability of visualization tools in terms of: 
	(a) Convenience and novice-friendliness.
	Natural language is a skill that is mastered by the public. By using natural language to interact with computers, V-NLI closes the tool-specific manipulations to users, as shown in Figure \ref{fig:pipeline}, facilitating the analysis flow for novices.
	(b) Intuitiveness and effectiveness. It is a consensus that visual analysis is most effective when users can focus on their data rather than manipulations on the interface of analysis tools\cite{Hoque2018}. With the help of V-NLI, users can express their analytic tasks in their terms.
	(c) Humanistic care. A sizeable amount of information we access nowadays is supported by visual means. V-NLI can be an innovative means for non-visual access, which promotes the inclusion of blind and low vision people.
	
\begin{figure*}[t]
		\setlength{\abovecaptionskip}{0cm}
		\setlength{\belowcaptionskip}{-0.1cm}
		\centering
		\includegraphics[width=\textwidth]{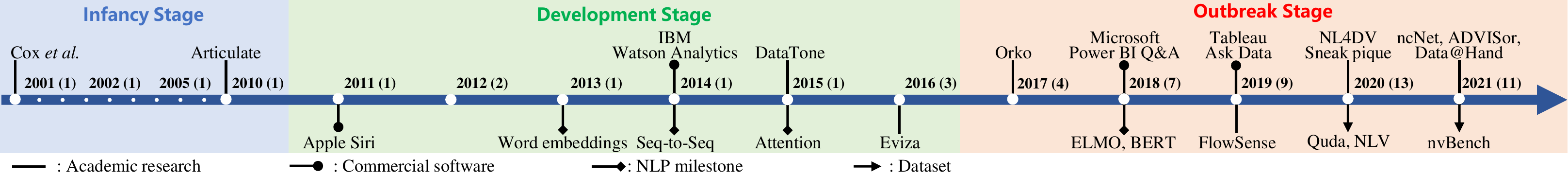}
		\caption{Timeline of V-NLI. We briefly divide the timeline into the infancy, development, and outbreak stages. The number of yearly published papers is attached. The timeline consists of four parts: academic research, commercial software, NLP milestone, and dataset.
		}
		\label{fig:timeline}
		\vspace{-0.5cm}
	\end{figure*}

	However, designing and implementing V-NLIs is challenging due to the ambiguous and underspecified nature of human language, the complexity of maintaining context in a conversational flow, and the lack of discoverability (communicating to users what the system can do). In the past two decades, to address the challenges, numerous systems have been developed in academic research and commercial software, especially in recent years.
	The timeline of V-NLI is shown in Figure \ref{fig:timeline}.
	Back in 2001, Cox \textit{et al.}\cite{Cox2001} presented an initial prototype of NLI for visualization,
	which can only accept well-structured queries. 
	Articulate\cite{Sun2010} introduced a two-step process to create visualizations from NL queries almost a decade later.
		It first extracts the user's analytic task and data attributes and then automatically determines the appropriate visualizations based on the information.
	Although the infancy-stage investigations were a promising start, as natural language was not yet a prevalent interaction modality, the V-NLI systems were restricted to simple prototypes. However, since Apple integrated Siri\cite{Sigtia2020} into the iPhone, NLIs began to attract more attention.
	Around 2013, the advent of word embeddings\cite{Mikolov2013} promoted the advances of neural networks for NLP, rekindling commercial interest in V-NLI. IBM firstly published their NL-based cognitive service, Watson Analytics\cite{IBM}, in 2014. Microsoft Power BI's Q\&A\cite{powerbi} and Tableau's Ask data\cite{askdata} were announced in 2018 and 2019, respectively, offering various features like autocompletion and context management.
	DataTone\cite{Gao2015b} first introduced ambiguity widgets, which are user interface widgets (e.g., dropdown and slider) to help users modify system-generated responses.
	As opposed to one-off commands, Eviza\cite{Setlur2016} enabled users to have interactive conversations with their data.
	After technological accumulation, the past five years have seen the outbreak of V-NLI (see the number of yearly published papers in Figure \ref{fig:timeline}).
	With the development of hardware devices, synergistic multimodal visualization interfaces gained notable interest. Orko\cite{Srinivasan2018} was the first system to combine touch and speech input on tablet devices, and Data@Hand\cite{Young-HoKim2021} focused on smartphones. 
	Sneak pique\cite{Setlur2020} explored how autocompletion can improve system discoverability while helping users formulate analytical questions.
	The pretrained language models obtained new state-of-the-art results on various NLP tasks from 2018, which provided great opportunities to improve the intelligence of V-NLI\cite{ELMo,Devlin2019}. ADVISor\cite{Liu2021b} and ncNet\cite{Luo2021a} were follow-up deep learning-based solutions. Quda\cite{Fu2018} and NLV\cite{Srinivasan2021} contributed datasets of NL queries for visual data analytics, and nvBench produced the first V-NLI benchmark\cite{Luo2021}.
	Beyond data exploration, FlowSense\cite{Yu2020b} augmented a data flow-based visualization system with V-NLI.
	The NL4DV\cite{Narechania2020} toolkit can be easily integrated into existing visualization systems to provide V-NLI service.
	
	
	\begin{table}[t]
		\setlength{\abovecaptionskip}{0cm}
		\setlength{\belowcaptionskip}{-0.1cm}
		\caption{Relevant venues}
		\label{tab:venue}
		\centering
		\renewcommand\arraystretch{1}
		\setlength{\tabcolsep}{0.5mm}{
			\begin{tabular}{m{3.4cm}|m{4.9cm}}
				\makecell[c]{\textbf{Field}} & \makecell[c]{\textbf{Venues}} \\ \hline
				\begin{tabular}[c]{@{}l@{}}Visualization\\(VIS)\end{tabular} & IEEE VIS (InfoVis, VAST, SciVis), EuroVis, PacificVis, TVCG, CGF, CGA \\ \hline
				Human-Computer Interaction (HCI) & CHI, UIST, IUI \\ \hline
				Natural Language Processing (NLP) & ACL, EMNLP, NAACL, COLING \\ \hline
				Data Mining and Management (DMM) & KDD, SIGMOD, ICDE, VLDB \\ \hline
		\end{tabular}}
		\vspace{-0.6cm}
	\end{table}
	
	Literature on V-NLI research is proliferating, covering aspects such as Visualization (VIS), Human-Computer Interaction (HCI), Natural Language Processing (NLP), and Data Mining and Management (DMM).
	As a result, there is an increasing need to better organize the research landscape, categorize current work, identify knowledge gaps, and assist people who are new to this growing area to understand the challenges and subtleties in the community. 
	For this purpose, there have been several prior efforts to summarize the advances in this area. For example, Srinivasan and Stasko (Short paper in EuroVis 2017\cite{Srinivasan2017}) conducted a simple examination of five existing V-NLI systems by comparing and contrasting them based on the tasks they allow users to perform.
	They (Journal paper in CGA 2020\cite{Srinivasan2020}) further highlighted critical challenges for evaluating V-NLIs and discussed benefits and considerations of three task framing strategies.
	Although the two surveys can provide valuable guidance for follow-up research, with the outbreak of V-NLI in recent years, considerable new works are to be covered and details to be discussed.
	To the best of our knowledge, this paper is the first step towards comprehensively reviewing V-NLI approaches in a systematic manner.
	
	This paper is structured as follows: First, we explain the scope and methodology of the survey in Section \ref{sec:Landscape}. What follows is the classification overview of existing works in Section \ref{sec:classification}. 
	Then, the comprehensive survey is presented in Sections \ref{sec:query understanding} - \ref{sec:presentation},
	corresponding to seven stages extracted in the information visualization pipeline\cite{Card1999}. 
	Finally, we shed light on several promising directions for future work in Section \ref{sec:future}. 

	\begin{table*}[bp]
		\setlength{\abovecaptionskip}{0.1cm}
		\setlength{\belowcaptionskip}{-0.1cm}
		\caption{Representative papers of related works to V-NLI. \textit{Related Stage} column lists the overlap with V-NLI stages in Figure \ref{fig:vispipeline} that apply similar technologies.}
		\label{tab:related works}
		\scriptsize
		\centering
		\renewcommand\arraystretch{1}
		\setlength{\tabcolsep}{1mm}{
			\begin{tabular}{m{5.8cm}m{1.8cm}m{1.4cm}m{8.1cm}}
				\hline
				\rowcolor[HTML]{EFEFEF} 
				\makecell[c]{\textbf{Topic}}& \makecell[c]{\textbf{Related Stage}} &\makecell[c]{\textbf{Survey}}& \makecell[c]{\textbf{Representative Paper}} \\ \hline
				Visualization Recommendation (VisRec)&B/C/D/E/G &\cite{Zhu2020,Qin2020} &
				\cite{Lee2021a,Kim2020,Zehrung2021a,Qian2021q,Zhou2020,Rossi,Lin2020a,Ma2020b,Moritz2019,Chen2020,Cui2019,Dibia2019b,Hu2019d,Wang2018b,Luo2018,Hu2018,Wongsuphasawat2017,Demiralp2017,Wongsuphasawat2016,Wongsuphasawat2016c,Mutlu2016b,Vartak2015b,Zeng2021,Kaur2015b,Choo2014b,Key2012,Gotz2009b,Mackinlay2007b,Seo2005b,Stolte2002b,Wu2021e,Li2021a,Vartak2014b}	
				\\ \hline
				Natural Language Interface for DataBase (NLI4DB)&A/B/E/F/G & \cite{Azcan2020,Affolter2019} &
				\cite{Baik2019,Basik2018,Bast2015,Bergamaschi2016,Blunschi2012,Eisenschlos2020,gu2016incorporating,guo2019complex,Gur2018,Herzig2020,inproceedings,Iyer2017,joshi2020spanbert,Kocijan_2019,Li2014,Li2014a,Li2016a,8029993,Narechania2021,Nihalani2011NaturalLI,Quamar2020,Saha2016,Sen2019,Shekarpour2015,Simitsis2007,miikkulainen:book93,9101534,xu2017,Yaghmazadeh2017,Yu2020,Zheng2017,zhong2017seq2sql}
				\\ \hline
				Conversational Interface for Visualization (CIV)&A/B/C/D/E/F/G &  &
				\cite{Bieliauskas2017,Seipel2019,Fast2017,Siddiqui2020,Hearst2019a,Nafari2013,Nafari2015,LeoJohn2017,Lee2021d}
				\\ \hline
				Natural Language Generation for Visualization (NLG4Vis)&B/E/G & & 
				\cite{Spreafico2020,Qian2021,Obeid2020,Moraes2014,Mittal1998a,Liu2020a,Kim1999,Ferres2010,Demir2012,Demir2008,Cui2019,Corio1999}
				\\ \hline
				Visual Question Answering (VQA)&A/B/E/F/G & \cite{Mathew2021} &
				\cite{Chaudhry2020,Kafle2018,Easoning2018,Liu2021b,Kim2020a,Singh2020,Wohlgenannt2019,Bylinskii,Madan2018,Yagcioglu2020}
				\\ \hline
				Annotation and Narrative Storytelling&A/B/C/D/E/F/G & \cite{Segel2010,Tong2018} &
				\cite{Chen2020a,Lai2020a,Shi2020,Lee,Ren,Wang2020,Chen,Hullman2013,Kim2020,Metoyer2018,Xia2020,Thompson2021,LiuTingting,Wang2020h,Bach2018,Lalle2019,Kong2012,Yuan2021,Wang2018,Kandogan,El-Dairi2019,Qian2020,Mitri2020,Chen2020,Bryan2017,Cui2020b,Kim2019,Fulda2016,Chena,Chen2019,Hopkins2020}
				\\ \hline
				
		\end{tabular}}
		\vspace{-0.3cm}
	\end{table*}

	\begin{table*}[]
		\setlength{\abovecaptionskip}{0.1cm}
		\setlength{\belowcaptionskip}{-0.1cm}
		\caption{Summary of representative works in V-NLI. 
			The first five columns list basic information, and the left columns are characteristics of V-NLI with categorical dimensions described in Section \ref{sec:classification}. Details of each column will be discussed in Sections \ref{sec:query understanding} - \ref{sec:presentation}.
		}
		\label{tab:vis_papers}
		\centering
		\scriptsize
		\begin{threeparttable}
			\setlength{\tabcolsep}{0.5mm}{
				\renewcommand\arraystretch{1}
				\rowcolors{2}{gray!25}{white}
				\begin{tabular}{lllll|cccc|c|ccc|c|cccc|c|ccccc}
					\multicolumn{1}{c}{} & 
					\multicolumn{1}{c}{} & 
					\multicolumn{1}{c}{} & 
					\multicolumn{1}{c}{} & 
					\multicolumn{1}{c}{} & 
					\multicolumn{4}{c}{\color{red}\textbf{\underline{\ref{sec:query understanding}}}} & 
					\multicolumn{1}{c}{\color{red}\textbf{\underline{\ref{sec:data transformation}}}} & 
					\multicolumn{3}{c}{\color{red}\textbf{\underline{\ref{sec:visual mapping}}}} & 
					\multicolumn{1}{c}{\color{red}\textbf{\underline{\ref{sec:view transformation}}}} & 
					\multicolumn{4}{c}{\color{red}\textbf{\underline{\ref{sec:human interaction}}}} & 
					\multicolumn{1}{c}{\color{red}\textbf{\underline{\ref{sec:context management}}}} & 
					\multicolumn{5}{c}{\color{red}\textbf{\underline{\ref{sec:presentation}}}} \\
					\textbf{Name} & \textbf{Publication} & \textbf{NLP Toolkit or Technology} & 
					\begin{tabular}[c]{@{}l@{}}\textbf{Visualization}\\\textbf{Type}\end{tabular} & 
					\begin{tabular}[c]{@{}l@{}}\textbf{Recommendation}\\\textbf{Algorithm}\end{tabular} & 
					\rotatebox{90}{\textbf{Semantic and Syntax Analysis} }& 
					\rotatebox{90}{\textbf{Task Inference} }& 
					\rotatebox{90}{\textbf{Data Attributes Inference} }& 
					\rotatebox{90}{\textbf{Default for Underspecified Utterances} }& 
					\rotatebox{90}{\textbf{Transformation for Data Insights} }& 
					\rotatebox{90}{\textbf{Spatial Substrate Mapping} }& 
					\rotatebox{90}{\textbf{Graphical Elements Mapping} }& 
					\rotatebox{90}{\textbf{Graphical Properties Mapping} }& 
					\rotatebox{90}{\textbf{View Transformation} }&
					\rotatebox{90}{\textbf{Ambiguity Widget} }& 
					\rotatebox{90}{\textbf{Autocompletion} }& 
					\rotatebox{90}{\textbf{Multimodal} }& 
					\rotatebox{90}{\textbf{WIMP} }& 
					\rotatebox{90}{\textbf{Dialogue Management} }& 
					\rotatebox{90}{\textbf{Visual Presentation} }& 
					\rotatebox{90}{\textbf{Annotation} }& 
					\rotatebox{90}{\textbf{Narrative Storytelling} }&
					\rotatebox{90}{\textbf{NL Description Generation} }&  
					\rotatebox{90}{\textbf{Visual Question Answering}} \\
					\midrule[1.5pt]
					Cox \textit{et   al.}\cite{Cox2001} & IJST'01 & Sisl & B/Ta & InfoStill & × & \checkmark & \checkmark & × & \checkmark & \checkmark & \checkmark & \checkmark & × & × & × & × & × & \checkmark & \checkmark & × & × & × & × \\
					Kato \textit{et al.}\cite{Kato2002} & COLING'02 & Logical form & B/L/P/A & Template & \checkmark & \checkmark & \checkmark & × & \checkmark & \checkmark & \checkmark & \checkmark & × & × & × & × & × & × & \checkmark & × & × & × & × \\
					RIA\cite{Zhou} & InfoVis'05 & - & I & Optimization & \checkmark & \checkmark & \checkmark & × & \checkmark & × & × & × & × & × & × & \checkmark & × & \checkmark & \checkmark & × & \checkmark & × & × \\
					Articulate\cite{Sun2010} & LNCS'10 & Stanford   Parser & S/R/B/L/P/Bo & Decision   tree & \checkmark & \checkmark & \checkmark & × & \checkmark & \checkmark & \checkmark & \checkmark & × & × & × & × & × & × & \checkmark & × & × & × & × \\
					Contextifier\cite{Hullman2013} & CHI'13 & NLTK & L & - & \checkmark & × & \checkmark & × & \checkmark & × & × & × & × & × & × & × & × & × & \checkmark & \checkmark & \checkmark & × & × \\
					NewsViews\cite{El-Dairi2019} & CHI'14 & OpenCalais & M & - & \checkmark & \checkmark & \checkmark & × & \checkmark & × & × & × & × & × & × & × & × & × & \checkmark & \checkmark & \checkmark & × & × \\
					Datatone\cite{Gao2015b} & UIST'15 & NLTK/Stanford   Parser & S/B/L & Template & \checkmark & \checkmark & \checkmark & × & \checkmark & \checkmark & \checkmark & \checkmark & × & \checkmark & × & × & \checkmark & × & \checkmark & × & × & × & × \\
					Articulate2\cite{Kumar2016,Kumar2017} & SIGDIAL'16 & ClearNLP/Stanford   Parser/OpenNLP & B/L/H & Decision   tree & \checkmark & \checkmark & \checkmark & × & \checkmark & \checkmark & \checkmark & \checkmark & \checkmark & × & × & \checkmark & × & \checkmark & \checkmark & × & × & × & × \\
					Eviza\cite{Setlur2016} & UIST'16 & ANTLR & M/L/B/S & Template & \checkmark & \checkmark & \checkmark & × & \checkmark & \checkmark & \checkmark & \checkmark & × & \checkmark & \checkmark & × & \checkmark & \checkmark & \checkmark & × & × & × & × \\
					TimeLineCurator\cite{Fulda2016} & TVCG'16 & TERNIP & Tl & Template & \checkmark & \checkmark & \checkmark & × & × & × & × & × & × & × & × & × & × & × & \checkmark & \checkmark & \checkmark & \checkmark & × \\
					Analyza\cite{Dhamdhere2017} & IUI'17 & Stanford   Parser & B/L & Template & \checkmark & \checkmark & \checkmark & × & \checkmark & \checkmark & \checkmark & \checkmark & × & \checkmark & \checkmark & × & \checkmark & \checkmark & \checkmark & × & × & × & × \\
					TSI\cite{Bryan2017} & TVCG'17 & Template & L/A/H & - & × & \checkmark & \checkmark & × & \checkmark & × & × & × & × & × & × & × & × & × & \checkmark & \checkmark & \checkmark & × & × \\
					Ava\cite{LeoJohn2017} & CIDR'17 & Controlled Natural Language & S/L/B & Template & \checkmark & \checkmark & \checkmark & × & \checkmark & \checkmark & \checkmark & \checkmark & × & × & × & × & × & \checkmark & \checkmark & × & × & × & \checkmark \\
					DeepEye\cite{Luo2018a} & SIGMOD'18 & OpenNLP & P/L/B/S & Template & \checkmark & × & \checkmark & × & \checkmark & \checkmark & \checkmark & \checkmark & × & × & × & × & \checkmark & × & \checkmark & × & × & × & × \\
					Evizeon\cite{Hoque2018} & TVCG'18 & CoreNLP & M/L/B/S & Template & \checkmark & \checkmark & \checkmark & × & \checkmark & \checkmark & \checkmark & \checkmark & × & \checkmark & \checkmark & × & \checkmark & \checkmark & \checkmark & × & × & × & × \\
					Iris\cite{Fast2017} & CHI'18 & Domain-specific language & S/T & Template & \checkmark & \checkmark & \checkmark & × & \checkmark & \checkmark & \checkmark & \checkmark & × & × & × & × & × & \checkmark & \checkmark & × & × & × & \checkmark \\
					Metoyer \textit{et   al.}\cite{Metoyer2018} & IUI'18 & CoreNLP & I & Template & \checkmark & × & \checkmark & × & \checkmark & × & × & × & × & × & × & × & \checkmark & × & \checkmark & × & \checkmark & × & \checkmark \\
					Orko\cite{Srinivasan2018} & TVCG'18 & CoreNLP/NLTK & N & - & \checkmark & \checkmark & \checkmark & × & \checkmark & \checkmark & × & \checkmark & \checkmark & \checkmark & × & \checkmark & \checkmark & \checkmark & \checkmark & × & × & × & × \\
					ShapeSearch\cite{Siddiqui2018b,Siddiqui2021} & VLDB'18 & Stanford   Parser & L & ShapeQuery & \checkmark & × & \checkmark & × & \checkmark & \checkmark & × & \checkmark & × & \checkmark & × & × & \checkmark & × & \checkmark & × & × & × & × \\
					Valletto\cite{Kassel2018} & CHI'18 & Spacy & B/S/M & Template & \checkmark & \checkmark & \checkmark & × & \checkmark & \checkmark & \checkmark & \checkmark & \checkmark & × & × & \checkmark & \checkmark & \checkmark & \checkmark & × & × & × & × \\
					ArkLang\cite{Setlur2019} & IUI'19 & Cocke-Kasami-Younger & B/G/L/M/P/S/Tr & Template & \checkmark & \checkmark & \checkmark & \checkmark & \checkmark & \checkmark & \checkmark & \checkmark & × & × & × & × & \checkmark & × & \checkmark & × & × & × & × \\
					Ask Data\cite{askdata,Tory2019b} & VAST'19 & Proprietary & B/G/L/M/P/S/T & Template & \checkmark & \checkmark & \checkmark & \checkmark & \checkmark & \checkmark & \checkmark & \checkmark & × & \checkmark & \checkmark & × & \checkmark & \checkmark & \checkmark & × & × & × & × \\
					Data2Vis\cite{Dibia2019b} & CGA'19 & Seq2Seq & B/A/L/S/T/St & Seq2Seq & × & \checkmark & × & × & \checkmark & \checkmark & \checkmark & \checkmark & × & × & × & × & \checkmark & × & \checkmark & × & × & × & × \\
					Hearst \textit{et   al.}\cite{Hearst2019} & VIS'19 & WordNet & - & - & \checkmark & × & \checkmark & \checkmark & \checkmark & × & × & × & × & × & × & × & × & × & × & × & × & × & × \\
					Voder\cite{Srinivasan2019a} & TVCG'19 & Stanford   Parser & St/Bo/B/S/D & Template & \checkmark & \checkmark & \checkmark & × & \checkmark & \checkmark & \checkmark & \checkmark & × & × & × & × & \checkmark & × & \checkmark & × & × & \checkmark & × \\
					AUDiaL\cite{Murillo-Morales2020} & LNCS'20 & CoreNLP & B/L & Knowledge   Base & \checkmark & \checkmark & \checkmark & × & \checkmark & \checkmark & \checkmark & \checkmark & × & \checkmark & × & × & \checkmark & × & \checkmark & × & × & × & × \\
					Bacci \textit{et   al.}\cite{Bacci2020} & LNCS'20 & Wit.ai & P/L/B/S & Template & \checkmark & \checkmark & \checkmark & × & \checkmark & \checkmark & \checkmark & \checkmark & × & × & \checkmark & × & \checkmark & × & \checkmark & × & × & × & × \\
					DataBreeze\cite{Srinivasan2020b} & TVCG'20 & CoreNLP/NLTK & S/Uc & Template & \checkmark & \checkmark & \checkmark & × & \checkmark & \checkmark & \checkmark & \checkmark & \checkmark & \checkmark & × & \checkmark & \checkmark & × & \checkmark & × & × & × & × \\
					FlowSense\cite{Yu2020b} & TVCG'20 & CoreNLP/SEMPRE & L/S/B/M/N/Ta & Template & \checkmark & \checkmark & \checkmark & × & \checkmark & \checkmark & \checkmark & \checkmark & × & × & \checkmark & × & \checkmark & × & \checkmark & × & × & × & × \\
					InChorus\cite{Srinivasan2020a} & CHI'20 & CoreNLP/NLTK & L/S/B & Template & \checkmark & \checkmark & \checkmark & × & \checkmark & \checkmark & \checkmark & \checkmark & \checkmark & × & × & \checkmark & \checkmark & × & \checkmark & × & × & × & × \\
					NL4DV\_Quda\cite{Fu2018} & arXiv'20 & CoreNLP & St/B/L/A/P/S/Bo/H & Template & \checkmark & \checkmark & \checkmark & × & \checkmark & \checkmark & \checkmark & \checkmark & × & \checkmark & × & \checkmark & \checkmark & × & \checkmark & × & × & × & × \\
					Sneak Pique\cite{Setlur2020} & UIST'20 & ANTLR & - & - & \checkmark & \checkmark & \checkmark & × & \checkmark & × & × & × & × & × & \checkmark & × & × & × & × & × & × & × & × \\
					Story Analyzer\cite{Mitri2020} & JCIS'20 & CoreNLP & Wc/C/T/Fg/B/M/I & Template & \checkmark & × & \checkmark & × & × & \checkmark & \checkmark & \checkmark & × & × & × & × & × & × & \checkmark & × & \checkmark & × & × \\
					Text-to-Viz\cite{Cui2020b} & TVCG'20 & Stanford   Parser & I & Template & \checkmark & \checkmark & \checkmark & × & \checkmark & × & × & \checkmark & × & × & × & × & × & × & \checkmark & × & \checkmark & \checkmark & × \\
					Vis-Annotator\cite{Lai2020a} & CHI'20 & Spacy/Mask-RCNN & B/L/P & - & \checkmark & × & \checkmark & × & \checkmark & × & × & × & × & × & × & × & × & × & \checkmark & \checkmark & × & × & × \\
					Sentifiers\cite{Setlur2020a} & VIS'20 & ANTLR & - & - & \checkmark & \checkmark & \checkmark & \checkmark & \checkmark & × & × & × & × & × & × & × & \checkmark & \checkmark & × & × & × & × & × \\
					NL4DV\cite{Narechania2020} & TVCG'21 & CoreNLP & St/B/L/A/P/S/Bo/H & Template & \checkmark & \checkmark & \checkmark & × & \checkmark & \checkmark & \checkmark & \checkmark & × & \checkmark & × & \checkmark & \checkmark & × & \checkmark & × & × & × & × \\
					Data@Hand\cite{Young-HoKim2021} & CHI'21 & Compromise/Chrono & B/L/S/Ra/Bo & Template & \checkmark & \checkmark & \checkmark & × & \checkmark & \checkmark & \checkmark & \checkmark & \checkmark & \checkmark & × & \checkmark & \checkmark & × & \checkmark & × & × & × & × \\
					Retrieve-Then-Adapt\cite{Qian2020} & TVCG'21 & Template & I & Template & \checkmark & × & × & × & × & × & × & × & × & × & × & × & \checkmark & × & \checkmark & \checkmark & \checkmark & \checkmark & × \\
					ADVISor\cite{Liu2021b} & PacificVis'21 & End-to-end network & B/L/S & Template & \checkmark & \checkmark & \checkmark & × & \checkmark & \checkmark & \checkmark & \checkmark & × & × & × & × & \checkmark & × & \checkmark & \checkmark & × & × & \checkmark \\
					Seq2Vis\cite{Luo2021} & SIGMOD'21 & Sep-to-Sep model & B/L/S/P & Seq2Seq & × & × & \checkmark & × & \checkmark & \checkmark & \checkmark & \checkmark & × & × & × & × & × & × & \checkmark & × & × & × & × \\
					ncNet\cite{Luo2021a} & VIS'21 & Transformer-based model& B/L/S/P & Template & × & × & \checkmark & × & \checkmark & \checkmark & \checkmark & \checkmark & × & × & × & × & × & × & \checkmark & × & × & × & ×\\
					GeoSneakPique\cite{Setlur} & VIS'21 & ANTLR & - & - & \checkmark & \checkmark & \checkmark & × & \checkmark & × & × & × & × & × & \checkmark & × & × & × & × & × & × & × & × \\
					Snowy\cite{Srinivasana} & UIST'21 & CoreNLP & - & - & \checkmark & \checkmark & \checkmark & × & \checkmark & × & × & × & × & × & \checkmark & × & × & \checkmark & × & × & × & × & × \\
					DT2VIS\cite{Jiang2021a} & CGA'21 & CoreNLP & B/L/S/P/Bo/R & Template & \checkmark & \checkmark & \checkmark & × & \checkmark & \checkmark & \checkmark & \checkmark & × & × & \checkmark & × & \checkmark & \checkmark & \checkmark & × & × & × & \checkmark \\
					\bottomrule[1.5pt]
			\end{tabular}}
			\begin{tablenotes}
				\scriptsize
				\item[*] Abbreviations for \textbf{\textit{Visualization Type}}: Bar(B), Table(Ta), Infographic(I), Scatter(S), Radar(R), Line(L), Pie(P), Boxplot(Bo), Icon(Ic), Map(M), Heatmap(H), Timeline(Tl), Area(A), Network(N), Tree(Tr), Strip(St), Donut(D), Gantt(G), Word clouds(Wc), Force graph(Fg), Range(Ra), Unit column charts(Uc), and Graphics(Gr).
			\end{tablenotes}
		\end{threeparttable}
		\vspace{-0.3cm}
	\end{table*}

	\vspace{-0.3cm}
	\section{Survey Landscape}\label{sec:Landscape}
	\subsection{Scope of the Survey}\label{sec:scope}
	
	In order to narrow the scope of the survey within a controllable range, we focused on visualization-oriented natural language interfaces, which accept natural language queries as input and output appropriate visualizations automatically.
		Users can input natural language in various ways, including typing with a keyboard (e.g., Datatone\cite{Gao2015b} in Figure \ref{fig:Datatone}), speech input via a microphone (e.g., InChorus\cite{Srinivasan2020a} in Figure \ref{fig:InChorus}), selecting text from articles (e.g., Metoyer \textit{et al.}\cite{Metoyer2018} in Figure \ref{fig:couple}), and inputting existing textual descriptions (e.g., Vis-Annotator\cite{Lai2020a} in Figure \ref{fig:Vis-Annotator}).

\begin{figure*}[t]
		\setlength{\abovecaptionskip}{0cm}
		\setlength{\belowcaptionskip}{-0.1cm}
		\centering
		\includegraphics[width=0.95\textwidth]{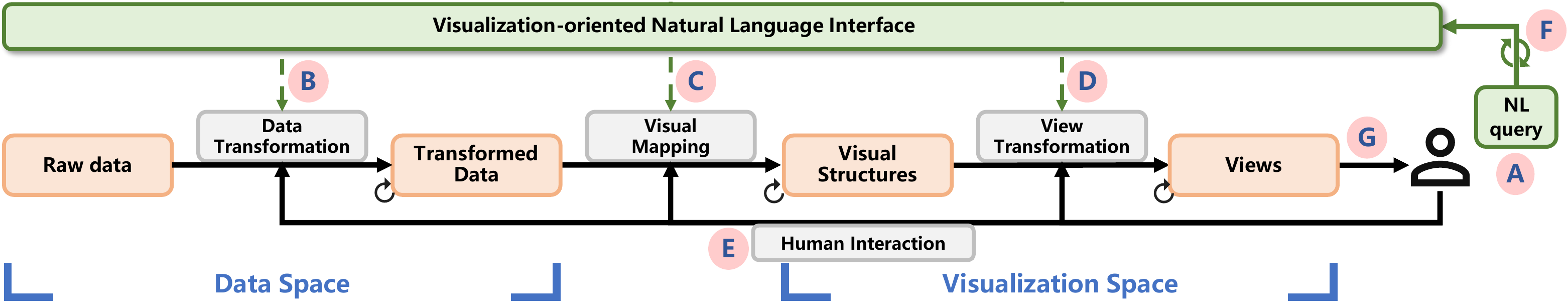}
		\caption{Extension of classic information visualization pipeline proposed by Card \textit{et al.}\cite{Card1999} with V-NLI. It depicts how V-NLI works to construct visualizations, which consists of the following seven stages: (A) Query Interpretation, (B) Data Transformation, (C) Visual Mapping, (D) View Transformation, (E) Human Interaction, (F) Dialogue Management, and (G) Presentation.}
		\label{fig:vispipeline}
		\vspace{-0.5cm}
	\end{figure*}

	In addition, several fields are closely related to V-NLI. As shown in Table \ref{tab:related works}, the \textit{Related Stage} column lists overlap stages with V-NLI in Figure \ref{fig:vispipeline} that apply similar technologies.
	In order to make the survey more comprehensive, when introducing V-NLI in the following sections, we will involve additional crucial discussions on the aforementioned related fields, with explanations of their importance and relationship with V-NLI.
	For example, Visualization Recommendation (VisRec)\cite{Zhu2020,Qin2020} acts as the back-end engine of V-NLI to recommend visualizations. Natural Language Interface for Database (NLI4DB)\cite{Affolter2019} and Conversational Interface for Visualization (CIV)\cite{Bieliauskas2017} share similar principles with V-NLI. Natural Language Generation for Visualization (NLG4Vis)\cite{Obeid2020,Liu2020a} and Visual Question Answering (VQA)\cite{Mathew2021,Kim2020a} complement visual data analysis with natural language as output. Annotation\cite{Ren,Shi2020} and Narrative Storytelling\cite{Tong2018} present fascinating visualizations by combining textual and visual elements. 

	\vspace{-0.3cm}
	\subsection{Survey Methodology}
	To comprehensively survey V-NLI, we performed an exhaustive review of relevant journals and conferences throughout the past twenty years (2000-2021), which broadly covers VIS, HCI, NLP, and DMM.
	The relevant venues are shown in Table \ref{tab:venue}.
	We began our survey by searching keywords (``natural language" AND ``visualization") in Google Scholar, resulting in 1436 papers (VIS: 455, HCI: 489, NLP: 289, and DMM: 203).
	We also searched for representative related works that appeared earlier or in the cited references of related papers. During the review process, we first examined the titles of papers from these publications to identify candidate papers. Next, abstracts of the candidate papers were browsed to determine whether they were related to V-NLI. The full text was reviewed to make a final decision if we could not obtain precise information from the title and abstract. Finally, we collected 57 papers about V-NLI that accepts natural language as input and outputs visualizations. Details of the systems are listed in Table \ref{tab:vis_papers}, including the 
	applied NLP technologies, chart types supported, visualization recommendation algorithm adopted, and various characteristics in V-NLI, which will be discussed in Sections \ref{sec:query understanding} - \ref{sec:presentation}. On this basis, we proceeded to a comprehensive analysis of the 57 papers to systematically understand the main research trends. We also collected 283 papers of related works described in Section \ref{sec:scope}. Table \ref{tab:related works} lists representative papers of each topic for reference, which will be selectively discussed with V-NLI characteristics in the subsequent sections. More information can be found at https://v-nlis.github.io/V-NLIs-survey/.

	\vspace{-0.3cm}
	\section{Classification Overview}\label{sec:classification}
	The information visualization pipeline presented by Card \textit{et al.}\cite{Card1999} (see Figure \ref{fig:vispipeline}) describes how the raw data transits into visualizations and interacts with the user.
	We extend this pipeline with an additional V-NLI layer (Green colored in Figure \ref{fig:vispipeline}). 
	On this basis, we move forward to develop categorical dimensions by focusing on how V-NLI facilitates visualization generation.
	Inspired by McNabb \textit{et al.}\cite{Mcnabb2017}, to facilitate the categorization process, the following stages in the pipeline are used:
	
	\begin{itemize}[leftmargin=*] 
		\item \textbf{Query Interpretation (A):}
		Since we add the V-NLI layer in the pipeline, query interpretation is a foundational stage. 
		Semantic and syntax analysis is generally performed first to discover hierarchical structures of the NL queries so that the system can parse relevant information in terms of data attributes and analytic tasks.
		Due to the vague nature of natural language, dealing with the underspecified utterances is another essential task in this stage.
		Details can be found in Section \ref{sec:query understanding}.
		
		\item \textbf{Data Transformation (B):}
		In the original pipeline, this stage mainly plays a role in transforming raw data into data tables along with various operations (e.g., aggregation and pivot). Since the majority of raw data to analyze is already in a tabular format, we rename it transformed data. 
		Data transformation is responsible for generating alternative data subsets or derivations for visualization. All operations on the data plane belong to this stage.
		Details can be found in Section \ref{sec:data transformation}.
		
		\item \textbf{Visual Mapping (C):}
		This stage focuses on mapping the information extracted from NL queries to visual structures.
		The three elements of visual mapping for information visualization are spatial substrate, graphical elements, and graphical properties\cite{Card1999}. Spatial substrate is the space to create visualizations, and it is essential to consider the layout configuration. Graphical elements are the visual elements appearing in the spatial substrate (e.g., points, lines, surfaces, and volumes). Graphical properties can be implemented on the graphical elements to make them more noticeable (e.g., size, orientation, color, textures, and shapes). Details can be found in Section \ref{sec:visual mapping}.
		
		\item \textbf{View Transformation (D):}
		View transformation (rendering) transforms visual structures into views by specifying graphical properties that turn these structures into pixels. Common forms include navigation, animation, and visual distortion (e.g., fisheye lens). 
		However, in the context of our survey, this stage is rarely involved in V-NLI. 
		Details can be found in Section \ref{sec:view transformation}.
		
		\item \textbf{Human Interaction (E):}
		Human interactions with the visualization interface feed back into the pipeline. A user can connect with a visualization manually by modifying or transforming a view state, or by reviewing the use, effectiveness, and knowledge on the visualization.
		Dimara \textit{et al.}\cite{Dimara2020} defined \textit{interaction for visualization} as: \textit{``The interplay between a person and a data interface involving a data-related intent, at least one action from the person and an interface reaction that is perceived as such.''} Describing interaction requires all the mandatory components: interplay, user, and interface.
		Details can be found in Section \ref{sec:human interaction}.
		
		\item \textbf{Dialogue Management} (F):
		Interpreting an utterance contextually is essential for visualization system intelligence, which is particularly evident in V-NLI. This stage involves each step in the pipeline and concentrates on facilitating a conversation with the system based on the current visualization state and previous utterances. Details can be found in Section \ref{sec:context management}.
		
		\item \textbf{Presentation (G):}
		We name this stage as \textit{Presentation}. The classic pipeline focuses on how to generate visualizations but ignores presenting them to the user. With natural language integrated into the pipeline, the vast majority of V-NLI systems accept natural language as input and directly display generated visualizations. Furthermore, complementing visualizations with natural language can provide additional surprises to the user.
		Details can be found in Section \ref{sec:presentation}.
		
	\end{itemize}
	
	The pipeline is meant to model work activities. A single utterance can cover multiple stages (e.g., data transformation, visual mapping, and human interaction), and users can iterate over any of these stages. An utterance can also be interpreted contextually. 
	The following seven sections will discuss the seven stages accordingly.

	\begin{table*}[t]
		\setlength{\abovecaptionskip}{0cm}
		\setlength{\belowcaptionskip}{-0.1cm}
		\caption{Comparison of commonly used NLP Toolkits}
		\label{tab:NLPtoolkits}
		\centering
		\small
		\setlength{\tabcolsep}{1.8mm}{
			\renewcommand\arraystretch{0.9}
			\rowcolors{2}{gray!25}{white}
			\begin{tabular}{cccccccc}
				\textbf{Toolkit} & \textbf{CoreNLP\cite{Manning2014}}  & \textbf{NLTK\cite{Bird2006}} & \textbf{OpenNLP\cite{opennlp}} & \textbf{SpaCy\cite{Honnibal2017}} & \textbf{Stanza\cite{stanza}} & \textbf{Flair\cite{Flair}} & \textbf{GoogleNLP\cite{googlenlp}} \\
				\midrule[1.3pt]
				Programming language & Java & Python & Java & Python & Python & Python & Python\\
				Tokenization & \checkmark & \checkmark & \checkmark & \checkmark & \checkmark & \checkmark & \checkmark \\
				Sentence segmentation & \checkmark & \checkmark & \checkmark & \checkmark & \checkmark & \checkmark & \checkmark \\
				Part-of-speech tagging & \checkmark & \checkmark & \checkmark & \checkmark & \checkmark & \checkmark & \checkmark \\ 
				Parsing & \checkmark & \checkmark & \checkmark & \checkmark & \checkmark & \checkmark & \checkmark  \\
				Lemmatization & \checkmark & \checkmark & \checkmark & \checkmark & \checkmark & \checkmark & \checkmark \\
				Named entities recognition & \checkmark & \checkmark & \checkmark & \checkmark & \checkmark & \checkmark & \checkmark \\
				Coreference resolution & \checkmark & \checkmark & \checkmark &-& \checkmark &-& - \\
				Entity linking & \checkmark & \checkmark &-& \checkmark & \checkmark &-& \checkmark \\
				Chunker &-&-& \checkmark & \checkmark &-&-& - \\
				Sentiment & \checkmark & - & - & - & \checkmark & - & \checkmark \\
				Text classification & - & \checkmark & - & \checkmark & - & \checkmark & \checkmark \\
				Train custom model & - & - & \checkmark & \checkmark & \checkmark & \checkmark & \checkmark \\
				\bottomrule[1.3pt]
		\end{tabular}}
		\vspace{-0.5cm}
	\end{table*}
	
	\vspace{-0.3cm}
	\section{Query Interpretation}\label{sec:query understanding}
	Query interpretation is the foundation of all subsequent stages. This section will discuss how to perform semantic and syntax analysis of the input natural language queries, infer analytic tasks of the user and data attributes to be analyzed, and make defaults for underspecified utterances.
	
	\vspace{-0.3cm}
	\subsection{Semantic and Syntax Analysis}\label{sec:Semantic and Syntax Analysis}
\begin{figure}[t]
		\setlength{\abovecaptionskip}{0cm}
		\setlength{\belowcaptionskip}{-0.1cm}
		\centering
		\includegraphics[width=\columnwidth]{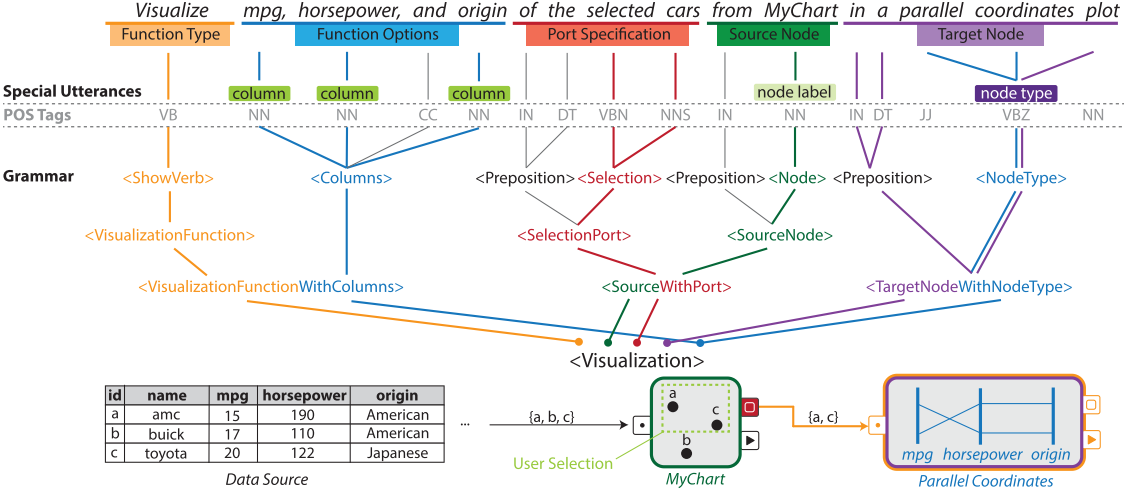}
		\caption{Semantics parsing in FlowSense\cite{Yu2020b}. The derivation of the query is shown as a parse tree.}
		\label{fig:FlowSense}
		\vspace{-0.4cm}
	\end{figure}

	Semantic and syntax analysis can be powerful to discover hierarchical structures and understand meanings in human language. 
	The semantic parser can conduct a series of NLP sub-tasks on the query string to extract valuable details that can be used to detect relevant phrases. The processing steps include tokenization, identifying parts-of-speech (POS) tags, recognizing name entities, removing stop words, performing stemming, creating a dependency tree, generating N-grams, etc. 
	For example, Flowsense\cite{Yu2020b} is a natural language interface designed for a dataflow visualization system\cite{Yu2017}. It applies a semantic parser with special utterances (table column names, node labels, node types, and dataset names) tagging and placeholders. Figure \ref{fig:FlowSense} displays a parse tree for the derivation of the user's query. The five major components of a query pattern and its related parts are highlighted by a unique color, and the expanded sub-diagram is illustrated at the bottom. 
	Flowsense is powered by the Stanford SEMPRE framework\cite{Zhang2017d} and CoreNLP toolkit\cite{Manning2014}. 
	Advanced NLP toolkits\cite{Manning2014,Bird2006,opennlp,Honnibal2017,stanza,Flair,googlenlp} allow developers to quickly integrate NLP services into their systems. As semantic and syntax analysis is a fundamental task for V-NLI, almost all systems support semantic parsing by directly using existing NLP toolkits, such as CoreNLP\cite{Manning2014}, NLTK\cite{Bird2006}, OpenNLP\cite{opennlp}, SpaCy\cite{Honnibal2017}, Stanza\cite{stanza}, Flair\cite{Flair}, and GoogleNLP\cite{googlenlp}, as listed in Table \ref{tab:vis_papers} (column \textit{NLP Toolkit or Technology}). We also sort out commonly used characteristics of existing NLP toolkits in Table \ref{tab:NLPtoolkits} for reference. 
	In addition, recently, there have been some systems that do not rely on these tools but leverage Language Models (LMs) to directly ``interpret" queries\cite{Liu2021b,Luo2021a}. They first generate a rich representation of the input by translating them into high-dimensional vectors and then adopt neural networks to enable smart visualization inference.
	
	\textbf{Discussion:}
	Most V-NLI systems rely on existing NLP toolkits for semantic and syntax analysis. So far, these tools are a good choice for query parsing and can be easily integrated into the system. However, they are trained on NLP datasets, thus lacking adequate consideration of visualization elements (e.g., mark, visual channel, and encoding property). A promising solution may be to develop a new NLP toolkit specifically for visualization.
	
	
	\begin{table*}[h]
		\setlength{\abovecaptionskip}{0cm}
		\setlength{\belowcaptionskip}{-0.1cm}
		\footnotesize
		\caption{ Ten low-level analytic tasks proposed by Amar \textit{et al.}\cite{Amar2005b}. They have been widely applied for automatic visualization creation.}
		\label{tab:task}
		\centering
		\setlength{\tabcolsep}{0.4mm}{
			\renewcommand\arraystretch{1}
			\begin{tabular}{m{2.9cm}m{8.7cm}m{6.2cm}}
				\hline
				\rowcolor[HTML]{EFEFEF} 
				\makecell[c]{\textbf{Task}} & \makecell[c]{\textbf{Description}}& \makecell[c]{\textbf{Representative papers}} \\ \hline		
				Characterize Distribution &  Given a set of data cases and a quantitative attribute of interest, characterize the distribution of that attribute’s values.& 
				\cite{Amar2005b,Saket2019b,Narechania2020,Sarikaya2018b,Wang2020h,Srinivasan2019a,Liu2021b,Fu2018,Yu2020b} \\ \hline
				
				Cluster & Given a dataset, find clusters of similar attribute values.&
				\cite{Amar2005b,Saket2019b,Shi2019,Srinivasan2019a,Fu2018,Yu2020b} \\ \hline
				
				Compute Derived Value & Given a dataset, compute an aggregate numeric representation of the data.&
				\cite{Amar2005b,Saket2019b,Wang2020h,Hu2018,Srinivasan2019a,Liu2021b,Fu2018,Yu2020b} \\ \hline
				
				Correlate & Given a dataset and two attributes, determine useful relationships between the values of those attributes.& \cite{Amar2005b,Saket2019b,Narechania2020,Sarikaya2018b,Shi2019,Wang2020h,Hu2018,Srinivasan2019a,Liu2021b,Fu2018,Yu2020b} \\ \hline
				
				Determine Range &  Given a dataset and a data attribute, find the span of values within the set.&
				\cite{Amar2005b,Saket2019b,Srinivasan2019a,Fu2018,Yu2020b} \\ \hline
				
				Filter & Find data cases satisfying the given concrete conditions on attribute values.&
				\cite{Amar2005b,Saket2019b,Narechania2020,Wang2020h,Srinivasan2019a,Fu2018,Yu2020b} \\ \hline
				
				Find Anomalies & Identify any anomalies within a given set of data cases with respect to a given relationship or expectation, e.g. statistical outliers.&
				\cite{Amar2005b,Saket2019b,Sarikaya2018b,Shi2019,Wang2020h,Hu2019e,Srinivasan2019a,Fu2018,Yu2020b} \\ \hline
				
				Find Extremum & Find data cases possessing an extreme value of an attribute over its range.&
				\cite{Amar2005b,Saket2019b,Kim2018b,Wang2020h,Hu2019e,Srinivasan2019a,Liu2021b,Fu2018,Yu2020b} \\ \hline		
				
				Retrieve Value & Given a set of specific cases, find attributes of those cases.&
				\cite{Amar2005b,Saket2019b,Narechania2020,Kim2018b,Wang2020h,Hu2019e,Srinivasan2019a,Fu2018,Yu2020b} \\ \hline
				
				Sort & Given a dataset, rank them according to some ordinal metric.&
				\cite{Amar2005b,Saket2019b,Wang2020h,Srinivasan2019a,Fu2018,Yu2020b} \\ \hline
				
				%
				%
				%
				%
				%
		\end{tabular}}
		\vspace{-0.6cm}
	\end{table*}
	
	\vspace{-0.2cm}
	\subsection{Task Inference}
	\subsubsection{Task modeling}
	\vspace{-0.1cm}
	A growing body of literature recognizes that the user's analytic task is vital for automatic visualization creation \cite{Brehmer2013,Rind2016,Kerracher2017,Kim2018b,Saket2019b}.
	As for task modeling, there have been many prior efforts to provide definitions of the analytic tasks. 
	For instance, Amar \textit{et al.}\cite{Amar2005b} proposed ten low-level analytic tasks that capture the user's activities while employing visualization tools for data exploration. The ten tasks are later extensively applied in numerous visualization systems\cite{Kim2018b,Saket2019b,Wang2020h,Moritz2019,Lin2020a,VandenElzen2013b,Shi2019,Srinivasan2019a,Jiang2021a}, as listed in Table \ref{tab:task}.
	Saket \textit{et al.}\cite{Saket2019b} evaluated the effectiveness of five commonly used chart types across the ten tasks\cite{Amar2005b} by a crowdsourced experiment. Furthermore, they derived chart type recommendations for different tasks.
	Kim \textit{et al.}\cite{Kim2018b} measured subject performance across task types derived from the ten tasks\cite{Amar2005b} and added a \textit{compare values} task. The analytic tasks are further grouped into two categories: \textit{value} tasks that just retrieve or compare individual values and \textit{summary} tasks that require identification or comparison of aggregate properties. 
	NL4DV\cite{Narechania2020} includes a \textit{Trend} task in addition.
	AutoBrief\cite{Kerpedjiev1997} enhances visualization systems by introducing domain-level tasks, while Matthew \textit{et al.}\cite{Brehmer2013} contributed a multi-level typology of visualization tasks.
	Deep into scatter charts, Sarikaya \textit{et al.}\cite{Sarikaya2018b} collected model tasks from a variety of sources in data visualization literature to formulate the seed for a scatterplot-specific analytic task list.
	Recently, Shen \textit{et al.}\cite{Shen2021} summarized 18 classical analytic tasks by a survey covering both academia and industry. 
	
	\vspace{-0.3cm}
	\subsubsection{Intent inference}\label{sec:task inference}
	Although considerable previous works focus on task modeling, few visualization systems have attempted to infer the user's analytic task before the emergence of natural language interfaces.
	Gotz and Wen\cite{Gotz2009b} monitored the user's click interactions for implicit signals of user intent. 
	Steichen \textit{et al.}\cite{Steichen2013} and Gingerich \textit{et al.}\cite{Gingerich2015} used the eye-gazing patterns of users while interacting with a given visualization to predict the user’s analytic task.
	Battle \textit{et al.}\cite{Battle2020b} investigated the relationship between latency, task complexity, and user performance. However, these behavior-based systems are limited to few pre-defined tasks; generalization for automatic visualization systems does not exist yet.
	
	Rather than inferring the analytic task through the user's behavior, systems supporting NL interaction depend on understanding the NL utterances to analyze the user's intent since they may hint at the user's analysis goals. 
	Most systems infer the analytic tasks by comparing the query tokens to a predefined list of task keywords\cite{Sun2010,Gao2015b,Hoque2018,Setlur2016,Yu2020b,Narechania2020}. 
	For example, NL4DV\cite{Narechania2020} identifies five low-level analytic tasks: \textit{Correlation}, \textit{Distribution}, \textit{Derived Value}, \textit{Trend}, and \textit{Filter}. A task keyword list is integrated internally (e.g., \textit{Correlation} task includes ‘correlate,’ ‘relationship,’ etc., \textit{Distribution} task includes ‘range,’ ‘spread,’ etc.). NL4DV also leverages POS tags and query parsing results to model the relationship between query phrases and populate task details. 
	For conversational analysis systems, Nicky\cite{Kincaid2017} interprets task information in conversations based on a domain-specific ontology. Evizeon\cite{Hoque2018} and Orko\cite{Srinivasan2018} support follow-up queries through conversational centering\cite{Grosz1986}, which is a model commonly used in linguistic conversational structure but only for the \textit{filter} task.
	ADE\cite{Cook2015} supports the analytic process via task recommendation invoked by inferences about user interactions in mixed-initiative environments. 
	The recommendation is also accompanied by a natural language explanation.
	Instead of simply matching keywords, Fu \textit{et al.}\cite{Fu2018} maintained a dataset of NL queries for visual data analytics and trained a multi-label task classification model based on BERT\cite{Devlin2019}, an advanced pre-trained NLP model.
	
	\textbf{Discussion:}
	Although most V-NLI systems support task inference, as shown in Table \ref{tab:vis_papers} (column \textit{Task Inference}), the tasks integrated in the system are limited (e.g., a subset of ten low-level tasks\cite{Amar2005b}). More tasks with hierarchical modeling can be considered to better cover the user's intent, as well as tasks for analyzing specific data types (e.g., tree\cite{Streeb2021}, map\cite{Duncan2021}, and graph\cite{Lee2006}). Besides, rule-based approaches account for the majority; various advanced learning models provide a great opportunity to infer analytic tasks in an unbiased and rigorous manner.
	
	\vspace{-0.3cm}
	\subsection{Data Attributes Inference} \label{sec:Data Attributes Inference}
	A dataset contains numerous data attributes. However, the user may only be interested in several certain data attributes in a query. The systems should be able to extract data attributes that are mentioned both explicitly (e.g., directly refer to attribute names) and implicitly (e.g., refer to an attribute’s values or alias). Illustrated by an example, NL4DV\cite{Narechania2020} maintains an alias map and allows developers to configure aliases (e.g., \textit{GDP} for \textit{Gross domestic product} and \textit{investment} for \textit{production budget}). It iterates through the generated N-grams discussed in Section \ref{sec:Semantic and Syntax Analysis}, checking for both syntactic and semantic similarity between N-grams and a lexicon composed of data attributes, aliases, and values. For syntactic similarity, NL4DV adopts cosine similarity function; for semantic similarity, it computes the Wu-Palmer similarity score based on WordNet\cite{Miller1995}. If the similarity score reaches the threshold, the corresponding data attributes will be extracted and further presented on the visualization. 
	Most V-NLI systems have taken a similar approach, and they just differ in the integrated rules. 
	Besides, Analyza\cite{Dhamdhere2017} utilizes additional heuristics to derive information from data attributes and expands the lexicon with a proprietary knowledge graph. 
	Recently, Liu \textit{et al.}\cite{Liu2021b} proposed a deep learning-based pipeline, ADVISor. It uses BERT\cite{Devlin2019} to generate embeddings of both NL queries and table headers, which are further used by a deep neural networks model to decide data attributes, filter conditions, and aggregation type. Similarly, Luo \textit{et al.}\cite{Luo2021a} presented an end-to-end solution using a transformer-based\cite{Furfaritony2002} sequence-to-sequence (seq2seq) model\cite{Bahdanau2015}.
	In the flow of visual analytical conversations, data attributes can also be extracted through co-reference resolution (Section \ref{sec:Co-reference resolution}).
	
	\textbf{Discussion:}
	During the survey and evaluation of state-of-the-art systems, we found that most V-NLIs are primarily geared to support specification-oriented queries like examples in their manual (e.g., ``Show me acceleration and horsepower across origins"), from which it is easy to extract data attributes. However, when it comes to underspecified or vague queries (e.g., include synonym, abbreviation, and terminology in different fields), these systems often fail to generate appropriate visualizations. So improving the robustness of V-NLIs should be one of the focuses in future work.

	\vspace{-0.3cm}
	\subsection{Default for Underspecified Utterances} \label{sec:Defaults for underspecified utterances}
\begin{figure}[t]
		\setlength{\abovecaptionskip}{0cm}
		\setlength{\belowcaptionskip}{-0.1cm}
		\centering
		\includegraphics[width=0.9\columnwidth]{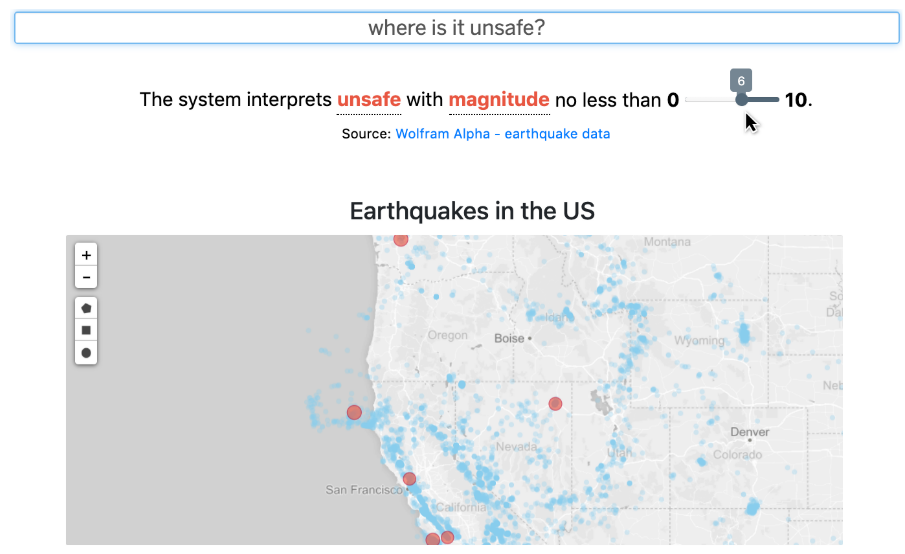}
		\caption{Vague modifier interpretation in Sentifiers\cite{Gao2015b}. The system associates the vague modifier ``unsafe'' with \textit{magnitude} attribute.}
		\label{fig:Sentifiers}
		\vspace{-0.5cm}
	\end{figure}

	Numerous systems support the \textit{filter} task\cite{Setlur2016,Gao2015b,Narechania2020,Yu2020b,Hoque2018}, aiming to select data attributes and ranges.
	An input query will be underspecified if it lacks enough information for the system to infer data attributes and perform filtering. Presenting defaults in the interface by inferring the underspecified utterances can be an effective option to address this issue. 
	However, the area has received relatively little attention, as shown in Table \ref{tab:vis_papers} (column \textit{Default for Underspecified Utterances}). Within the research, vague modifiers like ``tall'' and ``cheap'' are a prevalent kind of underspecification in human language. Hearst \textit{et al.}\cite{Hearst2019} made the first step toward design guidelines for dealing with vague modifiers based on existing cognitive linguistics research and a crowdsourcing study. 
	Tableau's Ask Data\cite{askdata} internally leverages lightweight and intermediate language, Arklang\cite{Setlur2019}, to infer underspecified details. It emphasizes how the linguistic context of previous utterances affects the interpretation of a new utterance. 
	Sentifiers\cite{Setlur2020a} can determine which data attributes and filter ranges to associate the vague predicates using word co-occurrence and sentiment polarities. As shown in Figure \ref{fig:Sentifiers}, when analyzing the earthquakes dataset, the user inputs ``where is it unsafe'' and Sentifiers automatically associates ``unsafe'' with the \textit{magnitude} attribute. A top N filter of \textit{magnitude} six and higher is applied, and similar negative sentiment polarities are marked red on the map. 
	
	\textbf{Discussion:}
	Although the aforementioned approaches are useful, some more complex interpretations are still not supported in current V-NLIs, such as combinations of vague modifiers. So when encountering ambiguity, apart from formulating a sensible default, human interaction (e.g., ambiguity widgets) is another effective method (see Section \ref{sec:ambiguity widget}).
	In addition, users’ explicit judgments for the same utterance can differ significantly. 
	A more adaptative solution may be to personalize the inferencing logic by identifying individuals’ unique analytic tasks.
	
	
	\vspace{-0.3cm}
	\section{Data Transformation}\label{sec:data transformation}
	After extracting data attributes, various data transformation operations can be made to transform raw data into focused data. This section will introduce how V-NLI systems 
	transform raw data for data insights, with additional discussions on a closely related topic, NLI for databases.
	
	\vspace{-0.3cm}
	\subsection{Transformation for Data Insights}
	A consensus is that the purpose of visualization is insights, not pictures. 
	A ``boring'' dataset may become ``interesting" after data transformations (e.g., aggregation, binning, and grouping).
	Therefore, identifying the data transformation information is another important characteristic in V-NLI. To describe related works, we categorize tabular data into four types: temporal, quantitative, nominal, and ordinal, which are widely adopted in the visualization community\cite{Moritz2019}. For temporal data, systems can support binning based on temporal units\cite{Luo2018a,Setlur2016,Fulda2016,Murillo-Morales2020,Mitri2020,Yu2020b,Narechania2020}. For instance, Setlur \textit{et al.}\cite{Setlur2016} developed a temporal analytics module based on temporal entities and expressions defined in TimeML\cite{Ingria2003}. To parse temporal expressions, this module incorporates the following temporal token types: Temporal Units (e.g., years, months, days, hours and minutes, and their abbreviations.), Temporal Prepositions (e.g., 'in,' 'during,' 'near,' and 'around.'), and Temporal Connectives (e.g., ‘before’ and ‘after’).
	For quantitative data, various aggregation functions (e.g., count, average, and sum) can be performed, binning and grouping operations are also commonly used\cite{Hoque2018,Cui2020b,Yu2020b,Gao2015b,Luo2018a,Dhamdhere2017,Setlur2016,Setlur2019,Srinivasan2019a,Oppermann,Kumar2016,Matsushita2004,Bacci2020,Zhou,Cook2015,Jiang2021a}. For example, DeepEye\cite{Luo2018a} includes various aggregation functions in the visualization language. Arklang\cite{Setlur2019}, which is an intermediate language to resolve NL utterances to formal queries, integrates a finite set of aggregation operators. Eviza\cite{Setlur2016} enables aggregation using regular spatial bins.
	Unlike continuous data, nominal and ordinal data are usually used as grouping metrics. 
	As an end-to-end system, ncNet\cite{Luo2021a} supports more complex data transformation types (e.g., relational join, GroupBY, and OrderBY) by applying Neural Machine Translation (NMT) models.
	After extracting relevant information, the commonly used visualization specification languages (e.g., Vega\cite{Satyanarayan2016}, Vega-Lite\cite{Satyanarayan2017}, and VizQL\cite{Stolte2002b}) all provide various data transformation primitives to realize.
	Besides, to our knowledge, most V-NLIs are designed to deal with tabular data, but there are also several systems that focus on specific data types (e.g., network\cite{Srinivasan2018} and map\cite{Setlur}). However, they mostly only display the raw data without data transformations.
	
	\textbf{Discussion:}
	The back-end engine can directly perform calculations and visualize the results if the data transformation information can be extracted from the user’s NL queries.
	Otherwise, despite advances in technologies for data management and analysis\cite{Battle2021,Ghosh2018}, it remains time-consuming to inspect a dataset and construct a visualization that allows meaningful analysis to begin. So a practical system should be in a position to automatically recommend data insights for users. 
	Fortunately, various systems have been designed to generate data facts in the context of visual data exploration, such as DataSite\cite{Cui2019}, Voder\cite{Srinivasan2019a}, Wrangler\cite{Kandel2011}, Quickinsights\cite{Ding2019}, Foresight\cite{Demiralp2017}, and SpotLight\cite{Harris}.
	They can serve as back-end engines for insight-driven recommendations.
	\vspace{-0.3cm}
	\subsection{NLI for Database}
	Natural Language Interface for Database (NLI4DB), or Text-to-SQL, is a task to automatically translate a user’s query text in natural language into an executable query for databases like SQL\cite{Affolter2019}. NLI4DB is strongly related to data transformation in V-NLI, and V-NLI can augment NLI4DB with effective visualizations of query results. 
	Besides, not all the queries about a dataset need a visualization as a response. For instance, one might ask, ``What was the GDP of the USA last year?'' or ``Which country won the most gold medals in the Tokyo Olympics?''. In such cases, NLI4DB can directly compute and present values in response to the user's questions instead of displaying through visualizations that require the user to interpret for answers. 
	Generally, there are three types of methods for NLI4DB: symbolic approach, statistical approach, and connectionist (neural network) approach\cite{Nihalani2011NaturalLI}. 
	The symbolic approach utilizes explicit linguistic rules and knowledge to process the user's query, which dominated the area in the early years. DISCERN\cite{miikkulainen:book93} is a representative example and integrates a model that extracts information from scripts, lexicon, and memory. The statistical approach exploits statistical learning methods such as hidden Markov models\cite{8029993} and probabilistic context-free grammars\cite{inproceedings}. From the view of the connectionist approach, the Text-to-SQL task is a kind of special Machine Translation (MT) problem. However, it is hard to be directly handled with standard end-to-end neural machine translation models due to the absence of detailed specification in the user's query\cite{guo2019complex}. Another problem is the  out-of-domain items mentioned in the user's query due to the user's unawareness of the ontology set. Aiming at the two problems, IRNet\cite{guo2019complex} proposes a tree-shaped intermediate representation synthesized from the user's query. It is later fed into an inference model to generate SQL statements with a domain knowledge base. Furthermore, to capture the special format of SQL statements, SQLnet\cite{xu2017} adopts a universal sketch as a template and predicts values for each slot. Wang \textit{et al.}\cite{9101534} employed a two-stage pipeline to subsequently predict the semantic structure and generate SQL statements with structure-enhanced query text. Recently, TaPas\cite{Herzig2020} extends BERT’s architecture and trains from weak supervision to answer questions over tables. For more details, please refer to the survey \cite{Azcan2020,Affolter2019} and related papers in Table \ref{tab:related works}.
	
	\textbf{Discussion:}
	V-NLI can augment NLI4DB with well-designed visualizations, and NLI4DB can complement V-NLI with exact value answers. 
	First, the two fields can draw on each other as they share similar principles. Second, we can combine the two technologies to generate visualizations that ``contain" answers. 
	Third, there are many benchmarks in the NLI4DB community like WikiSQL\cite{zhong2017seq2sql} and Spider\cite{Yu2020}, which can be utilized to further design V-NLI benchmarks.
	\vspace{-0.3cm}
	\section{Visual Mapping}\label{sec:visual mapping}
	This section will discuss how V-NLIs perform visual mapping in: spatial substrate, graphical elements, and graphical properties. 
	
	\vspace{-0.3cm}
	\subsection{Spatial Substrate}
	Spatial substrate is the space to create visualizations. It is important to determine the layout configurations to apply in the spatial substrate, such as which axes to use.
	Some V-NLI systems support explicitly specifying layout information like inputting ``show \textit{GDP} series at y axis and \textit{year} at x axis grouped by \textit{Country Code}''. If the mapping information is not clarified in the query, the search space will be very huge, in which some combinations of data attributes and visual encodings may not generate a valid visualization. For instance, the encoding type ``Y-axis'' is inappropriate for categorical attributes. 
	Fortunately, there are many design rules either from traditional wisdom or users to help prune meaningless visualizations. 
	These design rules are typically given by experts. Voyager\cite{Wongsuphasawat2016c,Wongsuphasawat2016,Wongsuphasawat2017}, DIVE\cite{Hu2018}, Show me\cite{Mackinlay2007b}, Polaris\cite{Stolte2002b}, Profiler\cite{Kandel2012b}, DeepEye\cite{Luo2018}, and Draco\cite{Moritz2019} all contribute to the enrichment of design rules with more data types. Besides, many systems\cite{Vartak2015b,Wongsuphasawat2017,Moritz2019} also allow users to specify their interested data attributes and assign them on certain axes, which is a more direct way to perform visual mapping. In Falx\cite{Wang2021a}, users can specify visualizations with examples of visual mapping, and the system automatically infers the visualization specification and transforms data to match the design. 
	With the development of machine learning in recent years, advanced models can be applied for more effective visual mapping. VizML\cite{Hu2019d} identifies five key design choices while creating visualizations, including mark type, X or Y column encoding, and whether X or Y is the single column represented along that axis or not. For a new dataset, 841 dataset-level features extracted are fed into neural network models to predict the design choices.
	Luo \textit{et al.}\cite{Luo2021a} proposed an end-to-end approach that applies the transformer-based seq2seq\cite{Bahdanau2015} model to directly map NL queries and data to chart templates.
	In addition to single chart, considering multiple visualizations, Evizeon\cite{Hoque2018} uses a grid-based layout algorithm to position new charts, while Voder\cite{Srinivasan2019a} allows the user to choose a slide or dashboard layout. 
	
	\textbf{Discussion:}
	When performing visual mappings in the spatial substrate, the vast majority of systems are limited to 2-dimension space (along x and y axes). However, considering more axes types, creating 3-dimensional (e.g., add z axes) and even hyper-dimensional (e.g., parallel coordinate) representations is possible.
	
	\begin{table}[t]
		\setlength{\abovecaptionskip}{0cm}
		\setlength{\belowcaptionskip}{-0.1cm}
		\caption{Mappings between data attribute combinations (N: Numeric, C: Categorical), analytic tasks, and chart types in Voder\cite{Srinivasan2019a}.}
		\label{tab:mark}
		\footnotesize
		\centering
		\setlength{\tabcolsep}{0.6mm}{
			\renewcommand\arraystretch{1}
			\begin{tabular}{p{1.4cm}p{4.9cm}p{2cm}}
				\hline
				\rowcolor[HTML]{EFEFEF} 
				\textbf{Attributes} & \textbf{Task(s)} & \textbf{Visualization} \\ \hline
				\multirow{6}{*}{N} & Find Extremum & Strip plot \\ \cline{2-3} 
				& \multirow{3}{*}{Characterize   Distribution} & Strip plot \\ \cline{3-3} 
				&& Box   plot \\ \cline{3-3} 
				&& Histogram \\ \cline{2-3} 
				& \multirow{2}{*}{Find Anomalies} & Strip plot \\ \cline{3-3} 
				&& Box   plot \\ \hline
				\multirow{2}{*}{C} & Find Extremum & \multirow{2}{*}{\begin{tabular}[c]{@{}l@{}}Bar chart \\ Domut chart\end{tabular}} \\ \cline{2-2}
				& Characterize Distribution &  \\ \hline
				\multirow{2}{*}{NxN} & Corelation & \multirow{2}{*}{Scatterplot} \\ \cline{2-2}
				& Characterize Distribution &  \\ \hline
				\multirow{3}{*}{CxN} & Characterize Distribution(+Derived Value) & \multirow{2}{*}{\begin{tabular}[c]{@{}l@{}}Bar chart \\ Domut chart\end{tabular}} \\ \cline{2-2}
				& Find Extremum(+Derived Value) &  \\ \cline{2-3} 
				& Find Extremum & \begin{tabular}[c]{@{}l@{}}Strip plot \\ Scatterplot\end{tabular} \\ \hline
				\multirow{2}{*}{CxC} & Find Extremum & \multirow{2}{*}{\begin{tabular}[c]{@{}l@{}}Stacked bar chart   \\ Scatterplot+Size\end{tabular}} \\ \cline{2-2}
				& Characterize Distribution(+Find   Extremum) &  \\ \hline
				NxNxN & Characterize Distribution & Scatterplot+Size \\ \hline
				\multirow{4}{*}{CxNxN} & Correlation & Scatterplot \\ \cline{2-3} 
				& Correlation(+Filter) & Scatterplot+Color \\ \cline{2-3} 
				& \multirow{2}{*}{Characterize   Distibution(+Filter)} & Scatterplot+Color \\ \cline{3-3} 
				&& Scatterplot+Size \\ \hline
				\multirow{3}{*}{CxCxN} & Find Extremum & \begin{tabular}[c]{@{}l@{}}Strip plot+Color   \\ Scatterplot+Color\end{tabular} \\ \cline{2-3} 
				& \multirow{2}{*}{Find Extremum(+Derived Value)} & \begin{tabular}[c]{@{}l@{}}Strip plot+Color   \\ Scatterplot+Color\end{tabular} \\ \cline{3-3} 
				&& Scatterplot+Size \\ \hline
		\end{tabular}}
		\vspace{-0.5cm}
	\end{table}
	
	\vspace{-0.3cm}
	\subsection{Graphical Elements}
	Graphical element (e.g., points, lines, surfaces, and volumes), which is usually named mark or chart type, is an important part of visualizations. Choosing an appropriate mark can convey information more efficiently. Similar to deciding the layout, some V-NLI systems allow users to specify marks like inputting ``create a scatterplot of \textit{mpg} and \textit{cylinders}'', which is a simple way to map mark information of visualizations. However, in most cases, the mark information is not accessible. So after parsing the NL queries, most systems integrate predefined rules or leverage Visualization Recommendation (VisRec) technologies to deal with the extracted elements.
	Voyager\cite{Wongsuphasawat2016c,Wongsuphasawat2017} applies visualization design best practices drawn from prior research\cite{Mackinlay1986b,Roth1994b,Seo2005b,Mackinlay2007b} and recommends mark types based on the data types of the x and y channels.
	Srinivasan \textit{et al.}\cite{Srinivasan2019a} developed the heuristics and mappings between data attribute combinations, analytic tasks, and chart types in Table \ref{tab:mark} through iterative informal feedback from fellow researchers and students.
	TaskVis\cite{Shen2021} integrates 18 classical low-level analysis tasks with their appropriate chart types by a survey both in academia and industry.
	Instead of considering common visualization types, Wang \textit{et al.}\cite{Wang2018b} developed rules for automatic selection of line graphs or scatter plots to visualize trends in time series, while Sarikaya \textit{et al.}\cite{Sarikaya2018b} deepened into scatter charts.
	To better benefit from design guidance provided by empirical studies, Moritz \textit{et al.}\cite{Moritz2019} proposed a formal framework that models visualization design knowledge as a collection of answer set programming constraints.
	
	\textbf{Discussion:}
	V-NLIs mostly maintain an alias map for data attribute inference (see Section \ref{sec:Data Attributes Inference}) but often ignore mark inference. Therefore, though they can explicitly interpret mark information from the user's NL queries, they may fail when the information is implicitly expressed (e.g., ``point" instead of ``scatterplot"). Besides, as shown in Table \ref{tab:vis_papers} (column \textit{Visualization Type}), existing works focus more on traditional chart types, more chart types, and even interactive visualizations can be extended in future work.

	\vspace{-0.3cm}
	\subsection{Graphical Properties}
	Jacques Bertin first identified seven ``retinal'' graphical properties (visual encoding properties) in visualizations: position, size, value, texture, color, orientation, and shape. Some other types are later expanded in the community, such as ``gestalt'' properties (e.g., connectivity and grouping) and animation\cite{Tversky2002,Mackinlay1986b}. 
	For visual mapping in V-NLI, color, size, and shape are most commonly applied to graphical elements, making them more noticeable. The rule-based approach is also dominant here, and human perceptual effectiveness metrics are usually considered. For example, the cardinality of variables in the color/size/shape channel should not be too high (e.g., 100). Otherwise, the chart would be too messy to distinguish. The aforementioned works also developed various design rules for graphical properties, especially for color, size, and shape channels in legend \cite{Wongsuphasawat2016c,Wongsuphasawat2017,Hu2018,Mackinlay2007b,Stolte2002b,Kandel2012b,Luo2018,Moritz2019,Shen2021}. Besides, Text-to-Viz\cite{Cui2020b} integrates a color module that aims to generate a set of color palettes for a specific infographic. Similarly, InfoColorizer\cite{Yuan2021} recommends color palettes for infographics in an interactive and dynamic manner. Wu \textit{et al.}\cite{Wu2021} proposed a learning-based method to automate layout parameter configurations, including orientation, bar bandwidth, max label length, etc. Liu \textit{et al.}\cite{LiuTingting} explored data-driven mark orientation for trend estimation in scatterplots. CAST\cite{Lee} and Data Animator\cite{Thompson2021} enable interactive creation of chart animations. 
	
	\textbf{Discussion:}
	For graphical properties, color, size, and shape account for the majority in current V-NLIs. Although some other forms have been investigated deeply, they are rarely used in V-NLI (e.g., orientation and texture). Besides, color is a tricky choice to use. On one hand, it is unfriendly to those who are color-blind. On the other hand, it may involve cultural issues in certain scenarios (e.g., red is a ``festive" color in China, whereas it may mean ``warning" in many Western societies). 
	So color needs to be used with the discretion of color vision deficiencies and cultural meaning.
	\begin{figure}[t]
		\setlength{\abovecaptionskip}{0cm}
		\setlength{\belowcaptionskip}{-0.1cm}
		\centering
		\includegraphics[width=0.9\columnwidth]{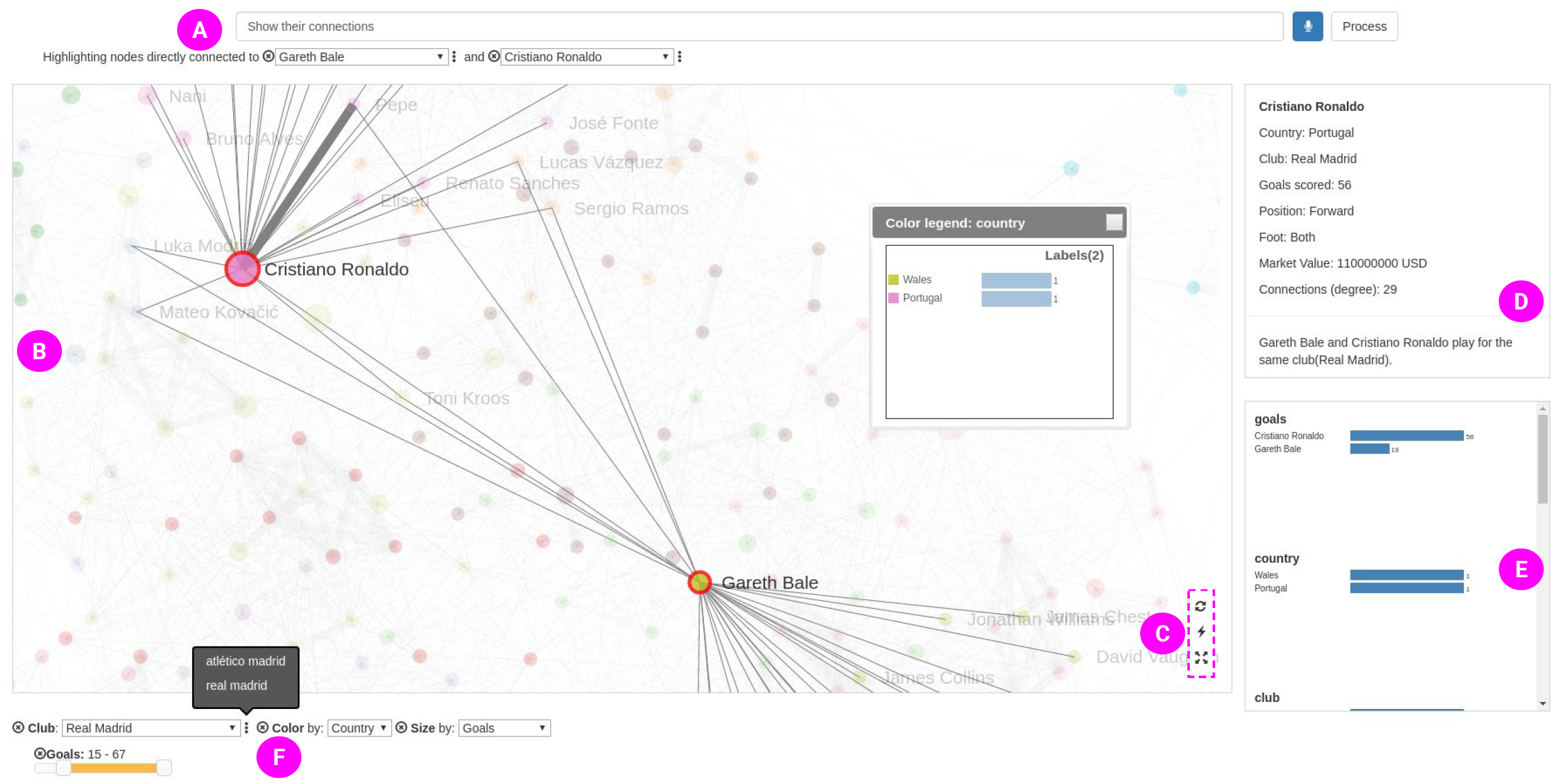}
		\caption{User interface of Orko\cite{Srinivasan2018} to explore a network of European soccer players.}
		\label{fig:Orko}
		\vspace{-0.5cm}
	\end{figure}

	\vspace{-0.3cm}
	\section{View Transformation}\label{sec:view transformation}
	After visual mapping, the generated visualization specifications can be rendered through a library (e.g., D3\cite{Bostock2011}). View transformations are also supported here.
	The three commonly used view transformation types are location probes which reveal additional information through locations, viewpoint controls that scale or translate a view, and distortions that modify the visual structure\cite{Card1999}. Surprisingly, this stage is rarely used in the context of our survey. To our knowledge, there have been few works that focus on natural language interaction for view transformation. It is usually realized through multimodal interaction, as shown in Table \ref{tab:vis_papers} (column \textit{View Transformation}). 
	For example, Orko\cite{Srinivasan2018} facilitates both natural language and direct manipulation input to explore network visualization. Figure \ref{fig:Orko} is the user interface of Orko, and it shows an example that explores a network of European soccer players. When the user says ``Show connection between Ronaldo and Bale'', the system will zoom in to display the details of nodes. It also allows users to perform view transformation (e.g., zooming and panning) through a pen or finger. Similarly, Valletto\cite{Kassel2018}, InChorus\cite{Srinivasan2020a}, Data@Hand\cite{Young-HoKim2021}, DataBreeze\cite{Srinivasan2020b}, Power BI's Q\&A\cite{powerbi}, and Tableau's Ask data\cite{askdata} all support standard view transformations.
	Besides, the software visualization community also presents several closely related works.
	Bieliauskas \textit{et al.}\cite{Bieliauskas2017} proposed an interaction approach with software visualizations based on a conversational interface. It can automatically display best fitting views according to meta information from natural language sentences. Seipel \textit{et al.}\cite{Seipel2019} explored natural language interaction for software visualization in Augmented Reality (AR) with Microsoft HoloLens device, which can provide various view transformation interactions through gesture, gaze, and speech.

	\textbf{Discussion:}
	View transformation has attracted less attention in the community. Only a few V-NLI systems support NL control over view transformations, and they are mainly limited to viewpoint navigation. Future improvements can focus on  other aspects (e.g., animation\cite{Thompson2021}, data-GIF\cite{Shu2020}, and visual distortions\cite{Nonato2019}).

	\vspace{-0.3cm}
	\section{Human Interaction}\label{sec:human interaction}
	This section will discuss how V-NLI allows users to provide feedback to visualization with human interactions. The primary purpose is to help users better express their intents. 
	

	\begin{figure}[t]
		\setlength{\abovecaptionskip}{0cm}
		\setlength{\belowcaptionskip}{-0.1cm}
		\centering
		\includegraphics[width=0.9\columnwidth]{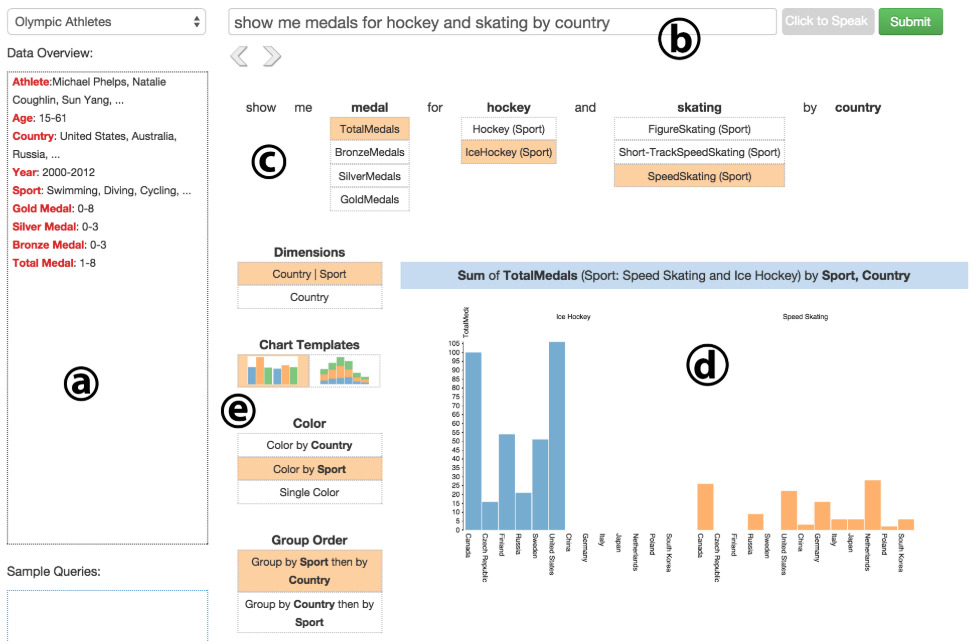}
		\caption{Ambiguity widgets in Datatone\cite{Gao2015b}. The user can correct the imprecise system decisions caused by ambiguity.}
		\label{fig:Datatone}
		\vspace{-0.5cm}
	\end{figure}
	
	\vspace{-0.3cm}
	\subsection{Ambiguity Widgets}\label{sec:ambiguity widget}
	Due to the vague nature of natural language, the system may fail to recognize the user intent and extract data attributes\cite{Metoyer2012}.
	Considerable research has been devoted to addressing the ambiguity and underspecification of NL queries. Mainstream approaches can be divided into two categories. One is to generate appropriate defaults by inferencing underspecified natural language utterances, discussed in Section \ref{sec:Defaults for underspecified utterances}. The other is to return the decision right to users through ambiguity widgets.
	Ambiguity widget is a kind of interactive widget that allows users to input through a mouse. Datatone\cite{Gao2015b} first integrates ambiguity widgets and presents a mixed-initiative approach to manage ambiguity in the user query. As shown in Figure \ref{fig:Datatone}, for the query in (b), DataTone detects data ambiguities about \textit{medal}, \textit{hockey}, and \textit{skating}. Three corresponding ambiguity widgets are presented for users to correct the system choices in (c). Additional design decision widgets like color and aggregation are also available in (e). Eviza\cite{Setlur2016}, Evizeon\cite{Hoque2018}, NL4DV\cite{Narechania2020}, and AUDiaL\cite{Murillo-Morales2020} all borrow the idea of DataTone and expand the ambiguity widgets to richer forms (see Table \ref{tab:vis_papers} (column \textit{Ambiguity Widgets})), such as maps and calendars. DIY\cite{Narechania2021} enables users to interactively assess the response from NLI4DB systems for correctness. 
	
	\textbf{Discussion:}
	Resolving underspecified queries through ambiguity widgets mainly consists of two steps: detecting ambiguity and presenting widgets to the user. For the first step, most systems now still leverage heuristics for algorithmic resolution of ambiguity. A general probabilistic framework may handle a broader range of ambiguities. For the second, richer interactive widgets can be adopted to enhance the user experience.
	
	\begin{figure}[t]
		\setlength{\abovecaptionskip}{0cm}
		\setlength{\belowcaptionskip}{-0.1cm}
		\centering
		\includegraphics[width=0.9\columnwidth]{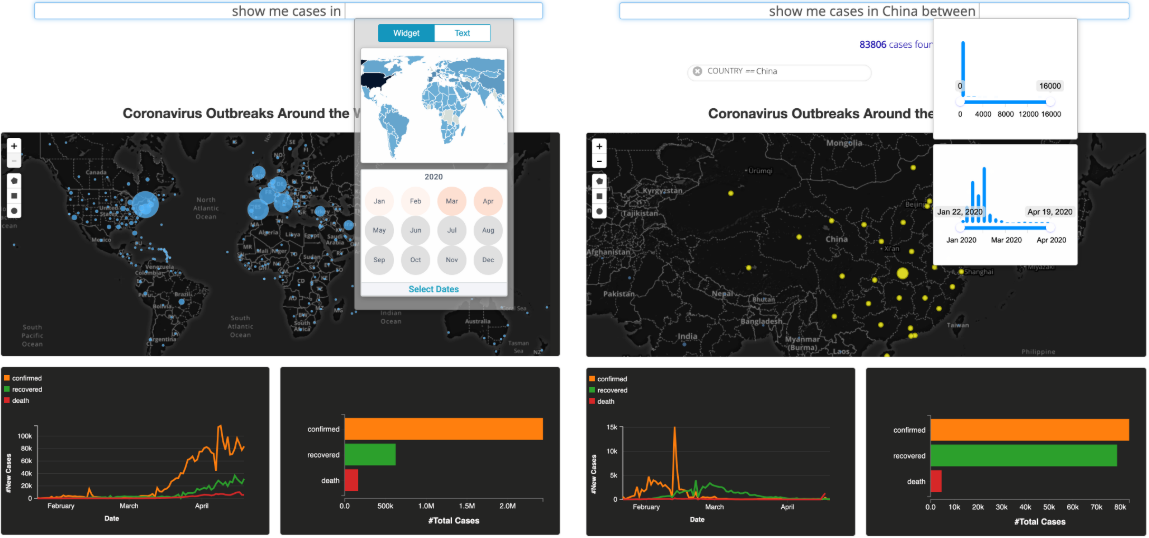}
		\caption{Autocompletion in Sneak pique\cite{Setlur2020}, The user is prompted with widgets that provide appropriate previews of the underlying data.}
		\label{fig:Sneak pique}
		\vspace{-0.4cm}
	\end{figure}

	\vspace{-0.3cm}
	\subsection{Autocompletion and Command Suggestion}\label{sec:Autocompletion}
	Users may be unaware of what operations the system can perform and whether a specific language structure is preferred in the system.
	Although advanced text understanding models give users the freedom to express their intents, system discoverability to help formulate analytical questions is still an indispensable part of V-NLI.
	Discoverability entails \textit{awareness}: making users aware of what operations the system can perform, and \textit{understanding}: educating users about how to phrase queries that can be interpreted correctly by the system\cite{Dontcheva}.
	Generally, current V-NLI systems pay relatively little attention to system discoverability. Major attempts include autocompletion and command suggestion.
	This characteristic offers interpretable hints or suggestions matching the visualizations and datasets, which is considered a fruitful interaction paradigm for information sense-making. 
	When using V-NLI of Tableau\cite{askdata} or Power BI\cite{powerbi}, the autocompleted content will be presented as we are typing, especially when there is a parsing error. They are mostly reminders of commonly used queries, such as data attributes and aggregation functions. Suggestions will not intrusively appear when the user has formulated a valid query. Similarly, Bacci \textit{et al.}\cite{Bacci2020} and Yu \textit{et al.}\cite{Yu2020b} both adopted template-based approaches to support autocompletion in their V-NLIs.
	In addition to text prompts, data previews can be more useful across all autocompletion variants.
	Deeper into this area, three crowdsourcing studies conducted by Setlur \textit{et al.}\cite{Setlur2020} indicated that users prefer widgets for previewing numerical, geospatial, and temporal data while textual autocompletion for hierarchical and categorical data. On this basis, they built a design probe, Sneak Pique. As shown in Figure \ref{fig:Sneak pique}, when analyzing a dataset of coronavirus cases, the user is prompted with widgets that provide appropriate previews of the underlying data in various types. The system also supports toggling from a widget to a corresponding text autocompletion dropdown.
	Most recently, Srinivasan \textit{et al.}\cite{Srinivasana} proposed Snowy, a prototype system that generates and recommends utterance recommendations for conversational visual analysis by suggesting data features while implicitly making users aware of which input the NLI supports.
	
	\textbf{Discussion:}
	Current utterance realization approaches are all template-based, which just works effectively for a small set of tasks and can hardly apply to large-scale systems.
	Going forward, another important consideration that needs further research is which commands to show and when and how the suggestions will be presented.
	In addition, existing technologies can only handle keyboard input.
	There remains an open area for the discoverability of speech-based V-NLI. 
		The tone of voice can also provide insights into the user’s sentiments\cite{Agosti2015}.

	\begin{figure}[t]
		\setlength{\abovecaptionskip}{0cm}
		\setlength{\belowcaptionskip}{-0.1cm}
		\centering
		\includegraphics[width=0.9\columnwidth]{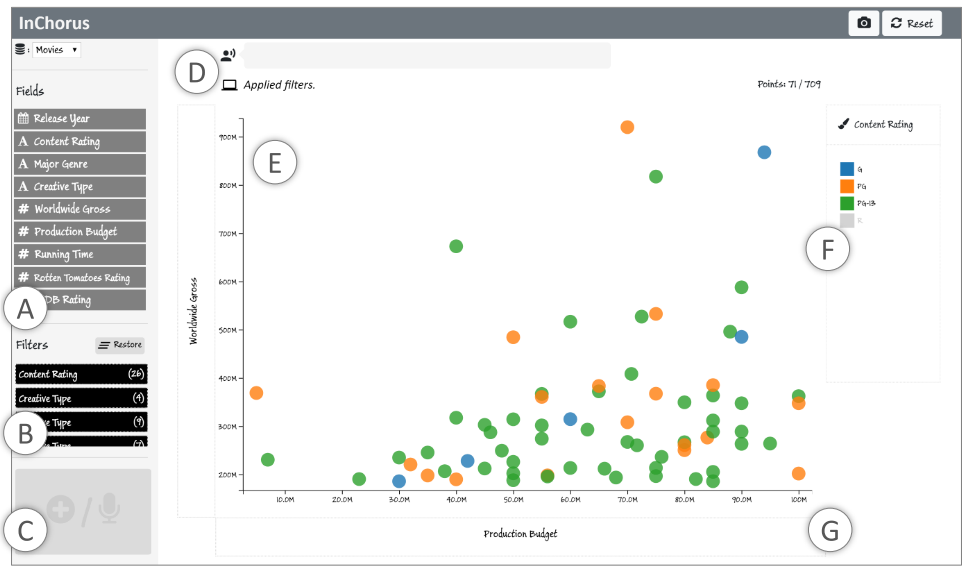}
		\caption{Multimodal interaction interface in InChorus\cite{Srinivasan2020a}, which allows pen, touch, and speech input on tablet devices.}
		\label{fig:InChorus}
		\vspace{-0.5cm}
	\end{figure}

	\vspace{-0.3cm}
	\subsection{Multimodal Interaction}
	Van Dam\cite{VanDam1997} envisioned post-WIMP user interfaces as ``one containing at least one interaction technique not dependent on classical 2D widgets such as menus and icons.'' With the advancements in hardware and software technologies that can be used to support novel interaction modalities, researchers are empowered to take a step closer to post-WIMP user interfaces, enabling users to focus more on their tasks\cite{Lee2021c}. 
	A qualitative user study conducted by Saktheeswaran \textit{et al.}\cite{Saktheeswaran2020} also found that participants strongly prefer multimodal input over unimodal input. 
	In recent years, many works have examined how multiple input forms (e.g., mouse, pen, touch, keyboard, and speech) can be combined to provide a more natural and engaging user experience. 
	For instance, 
	Orko\cite{Srinivasan2018} is a prototype visualization system that combines both natural language interface and direct manipulation to assist visual exploration and analysis of graph data. 
	Valletto\cite{Kassel2018} allows users to specify visualizations through a speech-based conversational interface, multitouch gestures, and a conventional GUI interface.
	InChorus\cite{Srinivasan2020a} is designed to maintain interaction consistency across different visualizations, and it supports pen, touch, and speech input on tablet devices. Its interface components are shown in Figure \ref{fig:InChorus}, including typical WIMP components (A, B, C), speech command display area (D), and content supporting pen and touch (E, F, G).
	DataBreeze\cite{Srinivasan2020b} couplings pen, touch, and speech-based multimodal interactions with flexible unit visualizations, enabling a novel data exploration experience. 
	RIA\cite{Zhou} is an intelligent multimodal conversation system to aid users in exploring large and complex datasets, which is powered by an optimization-based approach to visual context management.
	Data@Hand\cite{Young-HoKim2021} leverages the synergy of speech and touch input modalities for personal data exploration on the mobile phone.
	Srinivasan \textit{et al.}\cite{Dontcheva} proposed to leverage multimodal input to enhance discoverability and the human-computer interaction experience. 
	
	\textbf{Discussion:}
	V-NLI is an essential part in the context of post-WIMP interaction with visualization systems.
	Typing and speech are both commonly used modalities to input NL queries. However, speech has unique challenges from a system design aspect, such as triggering speech input, lack of assistive features like autocompletion, and transcription errors\cite{Lee2021c}, which deserve more attention in the future.
	Besides, limited research has incorporated human gesture recognition and tracking technology\cite{Jiang2021} to facilitate visualization creation, especially with large displays.

	\vspace{-0.3cm}
	\section{Dialogue Management}\label{sec:context management}
	This section will discuss dialogue management in V-NLI. Given the conversational nature of NLIs, users may frequently pose queries depending on their prior queries. By iterating upon their questions, the user can dive deep into their aspects of interest on a chart and refine existing visualizations.
	\begin{figure}[t]
		\setlength{\abovecaptionskip}{0cm}
		\setlength{\belowcaptionskip}{-0.1cm}
		\centering
		\includegraphics[width=\columnwidth,height=0.65\columnwidth]{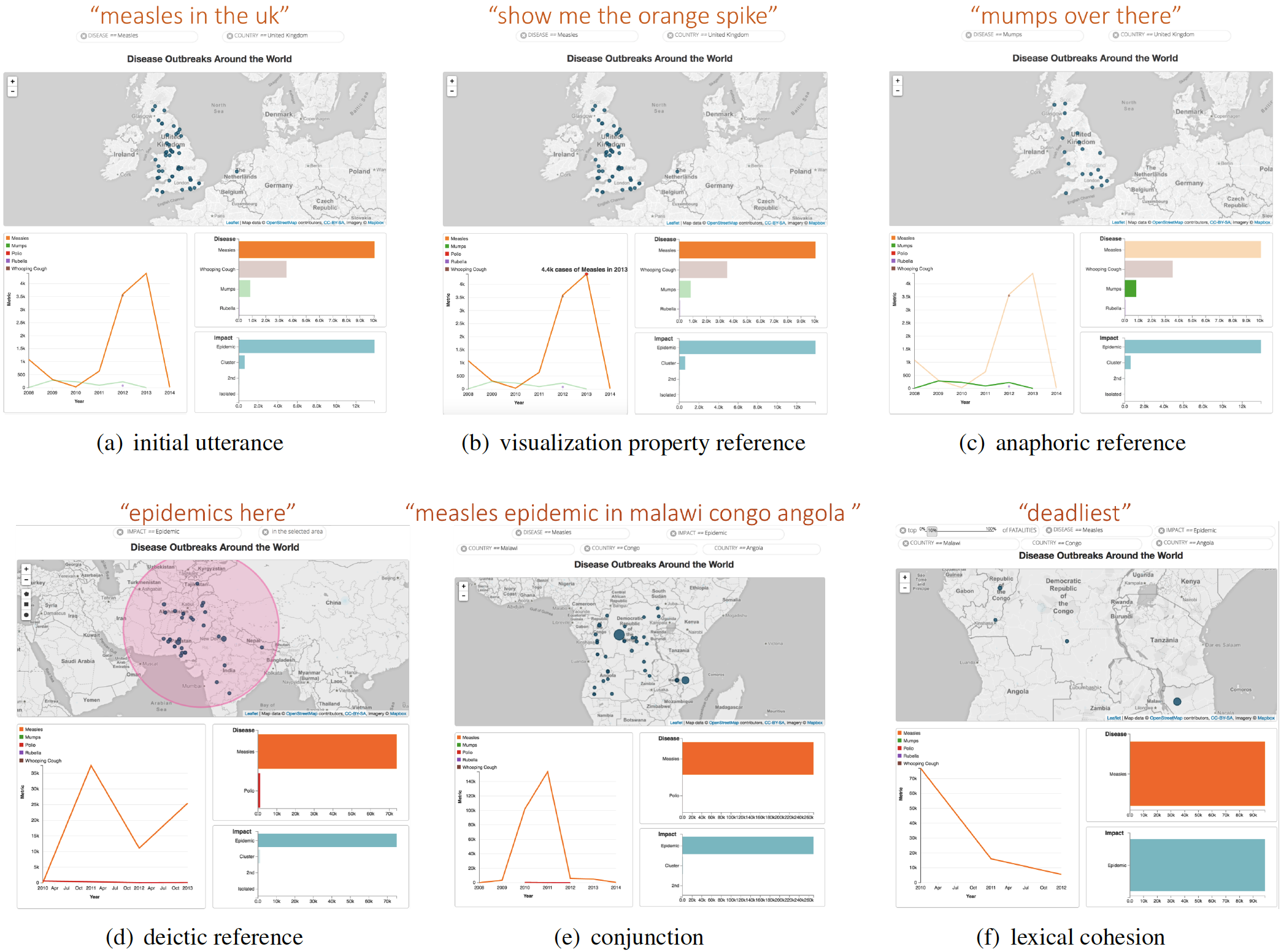}
		\caption{Conversational V-NLI in Evizeon\cite{Hoque2018}. The user has a back-and-forth information exchange with the system.}
		\label{fig:Evizeon}
		\vspace{-0.6cm}
	\end{figure}
	
	\vspace{-0.3cm}
	\subsection{Analytical Conversations}
	Conversional NLIs have been widely applied in many fields to prompt users to open-ended requests (e.g., recommender systems\cite{Condiff2017} and intelligent assistants\cite{Sigtia2020}). With conversational agents, users can also reinforce their understanding of how the system parses their queries. 
	In the visualization community, Eviza\cite{Setlur2016} and Evizeon\cite{Hoque2018} are two representative visualization systems that provide V-NLI for visual analysis cycles.
	Figure \ref{fig:Evizeon} shows a back-and-forth information exchange process between Evizeon\cite{Hoque2018} and the user when analyzing the measles dataset. The system supports various forms of NL interactions with a dashboard applying pragmatic principles. For example, when the user first types ``measles in the UK'', all the charts presented are filtered to cases of measles in the UK (Figure \ref{fig:Evizeon}(a)). Then, the user types ``show me the orange spike'', and Evizeon will add details to the spike in the orange line as it understands that this is a reference to visual properties of the line chart (Figure \ref{fig:Evizeon}(b)).
	Similarly, Articulate2\cite{Kumar2016} is intended to automatically transform the user queries into visualizations using a full-fledged conversational interface. 
	Analyza\cite{Dhamdhere2017} combines two-way conversation with a structured interface to enable effective data exploration. 
	Besides, Fast \textit{et al.}\cite{Fast2017} proposed a new conversational agent, Iris, which can generate visualizations from and combine the previous commands, even non-visualization related. Ava\cite{LeoJohn2017} uses controlled NL queries to program data science workflow.
	
	
	
\begin{figure}[t]
		\setlength{\abovecaptionskip}{0cm}
		\setlength{\belowcaptionskip}{-0.1cm}
		\centering
		\includegraphics[width=0.9\columnwidth]{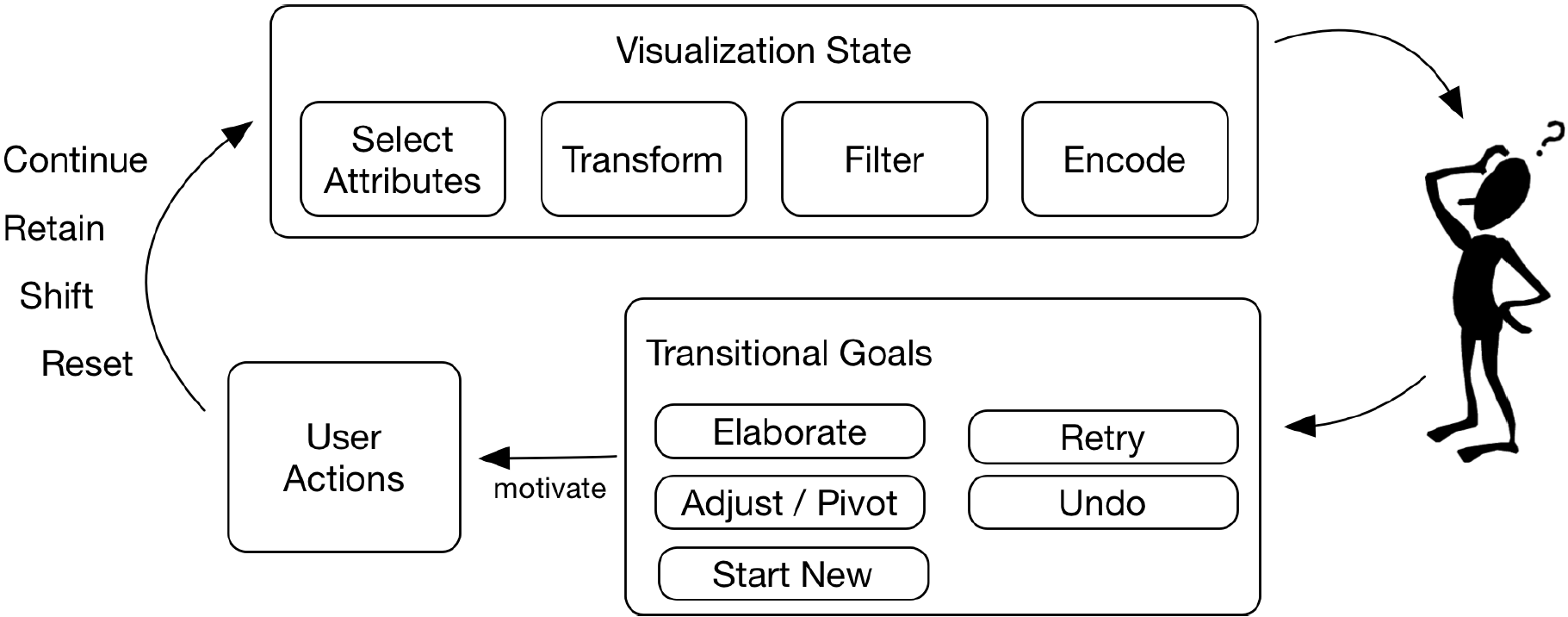}
		\caption{Conversational transitions model\cite{Tory2019b} that describes how to transition a visualization state during an analytical conversation.}
		\label{fig:Conversational}
		\vspace{-0.3cm}
	\end{figure}

	Conversational transitions model, which describes how to transition visualization states during the analytical conversation, is an essential part to enable the aforementioned interactive conversations. 
	In the early years, Grosz \textit{et al.}\cite{Grosz1986} explored how the context of a conversation adjusts over time to maintain coherence through transitional states (retain, shift, continue, and reset).
	On this basis, Tory and Setlur\cite{Tory2019b} introduced a conversational transitions model (see Figure \ref{fig:Conversational}) that emerged from their analysis in the design of Tableau's V-NLI feature, Ask data.
	After interpreting a generated visualization, a user may formulate a new NL query to continue the analytical conversation.
	The user’s transitional goal means how the user wishes to transform the existing visualization to answer a new question. The model contains the following transitional goals: elaborate, adjust/pivot, start new, retry, and undo, which in turn drive user actions. The visualization states (select attributes, transform, filter, and encode) are finally updated, and new visualizations are presented to the user.
	An essential insight during the analysis of Tory and Setlur was that applying transitions to filters alone (like Evizeon\cite{Hoque2018} and Orko\cite{Srinivasan2018}) is insufficient. The user's intent around transitions may apply to any aspects of a visualization state. 
	
	\textbf{Discussion:}
	Some V-NLI systems leverage a VisRec engine to generate visualizations. However, it is insufficient for analytical conversation. A more intelligent system should infer the user’s transitional goals based on their interactions and then respond to each visualization state accordingly.
	Besides, empirical studies\cite{Hearst2019a} have shown that users tend to prefer additional content beyond the exact answers when asking questions in such conversational interfaces, which had significant implications for designing insight-driven conversation interfaces in the future.

	\begin{table}
		\setlength{\abovecaptionskip}{0.cm}
		\setlength{\belowcaptionskip}{-0.1cm}
		\centering
		\scriptsize
		\caption{Examples of co-reference types}
		\label{tab:co-reference}
		\setlength{\tabcolsep}{0.9mm}{
			\renewcommand\arraystretch{1.1}
			\begin{tabular}{m{0.9cm}m{5.2cm}m{1.9cm}}
				\hline
				\rowcolor[HTML]{EFEFEF} 
				\makecell[c]{\textbf{Type}} & \makecell[c]{\textbf{Example}}& \makecell[c]{\textbf{Resolution}}\\\hline
				Pronoun-Noun & Show me the most popular \textbf{movie} in \textbf{New York} and \textbf{its} rating. & its $\rightarrow$ movie's\\ \hline
				Noun-Noun & How many \textbf{Wal-Mart stores} in \textbf{Seattle} and how's the annual profits for the \textbf{malls}? & malls $\rightarrow$ Wal-Mart\\\hline
				Pronoun-Pronoun & The \textbf{manchurian tigers} have been heavily hunted, which caused a dramatic drop of \textbf{their} existing number and \textbf{they} may finally get extinct. & \begin{tabular}[c]{@{}l@{}}they, their $\rightarrow$\\manchurian tigers\end{tabular}  \\\hline
				Pronoun-Pronoun & \textbf{We} are all the \textbf{Canadians} while \textbf{Lily} was born in the \textbf{U.S.} and \textbf{she} immigrated to \textbf{Canada} two years ago. & \begin{tabular}[c]{@{}l@{}}We $\rightarrow$ Canadians,\\she $\rightarrow$ Lily\end{tabular}  \\\hline
		\end{tabular}}
		\vspace{-0.4cm}
	\end{table}
	
	\vspace{-0.3cm}
	\subsection{Co-reference Resolution}\label{sec:Co-reference resolution}
	Co-reference Resolution (CR) aims to find linguistic expressions (called mentions) in a given text that refers to the same entity\cite{Stylianou2021}. CR is an important sub-task in dialogue management as the user usually uses pronouns to refer to certain visual elements. A typical scenario is to identify the entity to which a pronoun in the text refers. The task also involves recognizing co-reference relations between multiple noun phrases\cite{pradhan-etal-2012-conll}. Table \ref{tab:co-reference} shows examples of co-reference types. Before the booming development of deep learning-based language models, human-designed rules, knowledge, and features dominated CR tasks\cite{zhang2020brief}. Nowadays, most state-of-the-art CR models are neural networks and employ an end-to-end fashion, where the pipeline includes encoding context, generating representations for all potential mentions, and predicting co-reference relations\cite{lee2017endtoend}\cite{joshi2019bert}. Benefiting from the powerful understanding and predicting ability of language models like BERT\cite{Devlin2019} and GPT-3\cite{Brown}, reliance on annotated span mentions is becoming weaker\cite{kirstain2021coreference}. A CR task can even be treated as a token prediction task\cite{Kocijan_2019}. Several existing models exist to deal with this situation. Besides, the usage of structured knowledge bases\cite{zhang2019knowledgeaware}\cite{liu2016commonsense} and higher-order information\cite{lee2018higherorder} has been proven to be beneficial to the overall performance of CR tasks. 
	
	Co-reference resolution is also an essential task in the multimodal sense as users may even pose queries that are the follow-up to their direct manipulations on the interface.
	When the user faces a graphical representation, each element in the representation can become a referent. The system should have a declarative representation for all these potential referents and find the best match to make multimodal interaction more smooth. Articulate2\cite{Kumar2017} addresses this issue by leveraging Kinect to detect deictic gestures on the virtual touch screen in front of a large display.
	If referring expressions are detected, and Kinect has detected a gesture, information about any objects pointed to by the user will be stored. The system then can find the best match between properties of each relevant entity. The properties of visualizations and objects keeping track of include statistics and trends in the data, title, mark, and any more prominent objects within the visualization (e.g., hot-spots, street names, and bus stops).
	In multimodal systems\cite{Srinivasan2020b,Saktheeswaran2020,Srinivasan2018,Kassel2018,Srinivasan2020a,Young-HoKim2021,Dontcheva}, users can focus on specific visual elements in the visualization by selecting with mouse, pen, or finger. The follow-up queries can automatically link to related visual elements or data.
	
	\textbf{Discussion:}
	At the natural language level, existing V-NLIs mostly leverage NLP toolkits to perform co-reference resolution. Although useful, they lack detailed modeling of visualization elements, as discussed in Section \ref{sec:Semantic and Syntax Analysis}. At the multimodal level, in addition to commonly used modalities (e.g., mouse, pen, and finger), it may enable a better user experience by integrating more modalities like eye-gaze and gesture.
	
	
	\vspace{-0.3cm}
	\section{Presentation}\label{sec:presentation}
	In most cases, V-NLI accepts natural language as input and outputs well-designed visualizations. As shown in Table \ref{tab:vis_papers} (column \textit{Visualization Type}), the output is not limited to traditional charts but also involves other richer forms (e.g., maps, networks, and infographics).
	In addition to visual presentation, an emerging theme is to complement visualizations with natural language. The previous sections have discussed natural language as an input modality. This section will introduce using natural language as an output modality.
	
	
	
	\begin{figure}[t]
		\setlength{\abovecaptionskip}{0cm}
		\setlength{\belowcaptionskip}{-0.1cm}
		\centering
		\includegraphics[width=\columnwidth]{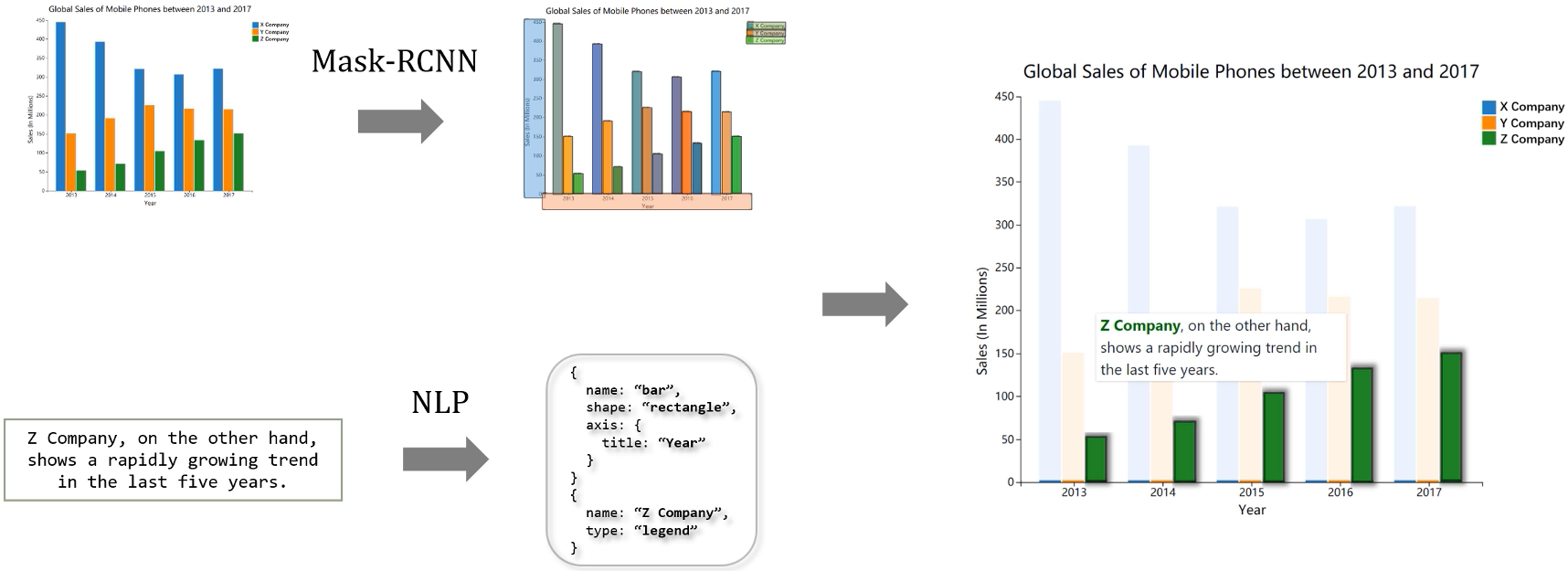}
		\caption{Visualizations with annotations generated by Vis-Annotator\cite{Lai2020a}.}
		\label{fig:Vis-Annotator}
		\vspace{-0.3cm}
	\end{figure}
	\vspace{-0.3cm}
	\subsection{Annotation}\label{sec:Annotation}
	Annotation plays a vital role in explaining and emphasizing key points in the dataset. The systems should generate valuable natural language statements and map the text to a visualization appropriately. 
	Annotation tools were first applied to complement news articles.
	Kandogan\cite{Kandogan} introduced the concept of just-in-time descriptive analytics that helps users easily understand the structure of data. 
	Given a piece of news, Contextifier\cite{Hullman2013} first computes clusters, outliers, and trends in line charts and then automatically produces annotated stock visualizations based on the information. Although it offers a promising example of how human-produced news visualizations can be complemented in a specific context, it is only suitable for stock sequence data.
	NewsViews\cite{El-Dairi2019} later extends Contextifier to support more types of appropriate data such as time series and georeferenced data. 
	In addition, several works incorporate the visual elements in the chart into annotations, synchronizing with textual descriptions.
	As shown in Figure \ref{fig:Vis-Annotator}, Vis-Annotator\cite{Lai2020a} leverages the Mask R-CNN model to identify visual elements in the target visualization, along with their visual properties. Textual descriptions of the chart are synchronously interpreted to generate visual search requests. Based on the identified information, each descriptive sentence is displayed beside the described focal areas as annotations.
	Similarly, Click2Annotate\cite{Chen} and Touch2Annotate\cite{Chena} are both semi-automatic annotations generators.
	To be interactive, Calliope\cite{Shi2020} and ChartAccent\cite{Ren} support creating stories via interactive annotation generation and placement.
	Most recently, ADVISor\cite{Liu2021b} can generate visualizations with annotations to answer the user's NL questions on tabular data.
	
	\textbf{Discussion:}
	Annotation can turn data into a story.
	However, the scalability of current works is weak as most of them are template-based.
	To improve the system usability, advanced NLG models\cite{Celikyilmaz2021} can be leveraged to assist generating appropriate textual annotations. Extending existing systems to support richer data types and annotation forms is also a meaningful direction. Besides, it would be helpful to devise an annotation specification language for additional flexibility.

	\begin{figure}[t]
		\setlength{\abovecaptionskip}{0cm}
		\setlength{\belowcaptionskip}{-0.1cm}
		\centering
		\includegraphics[width=\columnwidth,height=5cm]{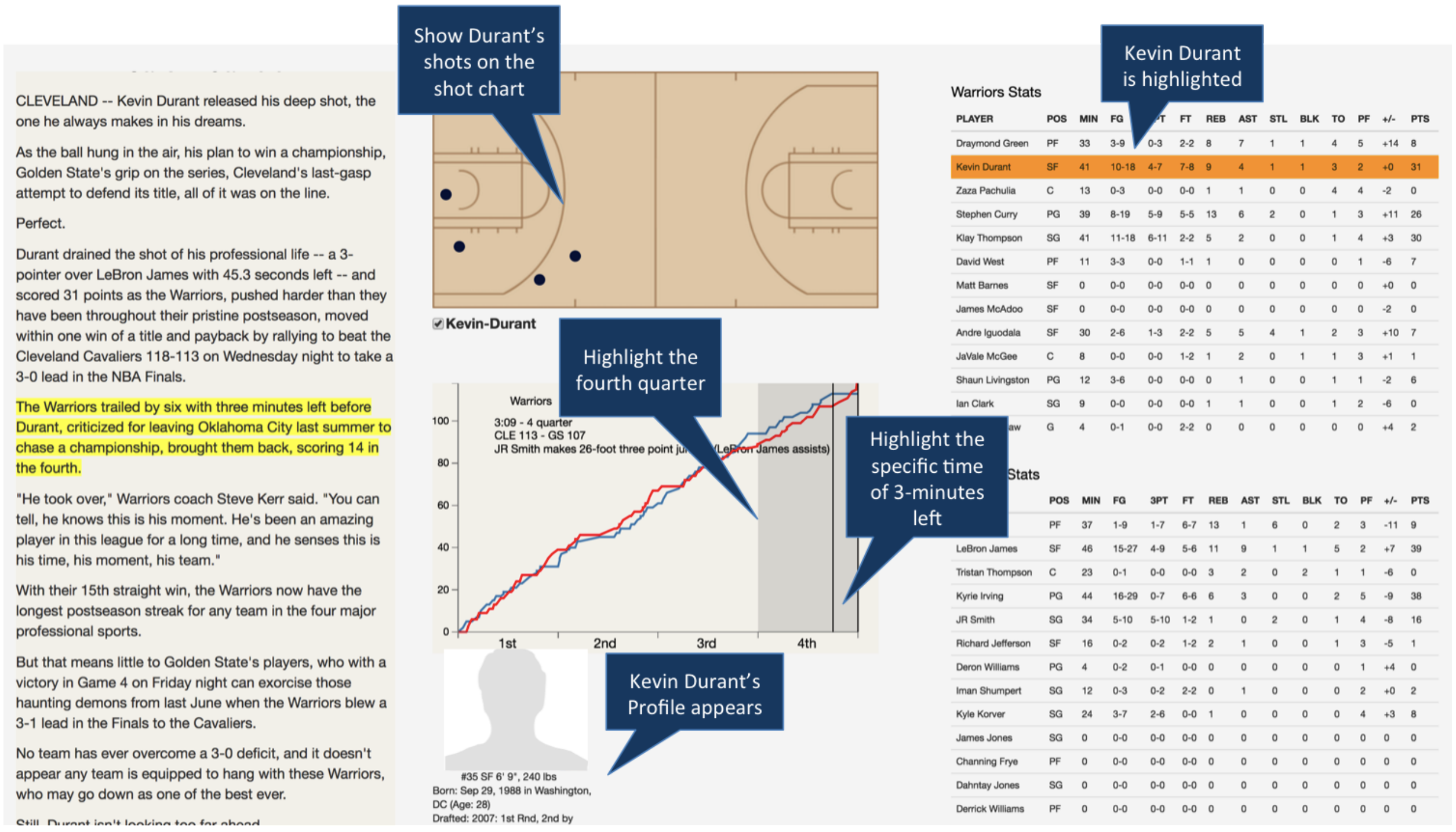}
		\caption{Coupling from the text about NBA game report to stories\cite{Metoyer2018}.}
		\label{fig:couple}
		\vspace{-0.4cm}
	\end{figure}
	\vspace{-0.3cm}
	\subsection{Narrative Storytelling}
	Narrative storytelling gives data a voice and helps communicate insights more effectively. A data story means a narrative to the data that depicts actors and their interactions, along with times, locations, and other entities\cite{Segel2010}. Constructing various visual and generated textual (natural language) elements in  visualizations is vital for narrative storytelling.
	To address this issue, the template is the most commonly applied method. For instance, Text-to-Viz\cite{Cui2020b} is designed to generate infographics from NL statements with proportion-related statistics. It builds 20 layout blueprints that describe the overall look of the resulting infographics.
	For each layout blueprint, Text-to-Viz enumerates all extracted segments by text analyzer and then generates all valid infographics.
	DataShot\cite{Wang2020h}, TSIs\cite{Bryan2017}, Chen \textit{et al.}\cite{Chen2019}, and Retrieve-Then-Adapt\cite{Qian2020} all use similar template-based approaches for narrative storytelling.
	With the advances of deep learning, some works leverage generative adversarial networks (GAN)\cite{NIPS2014_5ca3e9b1} to synthesize layouts\cite{Li2019f,Zheng2019a}.
	Those learning-based methods do not rely on handcrafted features. However, due to limited training data, they often fail to achieve comparable performance with rule-based methods.
	Likewise, storytelling has been extensively used to visualize narrative text produced by online writers and journalism media.
	Metoyer \textit{et al.}\cite{Metoyer2018} proposed a novel approach to generate data-rich stories. It first extracts narrative components (who, what, when, where) from the text and then generates narrative visualizations by integrating the supporting data evidence with the text. 
	As shown in Figure \ref{fig:couple}, given a highlighted NBA game report sentence, the system generates an initialized story dashboard based on the interpretation information.
	Similarly, Story Analyzer\cite{Mitri2020} extracts subjects, actions, and objects from the narrative text to produce interrelated and user-responsive story dashboards. 
	RIA\cite{Zhou} dynamically derives a set of layout constraints (e.g., ensure visual balance and avoid object occlusion) to tell spatial stories. 
	
	\textbf{Discussion:}
	Although the aforementioned systems are promising and work relatively well, they are restricted to one specific type of information respectively. More types of information (e.g., timeline and map) can be combined to create more expressive stories. Besides, similar to annotation, template-based methods also dominate narrative storytelling. It will be interesting to explore learning-based approaches to further improve the story quality as well as formulating datasets specially for this topic.
	

	
	\vspace{-0.3cm}
	\subsection{Natural Language Generation for Visualization}
	In the past years, there has been a body of research on automatically generating descriptions or captions for charts. 
	The majority of early works are rule-based\cite{Corio1999,Demir2008,Fasciano1997,Mittal1998a,Moraes2014}. While modern pipelines generally include a chart parsing module and a subsequent caption generation module. Chart parsing module\cite{Savva2011,Poco2017,Al-Zaidy2015} deconstructs the original charts and extracts valuable elements such as text, axis, and lines using relevant techniques like text localization, optical character recognition, and boundary tracing. 
	Deshpande \textit{et al.}\cite{Deshpande2020} proposed a novel method for chart parsing which adopts a question answering approach to query key elements in charts. The organized elements are subsequently fed into the caption generation module\cite{Liu2020a,Obeid2020} to output captions. A common shortcoming of the models above is the demand for manually designed procedures or features. For example, Al-Zaidy \textit{et al.}\cite{Al-Zaidy2015} relied on pre-defined templates to generate sentences, and Chart-to-Text\cite{Obeid2020} needs additional annotated tabular data to describe elements in a given chart. Benefiting from the development of deep learning, several end-to-end models have been proposed recently\cite{Qian2021,Spreafico2020,Obeid2020}. 
	FigJAM\cite{Qian2021} employs ResNet-50\cite{He2016} to encode a chart as a whole image and uses OCR to encode text to generate slot values. They are along with the image feature vectors as initialization of a LSTM network to generate captions.
	Spreafico \textit{et al.}\cite{Spreafico2020} exploited the encoder-decoder LSTM architecture to take time-series data as input and generate corresponding captions. 
	Obeid \textit{et al.}\cite{Obeid2020} applied a transformer-based model\cite{Furfaritony2002} for generating chart summaries.
	Most recently, Kim \textit{et al.}\cite{Kim1999} explored how readers gather takeaways when considering charts and captions together. Results suggest that the takeaways differ when the caption mentions visual features of differing prominence levels, which provides valuable guidance for future research. 
	
	\textbf{Discussion:}
	So far, most related works in NLG4Vis can only apply to simple charts (e.g., bar charts, scatterplots, and line charts). Further research can pay attention to improving the chart type coverage, as well as the language quality and descriptive ability. Besides, a more fundamental direction is to develop large datasets covering diverse domains and chart types.
	
	\begin{figure}[t]
		\setlength{\abovecaptionskip}{0cm}
		\setlength{\belowcaptionskip}{-0.1cm}
		\centering
		\includegraphics[width=0.9\columnwidth]{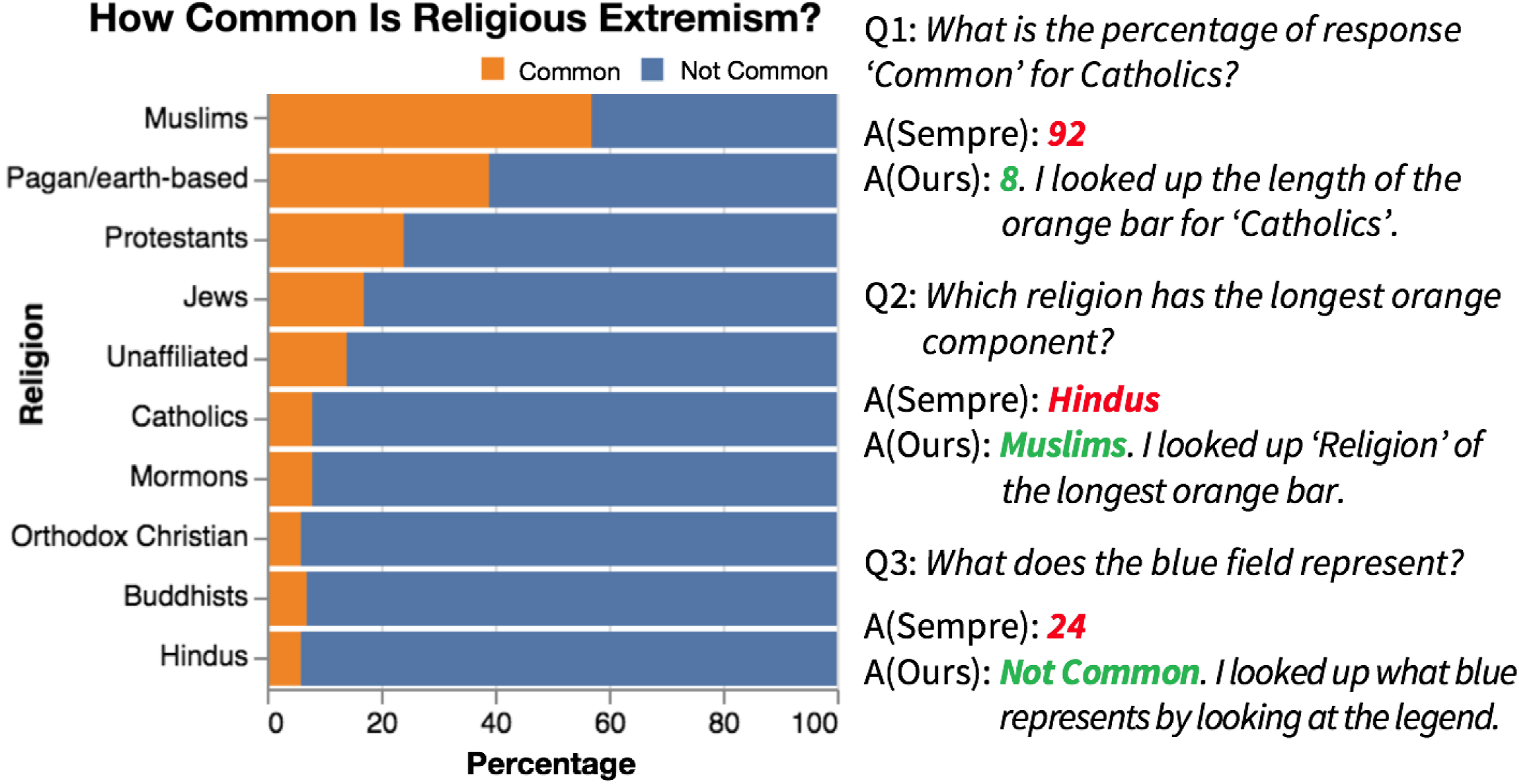}
		\caption{The automatic VQA pipeline\cite{Kim2020a} answers the three questions correctly (marked in green) and gives correct explanations.}
		\label{fig:VQA}
		\vspace{-0.3cm}
	\end{figure}
	
	\vspace{-0.3cm}
	\subsection{Visual Question Answering}
	Visual Question Answering (VQA) is a semantic understanding task that aims to answer questions based on a given visualization and possibly along with the informative text. In this subsection, we only focus on images with charts or infographics rather than images with natural objects and scenes in the computer vision community. Generally, visualizations and questions are respectively encoded and then fused to generate answers\cite{Mathew2021}. For visualizations, Kim \textit{et al.}\cite{Kim2020a} leveraged a semantic parsing model, Sempre\cite{Zhang2017d}, to develop an automatic chart question answering pipeline with visual explanations describing how the answer was produced. Figure \ref{fig:VQA} shows that the pipeline answers all three questions correctly (marked in green) and gives correct explanations of how it obtains the answer. 
	While for infographics, the OCR module additionally needs to identify and tokenize the text, which serves as facts for answer generation\cite{Mathew2021,Kafle2018,Easoning2018,Chaudhry2020,Singh2020}. Later models employ more sophisticated fusion modules, among which the attention mechanism has reached great success.
	LEAF-Net\cite{Chaudhry2020} comprises chart parsing, question, and answer encoding according to chart elements followed by an attention network is proposed.
	STL-CQA\cite{Singh2020} applies a structural transformer-based learning approach that emphasizes exploiting the structural properties of charts. 
	Recently, with the development of multimodal transformer\cite{Li2019,Su2019}, Mathew \textit{et al.}\cite{Mathew2021} proposed a pipeline where the question and infographic are mapped into the same vector space and simply added as the input to stacks of transformer layers. Besides, they also delivered a summary of VQA datasets for reference.
	
	\textbf{Discussion:}
	Some systems extend existing QA engines (e.g., Sempre) to answer questions about charts; as a result, the interpretation ability is limited as they are not designed for visualization. A more proper way is to develop specific models based on visualization datasets. However, the range of questions that these learning-based approaches can handle may be restricted to the diversity of datasets. So it may be promising to transfer existing models to other areas and iteratively extend the dataset.

	\vspace{-0.3cm}
	\section{Research Opportunity}\label{sec:future}
	By conducting a comprehensive survey under the guidance of an information visualization pipeline proposed by Card \textit{et al.}\cite{Card1999}, we found that, as a rapidly growing field, V-NLI faces many thorny challenges as well as open research opportunities. In this section, we organize these from a macro-perspective into five aspects: knowledge, model, interaction, presentation, and dataset, with additional discussions on applications. Our target is to cover the critical gaps and emerging topics that deserve more attention.
	
	\vspace{-0.3cm}
	\subsection{Domain Knowledge}
	A major limitation of most existing V-NLI systems is the absence of domain knowledge. We have conducted an evaluation on four state-of-the-art open-source V-NLIs, both academic\cite{Yu2020b,Narechania2020} and commercial\cite{askdata,powerbi}. We found that none of the systems can recognize that \textit{nitrogen dioxide} and \textit{$NO_2$} have the same meaning, nor can they recognize the relationship between \textit{age} and \textit{birth year}. Therefore, the support of domain knowledge is crucial, especially for extracting data attributes in the NL query.
	CogNet\cite{Wang}, a knowledge base dedicated to integrating various existing knowledge bases (e.g., FrameNet, YAGO, Freebase, DBpedia, Wikidata, and ConceptNet),
	may be useful for broadening the repertoire of utterances supported. Specifically, these knowledge bases can be wholly or partly used to create word embeddings. With the help of embedding results, the similarity between N-grams of the NL query and data attributes names can be computed at both the syntactic and semantic levels. Naturally, it is conceivable that data attributes with high similarity or whose similarity exceeds a certain threshold will be extracted.
	
	\vspace{-0.3cm}
	\subsection{NLP Model}
	\subsubsection{Application of more advanced NLP models}
	The performance of V-NLI depends to a great extent on NLP models. As shown in Table \ref{tab:vis_papers} (column \textit{NLP Toolkit or Technology} and column \textit{Recommendation Algorithm}), most of the existing V-NLI systems just apply hand-crafted grammar rules or typical NLP toolkits for convenience.
	Recently, several state-of-the-art NLP models have reached human performance in specific tasks,
	such as ELMO\cite{ELMo}, BERT\cite{Devlin2019}, GPT-3\cite{Brown}, and CPM-2\cite{Zhang2021a}.
	Some works have leveraged the advances for data visualization\cite{Wu2021d,Wang2020c}. However, few works have applied them in V-NLI.
	During our survey, we also found that existing works provided limited support for free-form NL input. To construct a more robust system, a promising direction is to apply more advanced NLP models to learn universal grammar patterns from a large corpus of conversations in visual analysis.
	In the process, we believe that the existing high-quality text datasets will help train and evaluate robust NLP models for V-NLI. 
	Besides, the community can pay more attention to the end-to-end approach due to its increased efficiency, cost cutting, ease of learning, and smarter visualization inference ability (e.g., chart type selection, data insight mining, and missing data prediction).
	To this end, advanced neural machine translation models\cite{Dabre2020} can be applied with various optimization schemes. 
	NLP models of multiple languages also deserve attention since all V-NLI systems we found can only support English. In fact, numerous NLP models can deal effectively with other languages (e.g., Chinese, French, and Spanish) and can be leveraged to develop V-NLI to serve more people who speak different languages.
	
	
	\begin{table*}[t]
		\setlength{\abovecaptionskip}{0cm}
		\setlength{\belowcaptionskip}{-0.1cm}
		\caption{Summary of existing V-NLI datasets.}
		\label{tab:datasets}
		\scriptsize
		\renewcommand\arraystretch{1}
		\begin{tabular}{cccccll}
			\hline
			\textbf{Name} & \textbf{Publication} & \textbf{NL queries} & \textbf{Data tables} & \textbf{Benchmark} & \multicolumn{1}{c}{\textbf{Other Contributions}} & \multicolumn{1}{c}{\textbf{Website}} \\ \hline
			Kim \textit{et al.}\cite{Kim2020a} & CHI'20 & 629 & 52 & × & VQA with explanations & https://github.com/dhkim16/VisQA-release \\
			Quda\cite{Fu2018} & arXiv'20 & 14,035 & 36 & × & Three Quda applications & https://freenli.github.io/quda/ \\
			NLV\cite{Srinivasan2021} & CHI'21 & 893 & 3 & \checkmark & Characterization of utterances & https://nlvcorpus.github.io/ \\
			nvBench\cite{Luo2021} & SIGMOD'21 & 25,750 & 780 & \checkmark & NL2SQL-to-NL2VIS and SEQ2VIS & https://github.com/TsinghuaDatabaseGroup/nvBench \\ \hline
		\end{tabular}
		\vspace{-0.5cm}
	\end{table*}

	\vspace{-0.3cm}
	\subsubsection{Deep interpretation of dataset semantics}
	Semantic information plays an important role in V-NLI. The state-of-the-art systems have considered leveraging semantic parsing toolkits like SEMPRE\cite{Zhang2017d} to parse NL queries\cite{Yu2020b,Kim2020a}. However, just considering the semantics of the NL query is limited, and the semantics of the dataset should also be taken into consideration. The existing technologies for data attribute matching only confine to the letter-matching level but do not go deep into the semantic-matching level, as described in Section \ref{sec:Data Attributes Inference}. For example, 
	when the analyzed dataset is about movies, a full-fledged system should be able to recognize that the \textit{Name} attribute in the dataset refers to the movie name and automatically associate it with other attributes appearing in the query.
	A promising way to augment semantic interpretation ability is to connect with recent semantic data type detection models for visualization like Sherlock\cite{Hulsebos2019}, Sato\cite{Zhang2019a}, ColNet\cite{Chen2019b},  DCOM\cite{Maji2021}, EXACTA\cite{Xian2021}, and Doduo\cite{Suhara2021}.
	Incorporating such connections will help better infer attribute types upon initialization and may also reduce the need for developers to manually configure attribute aliases. Besides, the aforementioned models are limited to a fixed set of semantic types. An additional essential direction is to extend the existing semantic data type detection models to support more semantic types. Additionally, with the contextually deep interpretation of dataset semantics, supporting queries for multi-table schema would be an interesting research hotspot.
	
	\vspace{-0.3cm}
	\subsection{Interaction}
	\subsubsection{Richer prompt to improve system discoverability}
	The user may not be aware of what input is valid and what chart types are supported by the system with open-ended textboxes. As described in Section \ref{sec:Autocompletion}, discoverability has received relatively little attention in current V-NLIs compared to other characteristics.
	The existing practices have revealed that real-time prompts can effectively help the user better understand system features and correct input errors in time when typing queries\cite{Setlur2020,Yu2020b,Dontcheva} (see Table \ref{tab:vis_papers} (column \textit{Autocompletion})).
	The most common method is template-based text autocompletion\cite{Yu2020b}. However, this method is relatively monotonous in practice and does not work for spoken commands. Several works offer the user prompt widgets of data previews to speed up the user's input speed\cite{Setlur2020}, but the supporting prompt widget types are limited. 
	We believe that richer interactive widgets with prompts and multimodal synergistic interaction can greatly improve the system's usability.
	Showing provenance of prompt behavior can also enhance the interpretability of visualization results.
	Concerning this issue, Setlur \textit{et al.}\cite{Setlur2020} conducted a crowdsourcing study regarding the efficacy of autocompletion suggestions. The insights drawn from the studies are of great value to inspire the future design of V-NLI. 
	
	\vspace{-0.2cm}
	\subsubsection{Take advantage of the user's interaction history}
	The intent behind NL queries can be satisfied by various types of charts. These visualization results are too broad that existing systems can hardly account for varying user preferences. Although several conversational V-NLI systems have been proposed\cite{Setlur2016, Hoque2018,Kumar2016} to analyze NL queries in context, few systems have taken the user's interaction history into account. Recently, Zehrung \textit{et al.}\cite{Zehrung2021a} conducted a crowdsourcing study analyzing trust in humans versus algorithmically generated visualization recommendations. Based on the results, they suggested that the recommendation system should be customized according to the specific user's information search strategy.
	Personalized information derived from historical user interaction and context can provide a richer model to satisfy the user's analytic tasks.
	A large number of innovative models in the recommendation system\cite{Zhang2019} can also be applied for reference. 
	Besides, Lee \textit{et al.}\cite{Lee2021a} recently deconstructed categorization in visualization recommendation. Several works\cite{Henkin2020,Sevastjanova2018} have studied the associations and expectations about verbalization and visualization reported by users. 
	Integrating the information for further modeling of the user's interaction history is another interesting research topic.


	\vspace{-0.3cm}
	\subsection{Presentation}
	We hypothesize that we would not truly have succeeded in democratizing access to visual analysis through natural language until we design systems that use NL both as input and output modality. For this vision, a promising direction is to combine all the aforementioned characteristics (e.g., VQA, NLG, and annotation) to design a hybrid system. In addition, existing V-NLI systems mostly only support 2D visualization. Nowadays, Immersive Analytics (IA) is a quickly evolving field that leverages immersive technologies for data analysis\cite{Fonnet2021,Ens2021}. In the visualization community, several works have augmented static visualizations with virtual content\cite{Chen2020a,Langner2021}. 
	Reinders \textit{et al.}\cite{Reinders2020} studied blind and low vision people’s preferences when exploring interactive 3D printed models (I3Ms). However, there have been no systems that support NLI for data visualization in an immersive way. A related work for reference is that Lee \textit{et al.}\cite{Lee2021} proposed the concept of \textit{data visceralization} and introduced a conceptual pipeline.
	It would be interesting to enrich this pipeline with the integration of V-NLI. 
	Besides, as described in Section \ref{sec:view transformation}, view transformations are rarely involved in existing systems. 
	With immersive technologies, more view transformations can be explored and integrated into V-NLI to provide an immersive interactive experience.

	\vspace{-0.3cm}
	\subsection{Dataset}
	There is a widely recognized consensus that large-scale data collection is an effective way to facilitate community development (e.g., ImageNet\cite{Deng2009} for image processing and GEM\cite{Gehrmann2021} for natural language generation). Although there are various datasets for general NLP tasks, they can hardly be directly applied to provide training samples for V-NLI models.  In the visualization community, several works have begun to collect large datasets for V-NLI, as shown in Table \ref{tab:datasets}. To assist the deployment of learning-based techniques for parsing human language, Fu \textit{et al.}\cite{Fu2018} proposed Quda, containing diverse user queries annotated with analytic tasks. 
	Srinivasan \textit{et al.}\cite{Srinivasan2021} conducted an online study to collect NL utterances and characterized them based on their phrasing type.
	VizNet\cite{Hu2019e} is a step forward in addressing the need for large-scale data and visualization repositories.
	Kim \textit{et al.}\cite{Kim2020a} collected questions people posed about various bar charts and line charts, along with their answers and explanations to the questions.
	Fully considering the user’s characteristics, Kassel \textit{et al.}\cite{Kassel2019} introduced a linguistically motivated two-dimensional answer space that varies in the level of both support and information to match the human language in visual analysis. 
	The major limitation of them is that the query types contained are not rich enough. 
	Besides, unfortunately, a benchmark or the ground truth of a given dataset is usually unavailable.
	Although Srinivasan \textit{et al.}\cite{Srinivasan2021} made an initial attempt to present a collection of NL utterances with the mapping visualizations, Luo \textit{et al.}\cite{Luo2021} synthesized NL-to-Vis benchmarks by piggybacking NL-to-SQL benchmarks and produced a NL-to-Vis benchmark, nvBench, the supporting visualization types and tasks are limited.
	Therefore, collecting large-scale datasets and creating new benchmarks that support more tasks, domains, and interactions will be an indispensable research direction in the future.

	\vspace{-0.3cm}
	\subsection{Application}
	Apart from facilitating data analysis, much attention has been paid to benefit real-world applications by integrating visual analysis and natural language interface. 
	Huang \textit{et al.}\cite{Huang2019c} proposed a query engine to convert, store and retrieve spatial uncertain mobile trajectories via intuitive NL input. 
	Leo John \textit{et al.}\cite{LeoJohn2018} proposed a novel V-NLI to promote medical imaging and biomedical research.
	With an audio travel podcast as input, Crosscast\cite{Xia2020a} identifies geographic locations and descriptive keywords within the podcast transcript through NLP and text mining techniques. The information is later used to select relevant photos from online repositories and synchronize their display to align with the audio narration.
	MeetingVis\cite{Shi2018} leverages ASR and NLP techniques to promote effective meeting summaries in team-based workplaces.
	PathViewer\cite{Wang2017c} leverages ideas from flow diagrams and NLP to visualize the sequences of intermediate steps that students take. 
	Since V-NLI can be easily integrated as a module into the visualization system, more applications can be explored in the future.
	
	
	\vspace{-0.3cm}
	\section{Conclusion}\label{sec:conclude}
	The past two decades have witnessed the rapid development of visualization-oriented natural language interfaces, which act as a complementary input modality for visual analytics.
	However, the community lacks a comprehensive survey of related works to guide follow-up research.
	We fill the gap, and the resulting survey gives a comprehensive overview of what characteristics are currently concerned and supported by V-NLI.
	We also propose several promising directions for future work.
	To our knowledge, this paper is the first step towards reviewing V-NLI in a novel and systematic manner.
	We hope that this paper can better guide future research and encourage the community to think further about NLI for data visualization.

	\vspace{-0.4cm}
	
	\section*{Acknowledgments}
	{
		\vspace{-0.1cm}
		The authors would like to thank all the reviewers for their valuable comments. They also thank all the authors of related papers, especially those who kindly provide images covered in this survey. 
		The work was supported by the National Natural Science Foundation of China (No. 71690231) and Beijing Key Laboratory of Industrial Bigdata System and Application.
		\vspace{-0.1cm}
	}
	
	\ifCLASSOPTIONcaptionsoff
	\newpage
	\fi

	\vspace{-0.3cm}
	\bibliographystyle{abbrv}
	\bibliography{lib_conference_abbrv}

\begin{thebibliography}{100}

\bibitem{opennlp}
{Apache OpenNLP. http://opennlp.apache.org/}.

\bibitem{askdata}
{Ask data. https://www.tableau.com/products/new-features/ask-data}.

\bibitem{googlenlp}
{Google NLP. https://cloud.google.com/natural-language/}.

\bibitem{IBM}
{IBM Watson Analytics. http://www.ibm. com/analytics/watson-analytics}.

\bibitem{powerbi}
{Microsoft Power BI.
  https://docs.microsoft.com/en-us/power-bi/create-reports/power-bi-tutorial-q-and-a}.

\bibitem{Affolter2019}
K.~Affolter, K.~Stockinger, and A.~Bernstein.
\newblock {A comparative survey of recent natural language interfaces for
  databases}.
\newblock {\em VLDB J.}, 28(5), 2019.

\bibitem{Flair}
A.~Akbik, D.~Blythe, and R.~Vollgraf.
\newblock {Contextual String Embeddings for Sequence Labeling}.
\newblock In {\em Proc. COLING'19}. ACM, 2018.

\bibitem{Al-Zaidy2015}
R.~A. Al-Zaidy and C.~L. Giles.
\newblock {Automatic Extraction of Data from Bar Charts}.
\newblock In {\em Proc. ICKC'15}. ACM, 2015.

\bibitem{Amar2005b}
R.~Amar, J.~Eagan, and J.~Stasko.
\newblock {Low-level components of analytic activity in information
  visualization}.
\newblock In {\em Proc. INFOVIS'05}. IEEE, 2005.

\bibitem{Bacci2020}
F.~Bacci, F.~M. Cau, and L.~D. Spano.
\newblock {Inspecting Data Using Natural Language Queries}.
\newblock {\em Lect. Notes Comput. Sci.}, 12254:771--782, 2020.

\bibitem{Bach2018}
B.~Bach, Z.~Wang, M.~Farinella, and et~al.
\newblock {Design Patterns for Data Comics}.
\newblock In {\em Proc. CHI'18}. ACM, 2018.

\bibitem{Bahdanau2015}
D.~Bahdanau, K.~Cho, and Y.~Bengio.
\newblock {Neural machine translation by jointly learning to align and
  translate}.
\newblock In {\em Proc. ICLR'15}, 2015.

\bibitem{Baik2019}
C.~Baik, H.~V. Jagadish, and Y.~Li.
\newblock {Bridging the semantic gap with SQL query logs in natural language
  interfaces to databases}.
\newblock In {\em Proc. ICDE'19}. IEEE, 2019.

\bibitem{Basik2018}
F.~Basik, B.~H{\"{a}}ttasch, A.~Ilkhechi, and et~al.
\newblock {DBPal: A learned NL-interface for databases}.
\newblock In {\em Proc. SIGMOD'18}. ACM, 2018.

\bibitem{Bast2015}
H.~Bast and E.~Haussmann.
\newblock {More Accurate Question Answering on Freebase}.
\newblock In {\em Proc. CIKM'15}. ACM, 2015.

\bibitem{Battle2020b}
L.~Battle, R.~J. Crouser, A.~Nakeshimana, and et~al.
\newblock {The Role of Latency and Task Complexity in Predicting Visual Search
  Behavior}.
\newblock {\em IEEE Trans. Vis. Comput. Graph.}, 26(1):1246--1255, 2020.

\bibitem{Battle2021}
L.~Battle and C.~Scheidegger.
\newblock {A Structured Review of Data Management Technology for Interactive
  Visualization and Analysis}.
\newblock {\em IEEE Trans. Vis. Comput. Graph.}, 27(2):1128--1138, 2021.

\bibitem{Belinkov2019}
Y.~Belinkov and J.~Glass.
\newblock {Analysis Methods in Neural Language Processing: A Survey}.
\newblock {\em Trans. Assoc. Comput. Linguist.}, 7:49--72, 2019.

\bibitem{Bergamaschi2016}
S.~Bergamaschi, F.~Guerra, M.~Interlandi, and et~al.
\newblock {Combining user and database perspective for solving keyword queries
  over relational databases}.
\newblock {\em Inf. Syst.}, 55:1--19, 2016.

\bibitem{Bieliauskas2017}
S.~Bieliauskas and A.~Schreiber.
\newblock {A Conversational User Interface for Software Visualization}.
\newblock In {\em Proc. VISSOFT'17}. IEEE, 2017.

\bibitem{Bird2006}
S.~Bird.
\newblock {NLTK: the natural language toolkit}.
\newblock In {\em Proc. COLING-ACL'06}. ACL, 2006.

\bibitem{Blunschi2012}
L.~Blunschi, C.~Jossen, D.~Kossmann, and et~al.
\newblock {SODA: Generating SQL for business users}.
\newblock {\em Proc. VLDB Endow.}, 5(10):932--943, 2012.

\bibitem{Bostock2011}
M.~Bostock, V.~Ogievetsky, and J.~Heer.
\newblock {D3: Data-Driven Documents}.
\newblock {\em IEEE Trans. Vis. Comput. Graph.}, 17(12):2301--2309, 2011.

\bibitem{Brehmer2013}
M.~Brehmer and T.~Munzner.
\newblock {A Multi-Level Typology of Abstract Visualization Tasks}.
\newblock {\em IEEE Trans. Vis. Comput. Graph.}, 19(12), 2013.

\bibitem{Brown}
T.~Brown, B.~Mann, N.~Ryder, and et~al.
\newblock {Language Models are Few-Shot Learners}.
\newblock In {\em Proc. NeurIPS'20}. MIT Press, 2020.

\bibitem{Bryan2017}
C.~Bryan, K.~L. Ma, and J.~Woodring.
\newblock {Temporal Summary Images: An Approach to Narrative Visualization via
  Interactive Annotation Generation and Placement}.
\newblock {\em IEEE Trans. Vis. Comput. Graph.}, 2017.

\bibitem{Bylinskii}
Z.~Bylinskii, S.~Alsheikh, S.~Madan, and et~al.
\newblock {Understanding Infographics through Textual and Visual Tag
  Prediction}.
\newblock {\em arXiv}, 2017.

\bibitem{Card1999}
S.~Card, J.~Mackinlay, and B.~Shneiderman.
\newblock {\em {Readings in Information Visualization: Using Vision to Think}}.
\newblock Morgan Kaufmann, 1999.

\bibitem{Celikyilmaz2021}
A.~Celikyilmaz, E.~Clark, and J.~Gao.
\newblock {Evaluation of Text Generation: A Survey}.
\newblock {\em arxiv}, pages 1--75, 2021.

\bibitem{Chaudhry2020}
R.~Chaudhry, S.~Shekhar, U.~Gupta, and et~al.
\newblock {LEAF-QA: Locate, encode attend for figure question answering}.
\newblock In {\em Proc. WACV'20}, 2020.

\bibitem{Chen2019b}
J.~Chen, E.~Jim{\'{e}}nez-Ruiz, I.~Horrocks, and C.~Sutton.
\newblock {ColNet: Embedding the Semantics of Web Tables for Column Type
  Prediction}.
\newblock In {\em Proc. AAAI'19}. AAAI, 2019.

\bibitem{Chen2020}
S.~Chen, J.~Li, G.~Andrienko, and et~al.
\newblock {Supporting Story Synthesis: Bridging the Gap between Visual
  Analytics and Storytelling}.
\newblock {\em IEEE Trans. Vis. Comput. Graph.}, 26(7):2499--2516, 2020.

\bibitem{Chen}
Y.~Chen, S.~Barlowe, and J.~Yang.
\newblock {Click2Annotate: Automated Insight Externalization with rich
  semantics}.
\newblock In {\em Proc. VAST'10}. IEEE, 2010.

\bibitem{Chena}
Y.~Chen, J.~Yang, S.~Barlowe, and D.~H. Jeong.
\newblock {Touch2Annotate: Generating better annotations with less human effort
  on multi-touch interfaces}.
\newblock In {\em Proc. CHI'10}. ACM, 2010.

\bibitem{Chen2020a}
Z.~Chen, W.~Tong, Q.~Wang, and et~al.
\newblock {Augmenting Static Visualizations with PapARVis Designer}.
\newblock In {\em Proc. CHI'20}. ACM, 2020.

\bibitem{Chen2019}
Z.~Chen, Y.~Wang, Q.~Wang, and et~al.
\newblock {Towards automated infographic design: Deep learning-based
  auto-extraction of extensible timeline}.
\newblock {\em IEEE Trans. Vis. Comput. Graph.}, 26(1):917--926, 2020.

\bibitem{Choo2014b}
J.~Choo, C.~Lee, H.~Kim, and et~al.
\newblock {VisIRR: Visual analytics for information retrieval and
  recommendation with large-scale document data}.
\newblock In {\em Proc. VAST'14}. IEEE, 2014.

\bibitem{Cook2015}
K.~Cook, N.~Cramer, D.~Israel, and et~al.
\newblock {Mixed-initiative visual analytics using task-driven
  recommendations}.
\newblock In {\em Proc. VAST'15}. IEEE, 2015.

\bibitem{Corio1999}
M.~Corio and G.~Lapalme.
\newblock {Generation of texts for information graphics}.
\newblock In {\em Proc. EWNLG'99}. ACL, 1999.

\bibitem{Cox2001}
K.~Cox, R.~E. Grinter, S.~L. Hibino, and et~al.
\newblock {A multi-modal natural language interface to an information
  visualization environment}.
\newblock {\em Int. J. Speech Technol.}, 4(3):297--314, 2001.

\bibitem{Cui2020b}
W.~Cui, X.~Zhang, Y.~Wang, and et~al.
\newblock {Text-to-Viz: Automatic Generation of Infographics from
  Proportion-Related Natural Language Statements}.
\newblock {\em IEEE Trans. Vis. Comput. Graph.}, 26(1):906--916, 2020.

\bibitem{Cui2019}
Z.~Cui, S.~K. Badam, M.~A. Yal{\c{c}}in, and N.~Elmqvist.
\newblock {DataSite: Proactive visual data exploration with computation of
  insight-based recommendations}.
\newblock {\em Inf. Vis.}, 18(2):251--267, 2019.

\bibitem{Dabre2020}
R.~Dabre, C.~Chu, and A.~Kunchukuttan.
\newblock {A Survey of Multilingual Neural Machine Translation}.
\newblock {\em ACM Comput. Surv.}, 53(5), 2020.

\bibitem{Demir2008}
S.~Demir, S.~Carberry, and K.~F. McCoy.
\newblock {Generating textual summaries of bar charts}.
\newblock In {\em Proc. INLG'08}. ACL, 2008.

\bibitem{Demir2012}
S.~Demir, S.~Carberry, and K.~F. McCoy.
\newblock {Summarizing Information Graphics Textually}.
\newblock {\em Comput. Linguist.}, 38(3):527--574, 2012.

\bibitem{Demiralp2017}
{\c{C}}.~Demiralp, P.~J. Haas, S.~Parthasarathy, and T.~Pedapati.
\newblock {Foresight: Recommending visual insights}.
\newblock {\em VLDB Endow.}, 10(12):1937--1940, 2017.

\bibitem{Deng2009}
J.~Deng, W.~Dong, R.~Socher, and et~al.
\newblock {ImageNet: A large-scale hierarchical image database}.
\newblock In {\em Proc. CVPR'09}. IEEE, 2009.

\bibitem{Deshpande2020}
A.~P. Deshpande and C.~N. Mahender.
\newblock {Summarization of Graph Using Question Answer Approach}.
\newblock In {\em Adv. Intell. Syst. Comput.}, pages 205--216. Springer, 2020.

\bibitem{Devlin2019}
J.~Devlin, M.-W. Chang, K.~Lee, and K.~Toutanova.
\newblock {BERT: Pre-training of Deep Bidirectional Transformers for Language
  Understanding}.
\newblock In {\em Proc. NAACL'19}. ACL, 2019.

\bibitem{Dhamdhere2017}
K.~Dhamdhere, K.~S. McCurley, R.~Nahmias, and et~al.
\newblock {Analyza: Exploring data with conversation}.
\newblock In {\em Proc. IUI'17}. ACM, 2017.

\bibitem{Dibia2019b}
V.~Dibia and C.~Demiralp.
\newblock {Data2Vis: Automatic Generation of Data Visualizations Using
  Sequence-to-Sequence Recurrent Neural Networks}.
\newblock {\em IEEE Comput. Graph. Appl.}, 39(5):33--46, 2019.

\bibitem{Dimara2020}
E.~Dimara and C.~Perin.
\newblock {What is Interaction for Data Visualization?}
\newblock {\em IEEE Trans. Vis. Comput. Graph.}, 26(1):119--129, 2020.

\bibitem{Ding2019}
R.~Ding, S.~Han, Y.~Xu, and et~al.
\newblock {Quickinsights: Quick and automatic discovery of insights from
  multi-dimensional data}.
\newblock In {\em Proc. SIGMOD'19}. ACM, 2019.

\bibitem{Duncan2021}
I.~K. Duncan, S.~Tingsheng, S.~T. Perrault, and M.~T. Gastner.
\newblock {Task-Based Effectiveness of Interactive Contiguous Area Cartograms}.
\newblock {\em IEEE Trans. Vis. Comput. Graph.}, 27(3):2136--2152, 2021.

\bibitem{Eisenschlos2020}
J.~Eisenschlos, S.~Krichene, and T.~M{\"{u}}ller.
\newblock {Understanding tables with intermediate pre-training}.
\newblock In {\em Find. Assoc. Comput. Linguist. EMNLP 2020}. ACL, 2020.

\bibitem{Ens2021}
B.~Ens, B.~Bach, M.~Cordeil, and et~al.
\newblock {Grand Challenges in Immersive Analytics}.
\newblock In {\em Proc. CHI'21}. ACM, 2021.

\bibitem{Fasciano1997}
M.~Fasciano and G.~Lapalme.
\newblock {PostGraphe: a system for the generation of statistical graphics and
  text Important factors in the generation process}.
\newblock In {\em Proc. INLG'96}. ACL, 1996.

\bibitem{Fast2017}
E.~Fast, B.~Chen, J.~Mendelsohn, and et~al.
\newblock {Iris: A conversational agent for complex tasks}.
\newblock In {\em Proc. CHI'18}. ACM, 2018.

\bibitem{Ferres2010}
L.~Ferres, G.~Lindgaard, and L.~Sumegi.
\newblock {Evaluating a tool for improving accessibility to charts and graphs}.
\newblock In {\em Proc. ASSETS'10}. ACM, 2010.

\bibitem{Fonnet2021}
A.~Fonnet and Y.~Prie.
\newblock {Survey of Immersive Analytics}.
\newblock {\em IEEE Trans. Vis. Comput. Graph.}, 27(3):2101--2122, 2019.

\bibitem{Fu2018}
S.~Fu, K.~Xiong, X.~Ge, and et~al.
\newblock {Quda: Natural Language Queries for Visual Data Analytics}.
\newblock {\em arXiv}.

\bibitem{Fulda2016}
J.~Fulda, M.~Brehmel, and T.~Munzner.
\newblock {TimeLineCurator: Interactive Authoring of Visual Timelines from
  Unstructured Text}.
\newblock {\em IEEE Trans. Vis. Comput. Graph.}, 22(1):300--309, 2016.

\bibitem{Gao2015b}
T.~Gao, M.~Dontcheva, E.~Adar, and et~al.
\newblock {Datatone: Managing ambiguity in natural language interfaces for data
  visualization}.
\newblock In {\em Proc. UIST'15}. ACM, 2015.

\bibitem{El-Dairi2019}
T.~Gao, J.~Hullman, E.~Adar, and et~al.
\newblock {NewsViews: An automated pipeline for creating custom
  geovisualizations for news}.
\newblock In {\em Proc. CHI'14}. ACM, 2014.

\bibitem{Lee}
T.~Ge, B.~Lee, and Y.~Wang.
\newblock {CAST: Authoring Data-Driven Chart Animations}.
\newblock In {\em Proc. CHI'21}. ACM, 2021.

\bibitem{Gehrmann2021}
S.~Gehrmann, T.~Adewumi, K.~Aggarwal, and et~al.
\newblock {The GEM Benchmark: Natural Language Generation, its Evaluation and
  Metrics}.
\newblock In {\em Proc. GEM'21}, pages 96--120. ACL, 2021.

\bibitem{Ghosh2018}
A.~Ghosh, M.~Nashaat, J.~Miller, and et~al.
\newblock {A comprehensive review of tools for exploratory analysis of tabular
  industrial datasets}.
\newblock {\em Vis. Informatics}, 2(4):235--253, 2018.

\bibitem{Gingerich2015}
M.~Gingerich and C.~Conati.
\newblock {Constructing models of user and task characteristics from eye gaze
  data for user-adaptive information highlighting}.
\newblock In {\em Proc. AAAI'15}. AAAI, 2015.

\bibitem{NIPS2014_5ca3e9b1}
I.~Goodfellow, J.~Pouget-Abadie, M.~Mirza, and et~al.
\newblock {Generative Adversarial Nets}.
\newblock In {\em Proc. NIPS'14}. MIT Press, 2014.

\bibitem{Gotz2009b}
D.~Gotz and Z.~Wen.
\newblock {Behavior-driven visualization recommendation}.
\newblock In {\em Proc. IUI'09}. ACM, 2009.

\bibitem{Grosz1986}
B.~J. Grosz and C.~L. Sidner.
\newblock {Attention, intentions, and the structure of discourse}.
\newblock {\em Comput. Linguist.}, 12(3):175--204, 1986.

\bibitem{gu2016incorporating}
J.~Gu, Z.~Lu, H.~Li, and V.~O. Li.
\newblock {Incorporating Copying Mechanism in Sequence-to-Sequence Learning}.
\newblock In {\em Proc. ACL'16}. ACL, 2016.

\bibitem{guo2019complex}
J.~Guo, Z.~Zhan, Y.~Gao, and et~al.
\newblock {Towards Complex Text-to-SQL in Cross-Domain Database with
  Intermediate Representation}.
\newblock In {\em Proc. ACL'19}. ACL, 2019.

\bibitem{Gur2018}
I.~Gur, S.~Yavuz, Y.~Su, and X.~Yan.
\newblock {DialSQL: Dialogue Based Structured Query Generation}.
\newblock In {\em Proc. ACL'18}. ACL, 2018.

\bibitem{Harris}
C.~Harris, R.~A. Rossi, S.~Malik, and et~al.
\newblock {Insight-centric Visualization Recommendation}.
\newblock {\em arXiv}, 2021.

\bibitem{He2016}
K.~He, X.~Zhang, S.~Ren, and J.~Sun.
\newblock {Deep Residual Learning for Image Recognition}.
\newblock In IEEE, editor, {\em Proc. CVPR'16}. IEEE, 2016.

\bibitem{Hearst2019a}
M.~Hearst and M.~Tory.
\newblock {Would You Like A Chart With That? Incorporating Visualizations into
  Conversational Interfaces}.
\newblock In {\em Proc. VIS'19}. IEEE, 2019.

\bibitem{Hearst2019}
M.~Hearst, M.~Tory, and V.~Setlur.
\newblock {Toward Interface Defaults for Vague Modifiers in Natural Language
  Interfaces for Visual Analysis}.
\newblock In {\em Proc. VIS'19}. IEEE, 2019.

\bibitem{Henkin2020}
R.~Henkin and C.~Turkay.
\newblock {Words of Estimative Correlation: Studying Verbalizations of
  Scatterplots}.
\newblock {\em IEEE Trans. Vis. Comput. Graph.}, 2020.

\bibitem{Herzig2020}
J.~Herzig, P.~K. Nowak, T.~M{\"{u}}ller, and et~al.
\newblock {TaPas: Weakly Supervised Table Parsing via Pre-training}.
\newblock In {\em Proc. ACL'20}. ACL, 2020.

\bibitem{Honnibal2017}
M.~Honnibal and I.~Montani.
\newblock {spacy 2: Natural language understanding with Bloom embeddings}.
\newblock {\em Convolutional Neural Networks Increm. Parsing}, 2017.

\bibitem{Hopkins2020}
A.~K. Hopkins, M.~Correll, and A.~Satyanarayan.
\newblock {VisuaLint: Sketchy In Situ Annotations of Chart Construction
  Errors}.
\newblock {\em Comput. Graph. Forum}, 39(3):219--228, 2020.

\bibitem{Hoque2018}
E.~Hoque, V.~Setlur, M.~Tory, and I.~Dykeman.
\newblock {Applying Pragmatics Principles for Interaction with Visual
  Analytics}.
\newblock {\em IEEE Trans. Vis. Comput. Graph.}, 24(1):309--318, 2018.

\bibitem{Hu2019d}
K.~Hu, M.~A. Bakker, S.~Li, and et~al.
\newblock {VizML: A machine learning approach to visualization recommendation}.
\newblock In {\em Proc. CHI'19}. ACM.

\bibitem{Hu2019e}
K.~Hu, S.~S. Gaikwad, M.~Hulsebos, and et~al.
\newblock {VizNet: Towards a large-scale visualization learning and
  benchmarking repository}.
\newblock In {\em Proc. CHI'19}. ACM, 2019.

\bibitem{Hu2018}
K.~Hu, D.~Orghian, and C.~Hidalgo.
\newblock {DIVE: A mixed-initiative system supporting integrated data
  exploration workflows}.
\newblock In {\em Proc. HILDA'2018}. ACM, 2018.

\bibitem{inproceedings}
B.~Huang, G.~Zhang, and P.~C.-Y. Sheu.
\newblock {A Natural Language Database Interface Based on a Probabilistic
  Context Free Grammar}.
\newblock In {\em Proc. WSCS'08}. IEEE, 2008.

\bibitem{Huang2019c}
Z.~Huang, Y.~Zhao, W.~Chen, and et~al.
\newblock {A Natural-language-based Visual Query Approach of Uncertain Human
  Trajectories}.
\newblock {\em IEEE Trans. Vis. Comput. Graph.}, 26(1):1--11, 2019.

\bibitem{Hullman2013}
J.~Hullman, N.~Diakopoulos, and E.~Adar.
\newblock {Contextifier: Automatic generation of annotated stock
  visualizations}.
\newblock In {\em Proc. CHI'13}. ACM.

\bibitem{Hulsebos2019}
M.~Hulsebos, A.~Satyanarayan, and et~al.
\newblock {Sherlock: A deep learning approach to semantic data type detection}.
\newblock In {\em Proc. KDD'19}. ACM, 2019.

\bibitem{Ingria2003}
R.~Ingria, R.~Sauri, J.~Pustejovsky, and et~al.
\newblock {TimeML: Robust Specification of Event and Temporal Expressions in
  Text}.
\newblock {\em New Dir. Quest. answering}, 3:28--34, 2003.

\bibitem{Iyer2017}
S.~Iyer, I.~Konstas, A.~Cheung, and et~al.
\newblock {Learning a Neural Semantic Parser from User Feedback}.
\newblock In {\em Proc. ACL'17}. ACL, 2017.

\bibitem{Jiang2021a}
Q.~Jiang, G.~Sun, Y.~Dong, and R.~Liang.
\newblock {DT2VIS: A Focus+Context Answer Generation System to Facilitate
  Visual Exploration of Tabular Data}.
\newblock {\em IEEE Comput. Graph. Appl.}, 41(5):45--56, 2021.

\bibitem{Jiang2021}
S.~Jiang, P.~Kang, X.~Song, and et~al.
\newblock {Emerging Wearable Interfaces and Algorithms for Hand Gesture
  Recognition: A Survey}.
\newblock {\em IEEE Rev. Biomed. Eng.}, pages 1--11, 2021.

\bibitem{joshi2020spanbert}
M.~Joshi, D.~Chen, Y.~Liu, and et~al.
\newblock {SpanBERT: Improving Pre-training by Representing and Predicting
  Spans}.
\newblock {\em Trans. Assoc. Comput. Linguist.}, 8:64--77, 2020.

\bibitem{joshi2019bert}
M.~Joshi, O.~Levy, L.~Zettlemoyer, and D.~Weld.
\newblock {BERT for Coreference Resolution: Baselines and Analysis}.
\newblock In {\em Proc. EMNLP'19}. ACL, 2019.

\bibitem{Kafle2018}
K.~Kafle, B.~Price, S.~Cohen, and C.~Kanan.
\newblock {DVQA: Understanding Data Visualizations via Question Answering}.
\newblock In {\em Proc. CVPR'18}. IEEE, 2018.

\bibitem{Easoning2018}
S.~E. Kahou, V.~Michalski, A.~Atkinson, and et~al.
\newblock {FigureQA: An Annotated Figure Dataset for Visual Reasoning}.
\newblock In {\em Proc. ICLR'18}, 2018.

\bibitem{Kandel2011}
S.~Kandel, A.~Paepcke, J.~Hellerstein, and J.~Heer.
\newblock {Wrangler: Interactive visual specification of data transformation
  scripts}.
\newblock In {\em Proc. CHI'11}. ACM, 2011.

\bibitem{Kandel2012b}
S.~Kandel, R.~Parikh, A.~Paepcke, and et~al.
\newblock {Profiler: Integrated statistical analysis and visualization for data
  quality assessment}.
\newblock In {\em Proc. AVI'12}. ACM, 2012.

\bibitem{Kandogan}
E.~Kandogan.
\newblock {Just-in-time annotation of clusters, outliers, and trends in
  point-based data visualizations}.
\newblock In {\em Proc. VAST'12}. IEEE, 2012.

\bibitem{Condiff2017}
J.~Kang, K.~Condiff, S.~Chang, and et~al.
\newblock {Understanding How People Use Natural Language to Ask for
  Recommendations}.
\newblock In {\em Proc. RecSys'17}. ACM, 2017.

\bibitem{Kassel2018}
J.~F. Kassel and M.~Rohs.
\newblock {Valletto: A multimodal interface for ubiquitous visual analytics}.
\newblock In {\em Proc. CHI EA'18}. ACM, 2018.

\bibitem{Kassel2019}
J.~F. Kassel and M.~Rohs.
\newblock {Talk to me intelligibly: Investigating an answer space to match the
  user's language in visual analysis}.
\newblock In {\em Proc. DIS'19}. ACM, 2019.

\bibitem{Kato2002}
T.~Kato, M.~Matsushita, and E.~Maeda.
\newblock {Answering it with charts: dialogue in natural language and charts}.
\newblock In {\em Proc. COLING'02}. ACM, 2002.

\bibitem{Kaur2015b}
P.~Kaur, M.~Owonibi, and B.~Koenig-Ries.
\newblock {Towards visualization recommendation-a semi-automated
  domain-specific learning approach}.
\newblock {\em CEUR Workshop Proc.}, 1366:30--35, 2015.

\bibitem{Kerpedjiev1997}
S.~Kerpedjiev, G.~Carenini, S.~F. Roth, and J.~D. Moore.
\newblock {AutoBrief: a multimedia presentation system for assisting data
  analysis}.
\newblock {\em Comput. Stand. Interfaces}, 18(6-7):583--593, 1997.

\bibitem{Kerracher2017}
N.~Kerracher and J.~Kennedy.
\newblock {Constructing and Evaluating Visualisation Task Classifications:
  Process and Considerations}.
\newblock {\em Comput. Graph. Forum}, 36(3):47--59, 2017.

\bibitem{Key2012}
A.~Key, B.~Howe, D.~Perry, and C.~Aragon.
\newblock {VizDeck: Self-organizing dashboards for visual analytics}.
\newblock In {\em Proc. SIGMOD'12}. ACM, 2012.

\bibitem{Kim2020a}
D.~H. Kim, E.~Hoque, and M.~Agrawala.
\newblock {Answering Questions about Charts and Generating Visual
  Explanations}.
\newblock In {\em Proc. CHI'20}. ACM.

\bibitem{Kim1999}
D.~H. Kim, V.~Setlur, and M.~Agrawala.
\newblock {Towards Understanding How Readers Integrate Charts and Captions: A
  Case Study with Line Charts}.
\newblock In {\em Proc. CHI'21}. ACM, 2021.

\bibitem{Kim2019}
H.~Kim, J.~Oh, Y.~Han, and et~al.
\newblock {Thumbnails for Data Stories: A Survey of Current Practices}.
\newblock In {\em Proc. VIS'19}. IEEE, 2019.

\bibitem{Kim2018b}
Y.~Kim and J.~Heer.
\newblock {Assessing Effects of Task and Data Distribution on the Effectiveness
  of Visual Encodings}.
\newblock {\em Comput. Graph. Forum}, 37(3):157--167, 2018.

\bibitem{Kim2020}
Y.~Kim and J.~Heer.
\newblock {Gemini: A Grammar and Recommender System for Animated Transitions in
  Statistical Graphics}.
\newblock {\em IEEE Trans. Vis. Comput. Graph.}, 27(2):485--494, 2021.

\bibitem{Young-HoKim2021}
Y.-H. Kim, B.~Lee, A.~Srinivasan, and E.~K. Choe.
\newblock {Data@Hand: Fostering Visual Exploration of Personal Data on
  Smartphones Leveraging Speech and Touch Interaction}.
\newblock In {\em Proc. CHI'21}. ACM, 2021.

\bibitem{Kincaid2017}
R.~Kincaid and G.~Pollock.
\newblock {Nicky: Toward a Virtual Assistant for Test and Measurement
  Instrument Recommendations}.
\newblock In {\em Proc. ICSC'17}. IEEE, 2017.

\bibitem{kirstain2021coreference}
Y.~Kirstain, O.~Ram, and O.~Levy.
\newblock {Coreference Resolution without Span Representations}.
\newblock {\em arXiv}, 2021.

\bibitem{Kocijan_2019}
V.~Kocijan, A.-M. Cretu, O.-M. Camburu, and et~al.
\newblock {A Surprisingly Robust Trick for the Winograd Schema Challenge}.
\newblock {\em Proc. 57th Annu. Meet. Assoc. Comput. Linguist. ACL'19}, pages
  4837--4842, 2019.

\bibitem{Kong2012}
N.~Kong and M.~Agrawala.
\newblock {Graphical Overlays: Using Layered Elements to Aid Chart Reading}.
\newblock {\em IEEE Trans. Vis. Comput. Graph.}, 18(12), 2012.

\bibitem{Kumar2017}
A.~Kumar, J.~Aurisano, B.~{Di Eugenio}, and et~al.
\newblock {Multimodal Coreference Resolution for Exploratory Data Visualization
  Dialogue: Context-Based Annotation and Gesture Identification}.
\newblock In {\em Proc. SEMDIAL'17}. ISCA.

\bibitem{Kumar2016}
A.~Kumar, J.~Aurisano, and et~al.
\newblock {Towards a dialogue system that supports rich visualizations of
  data}.
\newblock In {\em Proc. SIGDIAL'16}. ACL, 2016.

\bibitem{Lai2020a}
C.~Lai, Z.~Lin, R.~Jiang, and et~al.
\newblock {Automatic Annotation Synchronizing with Textual Description for
  Visualization}.
\newblock In {\em Proc. CHI'20}. ACM, 2020.

\bibitem{Lalle2019}
S.~Lalle, D.~Toker, and C.~Conati.
\newblock {Gaze-Driven Adaptive Interventions for Magazine-Style Narrative
  Visualizations}.
\newblock {\em IEEE Trans. Vis. Comput. Graph.}, 27(6):2941--2952, 2019.

\bibitem{Langner2021}
R.~Langner, M.~Satkowski, W.~B{\"{u}}schel, and R.~Dachselt.
\newblock {MARVIS: Combining Mobile Devices and Augmented Reality for Visual
  Data Analysis}.
\newblock In {\em Proc. CHI'21}. ACM, 2021.

\bibitem{Lee2021}
B.~Lee, D.~Brown, B.~Lee, and et~al.
\newblock {Data Visceralization: Enabling Deeper Understanding of Data Using
  Virtual Reality}.
\newblock {\em IEEE Trans. Vis. Comput. Graph.}, 27(2):1095--1105, 2021.

\bibitem{Lee2012}
B.~Lee, P.~Isenberg, N.~H. Riche, and S.~Carpendale.
\newblock {Beyond Mouse and Keyboard: Expanding Design Considerations for
  Information Visualization Interactions}.
\newblock {\em IEEE Trans. Vis. Comput. Graph.}, 2012.

\bibitem{Lee2006}
B.~Lee, C.~Plaisant, C.~S. Parr, and et~al.
\newblock {Task taxonomy for graph visualization}.
\newblock In {\em Proc. BELIV'06}. ACM, 2006.

\bibitem{Lee2021c}
B.~Lee, A.~Srinivasan, P.~Isenberg, and J.~Stasko.
\newblock {Post-wimp interaction for information visualization}.
\newblock {\em Found. Trends Human-Computer Interact.}, 14(1):1--95, 2021.

\bibitem{Lee2021d}
D.~J.~L. Lee, A.~Quamar, E.~Kandogan, and F.~{\"{O}}zcan.
\newblock {Boomerang: Proactive Insight-Based Recommendations for Guiding
  Conversational Data Analysis}.
\newblock In {\em Proc. SIGMOD'21 demo}, 2021.

\bibitem{Lee2021a}
D.~J.-L. Lee, V.~Setlur, M.~Tory, and et~al.
\newblock {Deconstructing Categorization in Visualization Recommendation: A
  Taxonomy and Comparative Study}.
\newblock {\em IEEE Trans. Vis. Comput. Graph.}, 2626:1--14, 2021.

\bibitem{lee2017endtoend}
K.~Lee, L.~He, M.~Lewis, and L.~Zettlemoyer.
\newblock {End-to-end Neural Coreference Resolution}.
\newblock In {\em Proc. EMNLP'17}. ACL, 2017.

\bibitem{lee2018higherorder}
K.~Lee, L.~He, and L.~Zettlemoyer.
\newblock {Higher-Order Coreference Resolution with Coarse-to-Fine Inference}.
\newblock In {\em Proc. NAACL'18}. ACL, 2018.

\bibitem{LeoJohn2018}
R.~J. {Leo John}, J.~M. Patel, A.~L. Alexander, and et~al.
\newblock {A Natural Language Interface for Dissemination of Reproducible
  Biomedical Data Science}.
\newblock {\em Lect. Notes Comput. Sci.}, 11073:197--205, 2018.

\bibitem{LeoJohn2017}
R.~J. {Leo John}, N.~Potti, and J.~M. Patel.
\newblock {Ava: From data to insights through conversation}.
\newblock In {\em Proc. CIDR'17}, 2017.

\bibitem{Li2014a}
F.~Li and H.~V. Jagadish.
\newblock {NaLIR: An interactive natural language interface for querying
  relational databases}.
\newblock In {\em Proc. SIGMOD'14}. ACM.

\bibitem{Li2014}
F.~Li and H.~V. Jagadish.
\newblock {Constructing an interactive natural language interface for
  relational databases}.
\newblock {\em Proc. VLDB Endow.}, 8(1):73--84, 2014.

\bibitem{Li2016a}
F.~Li and H.~V. Jagadish.
\newblock {Understanding Natural Language Queries over Relational Databases}.
\newblock {\em ACM SIGMOD Rec.}, 45(1):6--13, 2016.

\bibitem{Li2021a}
H.~Li, Y.~Wang, S.~Zhang, and et~al.
\newblock {KG4Vis: A Knowledge Graph-Based Approach for Visualization
  Recommendation}.
\newblock {\em IEEE Trans. Vis. Comput. Graph.}, pages 1--11, 2021.

\bibitem{Li2019f}
J.~Li, J.~Yang, A.~Hertzmann, and et~al.
\newblock {LayoutGAN: Generating Graphic Layouts with Wireframe
  Discriminators}.
\newblock {\em arXiv}, 2019.

\bibitem{Li2019}
L.~H. Li, M.~Yatskar, D.~Yin, and et~al.
\newblock {VisualBERT: A Simple and Performant Baseline for Vision and
  Language}.
\newblock {\em arXiv}, 2019.

\bibitem{8029993}
P.~Li, L.~Liu, J.~Xu, and et~al.
\newblock {Application of Hidden Markov Model in SQL Injection Detection}.
\newblock In {\em Proc. COMPSAC'17}. IEEE, 2017.

\bibitem{Lin2020a}
H.~Lin, D.~Moritz, and J.~Heer.
\newblock {Dziban: Balancing Agency {\&} Automation in Visualization Design via
  Anchored Recommendations}.
\newblock In {\em Proc. CHI'20}. ACM, 2020.

\bibitem{Liu2021b}
C.~Liu, Y.~Han, R.~Jiang, and X.~Yuan.
\newblock {ADVISor: Automatic Visualization Answer for Natural-Language
  Question on Tabular Data}.
\newblock In {\em Proc. PacificVis'21}. IEEE, 2021.

\bibitem{Liu2020a}
C.~Liu, L.~Xie, Y.~Han, and et~al.
\newblock {AutoCaption: An Approach to Generate Natural Language Description
  from Visualization Automatically}.
\newblock In {\em Proc. PacificVis'20}. IEEE, 2020.

\bibitem{liu2016commonsense}
Q.~Liu, H.~Jiang, Z.-H. Ling, and et~al.
\newblock {Commonsense Knowledge Enhanced Embeddings for Solving Pronoun
  Disambiguation Problems in Winograd Schema Challenge}.
\newblock {\em arXiv}, 2016.

\bibitem{LiuTingting}
T.~Liu, X.~Li, C.~Bao, and et~al.
\newblock {Data-Driven Mark Orientation for Trend Estimation in Scatterplots}.
\newblock In {\em Proc. CHI'21}. ACM, 2021.

\bibitem{Agosti2015}
G.~L{\'{o}}pez, L.~Quesada, and L.~A. Guerrero.
\newblock {Alexa vs. Siri vs. Cortana vs. Google Assistant: A Comparison of
  Speech-Based Natural User Interfaces}.
\newblock In {\em Proc. AHFE'17}. Springer, 2017.

\bibitem{Luo2018a}
Y.~Luo, X.~Qin, N.~Tang, and et~al.
\newblock {DeepEye: Creating good data visualizations by keyword search}.
\newblock In {\em Proc. SIGMOD'18}. ACM, 2018.

\bibitem{Luo2018}
Y.~Luo, X.~Qin, N.~Tang, and G.~Li.
\newblock {Deepeye: towards automatic data visualization}.
\newblock In {\em Proc. ICDE'18}. IEEE, 2018.

\bibitem{Luo2021a}
Y.~Luo, N.~Tang, G.~Li, and et~al.
\newblock {Natural Language to Visualization by Neural Machine Translation}.
\newblock In {\em Proc. VIS'21}. IEEE, 2021.

\bibitem{Luo2021}
Y.~Luo, N.~Tang, G.~Li, and et~al.
\newblock {Synthesizing Natural Language to Visualization (NL2VIS) Benchmarks
  from NL2SQL Benchmarks}.
\newblock In {\em Proc. SIGMOD'21}. ACM, 2021.

\bibitem{Ma2020b}
Y.~Ma, A.~K.~H. Tung, W.~Wang, and et~al.
\newblock {ScatterNet: A Deep Subjective Similarity Model for Visual Analysis
  of Scatterplots}.
\newblock {\em IEEE Trans. Vis. Comput. Graph.}, 26(3):1562--1576, 2020.

\bibitem{Mackinlay1986b}
J.~Mackinlay.
\newblock {Automating the design of graphical presentations of relational
  information}.
\newblock {\em ACM Trans. Graph.}, 5(2):110--141, 1986.

\bibitem{Mackinlay2007b}
J.~Mackinlay, P.~Hanrahan, and C.~Stolte.
\newblock {Show Me: Automatic Presentation for Visual Analysis}.
\newblock {\em IEEE Trans. Vis. Comput. Graph.}, 13(6):1137--1144, 2007.

\bibitem{Madan2018}
S.~Madan, Z.~Bylinskii, M.~Tancik, and et~al.
\newblock {Synthetically Trained Icon Proposals for Parsing and Summarizing
  Infographics}.
\newblock {\em arXiv}, 2018.

\bibitem{Maji2021}
S.~Maji, S.~S. Rout, and S.~Choudhary.
\newblock {DCoM: A Deep Column Mapper for Semantic Data Type Detection}.
\newblock {\em arXiv}, 2021.

\bibitem{Manning2014}
C.~Manning, M.~Surdeanu, J.~Bauer, and et~al.
\newblock {The Stanford CoreNLP Natural Language Processing Toolkit}.
\newblock In {\em Proc. ACL'14}. ACL, 2014.

\bibitem{Mathew2021}
M.~Mathew, V.~Bagal, and et~al.
\newblock {InfographicVQA}.
\newblock {\em arXiv}, 2021.

\bibitem{Matsushita2004}
M.~Matsushita, E.~Maeda, and T.~Kato.
\newblock {An interactive visualization method of numerical data based on
  natural language requirements}.
\newblock {\em Int. J. Hum. Comput. Stud.}, 60(4):469--488, 2004.

\bibitem{Mcnabb2017}
L.~McNabb and R.~S. Laramee.
\newblock {Survey of Surveys (SoS) ‐ Mapping The Landscape of Survey Papers
  in Information Visualization}.
\newblock {\em Comput. Graph. Forum}, 36(3):589--617, 2017.

\bibitem{Metoyer2012}
R.~Metoyer, B.~Lee, N.~{Henry Riche}, and M.~Czerwinski.
\newblock {Understanding the verbal language and structure of end-user
  descriptions of data visualizations}.
\newblock In {\em Proc. CHI'12}. ACM, 2012.

\bibitem{Metoyer2018}
R.~Metoyer, Q.~Zhi, B.~Janczuk, and W.~Scheirer.
\newblock {Coupling story to visualization: Using textual analysis as a bridge
  between data and interpretation}.
\newblock In {\em Proc. IUI'18}. ACM, 2018.

\bibitem{Mikolov2013}
T.~Mikolov, K.~Chen, G.~Corrado, and J.~Dean.
\newblock {Efficient estimation of word representations in vector space}.
\newblock In {\em Proc. ICLR'13}, 2013.

\bibitem{Miller1995}
G.~A. Miller.
\newblock {WordNet: A Lexical Database for English}.
\newblock {\em Commun. ACM}, 38(11):39--41, 1995.

\bibitem{Mitri2020}
M.~Mitri.
\newblock {Story Analysis Using Natural Language Processing and Interactive
  Dashboards}.
\newblock {\em J. Comput. Inf. Syst.}, pages 1--11, 2020.

\bibitem{Mittal1998a}
V.~O. Mittal, J.~D. Moore, G.~Carenini, and S.~Roth.
\newblock {Describing Complex Charts in Natural Language: A Caption Generation
  System}.
\newblock {\em Comput. Linguist.}, 24(3):431--467, 1998.

\bibitem{Moraes2014}
P.~Moraes, G.~Sina, K.~McCoy, and S.~Carberry.
\newblock {Generating Summaries of Line Graphs}.
\newblock In {\em Proc. INLG'14}. ACL, 2014.

\bibitem{Moritz2019}
D.~Moritz, C.~Wang, G.~L. Nelson, and et~al.
\newblock {Formalizing Visualization Design Knowledge as Constraints:
  Actionable and Extensible Models in Draco}.
\newblock {\em IEEE Trans. Vis. Comput. Graph.}, 25(1):438--448, 2019.

\bibitem{Murillo-Morales2020}
T.~Murillo-Morales and K.~Miesenberger.
\newblock {AUDiaL: A Natural Language Interface to Make Statistical Charts
  Accessible to Blind Persons}.
\newblock In {\em Lect. Notes Comput. Sci.}, pages 373--384. Springer, 2020.

\bibitem{Mutlu2016b}
B.~Mutlu, E.~Veas, and C.~Trattner.
\newblock {VizRec: Recommending personalized visualizations}.
\newblock {\em ACM Trans. Interact. Intell. Syst.}, 6(4):1--39, 2016.

\bibitem{Nafari2013}
M.~Nafari and C.~Weaver.
\newblock {Augmenting Visualization with Natural Language Translation of
  Interaction: A Usability Study}.
\newblock {\em Comput. Graph. Forum}, 32(3):391--400, 2013.

\bibitem{Nafari2015}
M.~Nafari and C.~Weaver.
\newblock {Query2Question: Translating visualization interaction into natural
  language}.
\newblock {\em IEEE Trans. Vis. Comput. Graph.}, 21(6):756--769, 2015.

\bibitem{Narechania2021}
A.~Narechania, A.~Fourney, and et~al.
\newblock {DIY: Assessing the Correctness of Natural Language to SQL Systems}.
\newblock In {\em Proc. IUI'21}. ACM, 2021.

\bibitem{Narechania2020}
A.~Narechania, A.~Srinivasan, and J.~Stasko.
\newblock {NL4DV: A Toolkit for Generating Analytic Specifications for Data
  Visualization from Natural Language Queries}.
\newblock {\em IEEE Trans. Vis. Comput. Graph.}, 27(2), 2021.

\bibitem{Nihalani2011NaturalLI}
N.~Nihalani, S.~Silakari, and M.~Motwani.
\newblock {Natural language Interface for Database: A Brief review}.
\newblock {\em Int. J. Comput. Sci.}, 8(2):600--608, 2011.

\bibitem{Nonato2019}
L.~G. Nonato and M.~Aupetit.
\newblock {Multidimensional Projection for Visual Analytics: Linking Techniques
  with Distortions, Tasks, and Layout Enrichment}.
\newblock {\em IEEE Trans. Vis. Comput. Graph.}, 25(8):2650--2673, 2019.

\bibitem{Obeid2020}
J.~Obeid and E.~Hoque.
\newblock {Chart-to-Text: Generating Natural Language Descriptions for Charts
  by Adapting the Transformer Model}.
\newblock In {\em Proc. INLG'20}. ACL, 2020.

\bibitem{Oppermann}
M.~Oppermann, R.~Kincaid, and T.~Munzner.
\newblock {VizCommender: Computing Text-Based Similarity in Visualization
  Repositories for Content-Based Recommendations}.
\newblock {\em IEEE Trans. Vis. Comput. Graph.}, 2021.

\bibitem{ELMo}
M.~Peters, M.~Neumann, M.~Iyyer, and et~al.
\newblock {Deep Contextualized Word Representations}.
\newblock In {\em Proc. NAACL'18}. ACL, 2018.

\bibitem{Poco2017}
J.~Poco and J.~Heer.
\newblock {Reverse‐Engineering Visualizations: Recovering Visual Encodings
  from Chart Images}.
\newblock {\em Comput. Graph. Forum}, 2017.

\bibitem{pradhan-etal-2012-conll}
S.~Pradhan, A.~Moschitti, N.~Xue, and et~al.
\newblock {CoNLL-2012 Shared Task: Modeling Multilingual Unrestricted
  Coreference in OntoNotes}.
\newblock In {\em Proc. EMNLP'12}. ACL, 2012.

\bibitem{stanza}
P.~Qi, Y.~Zhang, Y.~Zhang, and et~al.
\newblock {Stanza: A Python Natural Language Processing Toolkit for Many Human
  Languages}.
\newblock In {\em Proc. ACL'20}. ACL.

\bibitem{Qian2020}
C.~Qian, S.~Sun, W.~Cui, and et~al.
\newblock {Retrieve-Then-Adapt: Example-based Automatic Generation for
  Proportion-related Infographics}.
\newblock {\em IEEE Trans. Vis. Comput. Graph.}, 27(2):443--452, 2021.

\bibitem{Qian2021}
X.~Qian, E.~Koh, F.~Du, and et~al.
\newblock {Generating Accurate Caption Units for Figure Captioning}.
\newblock In {\em Proc. WWW'21}. ACM, 2021.

\bibitem{Rossi}
X.~Qian, R.~A. Rossi, F.~Du, and et~al.
\newblock {Learning to Recommend Visualizations from Data}.
\newblock In {\em Proc. KDD'21}. ACM, 2021.

\bibitem{Qian2021q}
X.~Qian, R.~A. Rossi, F.~Du, and et~al.
\newblock {Personalized Visualization Recommendation}.
\newblock {\em arXiv}, 2021.

\bibitem{Qin2020}
X.~Qin, Y.~Luo, N.~Tang, and G.~Li.
\newblock {Making data visualization more efficient and effective: a survey}.
\newblock {\em VLDB J.}, 29(1):93--117, 2020.

\bibitem{Quamar2020}
A.~Quamar, C.~Lei, D.~Miller, and et~al.
\newblock {An Ontology-Based Conversation System for Knowledge Bases}.
\newblock In {\em Proc. SIGMOD'20}. ACM, 2020.

\bibitem{Reinders2020}
S.~Reinders, M.~Butler, and K.~Marriott.
\newblock {"Hey Model!" - Natural User Interactions and Agency in Accessible
  Interactive 3D Models}.
\newblock In {\em Proc. CHI'20}. ACM, 2020.

\bibitem{Ren}
D.~Ren, M.~Brehmer, {Bongshin Lee}, and et~al.
\newblock {ChartAccent: Annotation for data-driven storytelling}.
\newblock In {\em Proc. PacificVis'17}. IEEE, 2017.

\bibitem{Rind2016}
A.~Rind, W.~Aigner, M.~Wagner, and et~al.
\newblock {Task Cube: A three-dimensional conceptual space of user tasks in
  visualization design and evaluation}.
\newblock {\em Inf. Vis.}, 15(4):288--300, 2016.

\bibitem{Roth1994b}
S.~F. Roth, J.~Kolojejchick, J.~Mattis, and J.~Goldstein.
\newblock {Interactive graphic design using automatic presentation knowledge}.
\newblock In {\em Proc. CHI'94}. ACM.

\bibitem{Saha2016}
D.~Saha, A.~Floratou, K.~Sankaranarayanan, and et~al.
\newblock {ATHENA: An ontologydriven system for natural language querying over
  relational data stores}.
\newblock {\em Proc. VLDB Endow.}, 9(12):1209--1220, 2016.

\bibitem{Saket2019b}
B.~Saket, A.~Endert, and C.~Demiralp.
\newblock {Task-Based Effectiveness of Basic Visualizations}.
\newblock {\em IEEE Trans. Vis. Comput. Graph.}, 25(7), 2019.

\bibitem{Saktheeswaran2020}
A.~Saktheeswaran, A.~Srinivasan, and J.~Stasko.
\newblock {Touch? Speech? or Touch and Speech? Investigating Multimodal
  Interaction for Visual Network Exploration and Analysis}.
\newblock {\em IEEE Trans. Vis. Comput. Graph.}, 26(6):2168--2179, 2020.

\bibitem{Sarikaya2018b}
A.~Sarikaya and M.~Gleicher.
\newblock {Scatterplots: Tasks, Data, and Designs}.
\newblock {\em IEEE Trans. Vis. Comput. Graph.}, 24(1):402--412, 2018.

\bibitem{Satyanarayan2017}
A.~Satyanarayan, D.~Moritz, K.~Wongsuphasawat, and J.~Heer.
\newblock {Vega-Lite: A Grammar of Interactive Graphics}.
\newblock {\em IEEE Trans. Vis. Comput. Graph.}, 23(1):341--350, 2017.

\bibitem{Satyanarayan2016}
A.~Satyanarayan, R.~Russell, J.~Hoffswell, and J.~Heer.
\newblock {Reactive Vega: A Streaming Dataflow Architecture for Declarative
  Interactive Visualization}.
\newblock {\em IEEE Trans. Vis. Comput. Graph.}, 22(1):659--668, 2016.

\bibitem{Savva2011}
M.~Savva, N.~Kong, A.~Chhajta, and et~al.
\newblock {ReVision: Automated classification, analysis and redesign of chart
  images}.
\newblock In {\em Proc. UIST'11}. ACM, 2011.

\bibitem{Segel2010}
E.~Segel and J.~Heer.
\newblock {Narrative visualization: Telling stories with data}.
\newblock {\em IEEE Trans. Vis. Comput. Graph.}, 16(6):1139--1148, 2010.

\bibitem{Seipel2019}
P.~Seipel, A.~Stock, S.~Santhanam, and et~al.
\newblock {Speak to your Software Visualization—Exploring Component-Based
  Software Architectures in Augmented Reality with a Conversational Interface}.
\newblock In {\em Proc. VISSOFT'19}. IEEE, 2019.

\bibitem{Sen2019}
J.~Sen, F.~Ozcan, and et~al.
\newblock {Natural Language Querying of Complex Business Intelligence Queries}.
\newblock In {\em Proc. SIGMOD'19}. ACM, 2019.

\bibitem{Seo2005b}
J.~Seo and B.~Shneiderman.
\newblock {A Rank-by-Feature Framework for Interactive Exploration of
  Multidimensional Data}.
\newblock {\em Inf. Vis.}, 2005.

\bibitem{Setlur}
V.~Setlur, S.~Battersby, and T.~Wong.
\newblock {GeoSneakPique : Visual Autocompletion for Geospatial Queries}.
\newblock In {\em Proc. VIS'21}. IEEE, 2021.

\bibitem{Setlur2016}
V.~Setlur, S.~E. Battersby, M.~Tory, and et~al.
\newblock {Eviza: A natural language interface for visual analysis}.
\newblock In {\em Proc. UIST'16}. ACM, 2016.

\bibitem{Setlur2020}
V.~Setlur, E.~Hoque, D.~H. Kim, and A.~X. Chang.
\newblock {Sneak pique: Exploring autocompletion as a data discovery scaffold
  for supporting visual analysis}.
\newblock In {\em Proc. UIST'20}. ACM, 2020.

\bibitem{Setlur2020a}
V.~Setlur and A.~Kumar.
\newblock {Sentifiers: Interpreting Vague Intent Modifiers in Visual Analysis
  using Word Co-occurrence and Sentiment Analysis}.
\newblock In {\em Proc. VIS'20}. IEEE, 2020.

\bibitem{Setlur2019}
V.~Setlur, M.~Tory, and A.~Djalali.
\newblock {Inferencing underspecified natural language utterances in visual
  analysis}.
\newblock In {\em Proc. IUI'19}. ACM, 2019.

\bibitem{Sevastjanova2018}
R.~Sevastjanova, F.~Beck, B.~Ell, and et~al.
\newblock {Going beyond Visualization: Verbalization as Complementary Medium to
  Explain Machine Learning Models}.
\newblock In {\em Proc. VISxAI'18}. IEEE, 2018.

\bibitem{Shekarpour2015}
S.~Shekarpour, E.~Marx, A.-C. {Ngonga Ngomo}, and S.~Auer.
\newblock {SINA: Semantic interpretation of user queries for question answering
  on interlinked data}.
\newblock {\em J. Web Semant.}, 30:39--51, 2015.

\bibitem{Shen2021}
L.~Shen, E.~Shen, Z.~Tai, and et~al.
\newblock {TaskVis : Task-oriented Visualization Recommendation}.
\newblock In {\em Proc. EuroVis'21}. Eurographics, 2021.

\bibitem{Shi2019}
D.~Shi, Y.~Shi, X.~Xu, and et~al.
\newblock {Task-Oriented Optimal Sequencing of Visualization Charts}.
\newblock In {\em Proc. VDS'19}. IEEE, 2019.

\bibitem{Shi2020}
D.~Shi, X.~Xu, F.~Sun, and et~al.
\newblock {Calliope: Automatic Visual Data Story Generation from a
  Spreadsheet}.
\newblock {\em IEEE Trans. Vis. Comput. Graph.}, 27(2):453--463, 2021.

\bibitem{Shi2018}
Y.~Shi, C.~Bryan, S.~Bhamidipati, and et~al.
\newblock {MeetingVis: Visual Narratives to Assist in Recalling Meeting Context
  and Content}.
\newblock {\em IEEE Trans. Vis. Comput. Graph.}, 24(6):1918--1929, 2018.

\bibitem{Shu2020}
X.~Shu, A.~Wu, J.~Tang, and et~al.
\newblock {What Makes a Data-GIF Understandable?}
\newblock {\em IEEE Trans. Vis. Comput. Graph.}, 27(2):1492--1502, 2021.

\bibitem{Siddiqui2020}
N.~Siddiqui and E.~Hoque.
\newblock {ConVisQA: A Natural Language Interface for Visually Exploring Online
  Conversations}.
\newblock In {\em Proc. IV'20}. IEEE, 2020.

\bibitem{Siddiqui2018b}
T.~Siddiqui, P.~Luh, Z.~Wang, and et~al.
\newblock {ShapeSearch: Flexible patternbased querying of trend line
  visualizations}.
\newblock {\em Proc. VLDB Endow.}, 11(12):1962--1965, 2018.

\bibitem{Siddiqui2021}
T.~Siddiqui, P.~Luh, Z.~Wang, and et~al.
\newblock {From Sketching to Natural Language: Expressive Visual Querying for
  Accelerating Insight}.
\newblock {\em SIGMOD Rec.}, 50(1):51--58, 2021.

\bibitem{Sigtia2020}
S.~Sigtia, E.~Marchi, S.~Kajarekar, and et~al.
\newblock {Multi-Task Learning for Speaker Verification and Voice Trigger
  Detection}.
\newblock In {\em Proc. ICASSP'20}. IEEE, 2020.

\bibitem{Simitsis2007}
A.~Simitsis, G.~Koutrika, and Y.~Ioannidis.
\newblock {Pr{\'{e}}cis: from unstructured keywords as queries to structured
  databases as answers}.
\newblock {\em VLDB J.}, 17(1):117--149, 2007.

\bibitem{miikkulainen:book93}
S.~Sinclair, R.~Miikkulainen, and S.~Sinclair.
\newblock {\em {Subsymbolic Natural Language Processing: An Integrated Model of
  Scripts, Lexicon, and Memory}}, volume~73.
\newblock MIT Press, 1997.

\bibitem{Singh2020}
H.~Singh and S.~Shekhar.
\newblock {STL-CQA: Structure-based Transformers with Localization and Encoding
  for Chart Question Answering}.
\newblock In {\em Proc. EMNLP'20}. ACL, 2020.

\bibitem{Spreafico2020}
A.~Spreafico and G.~Carenini.
\newblock {Neural Data-Driven Captioning of Time-Series Line Charts}.
\newblock In {\em Proc. AVI'20}. ACM, 2020.

\bibitem{Dontcheva}
A.~Srinivasan, M.~Dontcheva, E.~Adar, and S.~Walker.
\newblock {Discovering natural language commands in multimodal interfaces}.
\newblock In {\em Proc. IUI'19}. ACM, 2019.

\bibitem{Srinivasan2019a}
A.~Srinivasan, S.~M. Drucker, A.~Endert, and J.~Stasko.
\newblock {Augmenting visualizations with interactive data facts to facilitate
  interpretation and communication}.
\newblock {\em IEEE Trans. Vis. Comput. Graph.}, 25(1):672--681, 2019.

\bibitem{Srinivasan2020a}
A.~Srinivasan, B.~Lee, N.~{Henry Riche}, and et~al.
\newblock {InChorus: Designing Consistent Multimodal Interactions for Data
  Visualization on Tablet Devices}.
\newblock In {\em Proc. CHI'20}. ACM, 2020.

\bibitem{Srinivasan2020b}
A.~Srinivasan, B.~Lee, and J.~T. Stasko.
\newblock {Interweaving Multimodal Interaction with Flexible Unit
  Visualizations for Data Exploration}.
\newblock {\em IEEE Trans. Vis. Comput. Graph.}, 14(8):1--15, 2020.

\bibitem{Srinivasan2021}
A.~Srinivasan, N.~Nyapathy, B.~Lee, and et~al.
\newblock {Collecting and Characterizing Natural Language Utterances for
  Specifying Data Visualizations}.
\newblock In {\em Proc. CHI'21}, 2021.

\bibitem{Srinivasana}
A.~Srinivasan and V.~Setlur.
\newblock {Snowy:Recommending Utterances for Conversational Visual Analysis}.
\newblock In {\em Proc. UIST'21}. ACM, 2021.

\bibitem{Srinivasan2017}
A.~Srinivasan and J.~Stasko.
\newblock {Natural Language Interfaces for Data Analysis with Visualization:
  Considering What Has and Could Be Asked}.
\newblock In {\em Proc. EuroVis'17}. Eurographics, 2017.

\bibitem{Srinivasan2018}
A.~Srinivasan and J.~Stasko.
\newblock {Orko: Facilitating Multimodal Interaction for Visual Exploration and
  Analysis of Networks}.
\newblock {\em IEEE Trans. Vis. Comput. Graph.}, 24(1):511--521, 2018.

\bibitem{Srinivasan2020}
A.~Srinivasan and J.~Stasko.
\newblock {How to ask what to say?: Strategies for evaluating natural language
  interfaces for data visualization}.
\newblock {\em IEEE Comput. Graph. Appl.}, 40(4):96--103, 2020.

\bibitem{Steichen2013}
B.~Steichen, G.~Carenini, and C.~Conati.
\newblock {User-adaptive information visualization - Using eye gaze data to
  infer visualization tasks and user cognitive abilities}.
\newblock In {\em Proc. IUI'13}. ACM, 2013.

\bibitem{Stolte2002b}
C.~Stolte, D.~Tang, and P.~Hanrahan.
\newblock {Polaris: a system for query, analysis, and visualization of
  multidimensional relational databases}.
\newblock {\em IEEE Trans. Vis. Comput. Graph.}, 8(1):52--65, 2002.

\bibitem{Streeb2021}
D.~Streeb, Y.~Metz, U.~Schlegel, and et~al.
\newblock {Task-based Visual Interactive Modeling: Decision Trees and
  Rule-based Classifiers}.
\newblock {\em IEEE Trans. Vis. Comput. Graph.}, XXX(XXX):1--18, 2021.

\bibitem{Stylianou2021}
N.~Stylianou and I.~Vlahavas.
\newblock {A neural Entity Coreference Resolution review}.
\newblock {\em Expert Syst. Appl.}, 168(114466):1--20, 2021.

\bibitem{Su2019}
W.~Su, X.~Zhu, Y.~Cao, and et~al.
\newblock {VL-BERT: Pre-training of Generic Visual-Linguistic Representations}.
\newblock {\em arXiv}, 2019.

\bibitem{Suhara2021}
Y.~Suhara, J.~Li, Y.~Li, and et~al.
\newblock {Annotating Columns with Pre-trained Language Models}.
\newblock {\em arXiv}, 2021.

\bibitem{Sun2010}
Y.~Sun, J.~Leigh, A.~Johnson, and S.~Lee.
\newblock {Articulate: A Semi-automated Model for Translating Natural Language
  Queries into Meaningful Visualizations}.
\newblock In {\em Lect. Notes Comput. Sci.} Springer, 2010.

\bibitem{Thompson2021}
J.~R. Thompson, Z.~Liu, and J.~Stasko.
\newblock {Data Animator: Authoring Expressive Animated Data Graphics}.
\newblock In {\em Proc. CHI'21}. ACM, 2021.

\bibitem{Tong2018}
C.~Tong, R.~Roberts, R.~Borgo, and et~al.
\newblock {Storytelling and visualization: An extended survey}.
\newblock {\em Inf.}, 9(3):1--42, 2018.

\bibitem{Tory2019b}
M.~Tory and V.~Setlur.
\newblock {Do What I Mean, Not What I Say! Design Considerations for Supporting
  Intent and Context in Analytical Conversation}.
\newblock In {\em Proc. VAST'19}. IEEE, 2019.

\bibitem{Tversky2002}
B.~Tversky, J.~B. Morrison, and M.~Betrancourt.
\newblock {Animation: Can it facilitate?}
\newblock {\em Int. J. Hum. Comput. Stud.}, 57(4):247--262, 2002.

\bibitem{VanDam1997}
A.~{Van Dam}.
\newblock {Post-WIMP User Interfaces}.
\newblock {\em Commun. ACM}, 40(2), 1997.

\bibitem{VandenElzen2013b}
S.~{Van Den Elzen} and J.~J. {Van Wijk}.
\newblock {Small multiples, large singles: A new approach for visual data
  exploration}.
\newblock {\em Comput. Graph. Forum}, 32(3 PART2):191--200, 2013.

\bibitem{Vartak2014b}
M.~Vartak, S.~Madden, A.~Parameswaran, and N.~Polyzotis.
\newblock {SEEDB: Automatically generating query visualizations}.
\newblock {\em Proc. VLDB Endow.}, 7(13):1581--1584, 2014.

\bibitem{Vartak2015b}
M.~Vartak, S.~Rahman, S.~Madden, and et~al.
\newblock {SEEDB: Efficient data-driven visualization recommendations to
  support visual analytics}.
\newblock {\em Proc. VLDB Endow.}, 8(13):2182--2193, 2015.

\bibitem{Furfaritony2002}
A.~Vaswani, N.~Shazeer, N.~Parmar, and et~al.
\newblock {Attention Is All You Need}.
\newblock In {\em Proc. NIPS'17}. NIPS, 2017.

\bibitem{Wang}
C.~Wang, Y.~Chen, Z.~Xue, and et~al.
\newblock {CogNet: Bridging Linguistic Knowledge, World Knowledge
  andCommonsense Knowledge}.
\newblock In {\em Proc. AAAI'21}. AAAI, 2020.

\bibitem{Wang2021a}
C.~Wang, Y.~Feng, R.~Bodik, and et~al.
\newblock {Falx: Synthesis-Powered Visualization Authoring}.
\newblock In {\em Proc. CHI'21}. ACM, 2021.

\bibitem{Wang2020c}
Q.~Wang, Z.~Chen, Y.~Wang, and H.~Qu.
\newblock {A Survey on ML4VIS: Applying Machine Learning Advances to Data
  Visualization}.
\newblock {\em arXiv}, 2021.

\bibitem{9101534}
W.~Wang, Y.~Tian, H.~Wang, and W.-S. Ku.
\newblock {A Natural Language Interface for Database: Achieving
  Transfer-learnability Using Adversarial Method for Question Understanding}.
\newblock In {\em Proc. ICDE'20}. IEEE, 2020.

\bibitem{Wang2018b}
Y.~Wang, F.~Han, L.~Zhu, and et~al.
\newblock {Line Graph or Scatter Plot? Automatic Selection of Methods for
  Visualizing Trends in Time Series}.
\newblock {\em IEEE Trans. Vis. Comput. Graph.}, 24(2):1141--1154, 2018.

\bibitem{Wang2020h}
Y.~Wang, Z.~Sun, H.~Zhang, and et~al.
\newblock {DataShot: Automatic Generation of Fact Sheets from Tabular Data}.
\newblock {\em IEEE Trans. Vis. Comput. Graph.}, 26(1):895--905, 2020.

\bibitem{Wang2017c}
Y.~Wang, W.~M. White, and E.~Andersen.
\newblock {PathViewer: Visualizing Pathways through Student Data}.
\newblock In {\em Proc. CHI'17}. ACM, 2017.

\bibitem{Wang2018}
Y.~Wang, H.~Zhang, H.~Huang, and et~al.
\newblock {InfoNice: Easy Creation of Information Graphics}.
\newblock In {\em Proc. CHI'18}. ACM, 2018.

\bibitem{Wang2020}
Z.~Wang, L.~Sundin, D.~Murray-Rust, and B.~Bach.
\newblock {Cheat Sheets for Data Visualization Techniques}.
\newblock In {\em Proc. CHI'20}. ACM, 2020.

\bibitem{Wohlgenannt2019}
G.~Wohlgenannt, D.~Mouromtsev, D.~Pavlov, and et~al.
\newblock {A Comparative Evaluation of Visual and Natural Language Question
  Answering over Linked Data}.
\newblock In {\em Proc.K'19}. SCITEPRESS, 2019.

\bibitem{Wongsuphasawat2016}
K.~Wongsuphasawat, D.~Moritz, A.~Anand, and et~al.
\newblock {Towards a general-purpose query language for visualization
  recommendation}.
\newblock In {\em Proc. HILDA'16}. ACM, 2016.

\bibitem{Wongsuphasawat2016c}
K.~Wongsuphasawat, D.~Moritz, A.~Anand, and et~al.
\newblock {Voyager: Exploratory Analysis via Faceted Browsing of Visualization
  Recommendations}.
\newblock {\em IEEE Trans. Vis. Comput. Graph.}, 22(1):649--658, 2016.

\bibitem{Wongsuphasawat2017}
K.~Wongsuphasawat, Z.~Qu, D.~Moritz, and et~al.
\newblock {Voyager 2: Augmenting visual analysis with partial view
  specifications}.
\newblock In {\em Proc. CHI'17}. ACM.

\bibitem{Wu2021d}
A.~Wu, Y.~Wang, X.~Shu, and et~al.
\newblock {AI4VIS: Survey on Artificial Intelligence Approaches for Data
  Visualization}.
\newblock {\em IEEE Trans. Vis. Comput. Graph.}, pages 1--20, 2021.

\bibitem{Wu2021e}
A.~Wu, Y.~Wang, M.~Zhou, and et~al.
\newblock {MultiVision: Designing Analytical Dashboards with Deep Learning
  Based Recommendation}.
\newblock {\em IEEE Trans. Vis. Comput. Graph.}, pages 1--11, 2021.

\bibitem{Wu2021}
A.~Wu, L.~Xie, B.~Lee, and et~al.
\newblock {Learning to Automate Chart Layout Configurations Using Crowdsourced
  Paired Comparison}.
\newblock In {\em Proc. CHI'21}. ACM, 2021.

\bibitem{Xia2020}
H.~Xia.
\newblock {Crosspower: Bridging graphics and linguistics}.
\newblock In {\em Proc. UIST'20}. ACM, 2020.

\bibitem{Xia2020a}
H.~Xia, J.~Jacobs, and M.~Agrawala.
\newblock {Crosscast: Adding Visuals to Audio Travel Podcasts}.
\newblock In {\em Proc. UIST'20}. ACM, 2020.

\bibitem{Xian2021}
Y.~Xian, H.~Zhao, T.~Y. Lee, and et~al.
\newblock {EXACTA: Explainable Column Annotation}.
\newblock In {\em Proc. KDD'21}. ACM, 2021.

\bibitem{xu2017}
X.~Xu, C.~Liu, and D.~Song.
\newblock {SQLNet: Generating Structured Queries From Natural Language Without
  Reinforcement Learning}.
\newblock {\em arXiv}, 2017.

\bibitem{Yagcioglu2020}
S.~Yagcioglu, A.~Erdem, E.~Erdem, and N.~Ikizler-Cinbis.
\newblock {RecipeQA: A Challenge Dataset for Multimodal Comprehension of
  Cooking Recipes}.
\newblock In {\em Proc. EMNLP'18}. ACL, 2018.

\bibitem{Yaghmazadeh2017}
N.~Yaghmazadeh, Y.~Wang, I.~Dillig, and T.~Dillig.
\newblock {SQLizer: query synthesis from natural language}.
\newblock {\em Proc. ACM Program. Lang.}, 2017.

\bibitem{Young2018}
T.~Young, D.~Hazarika, S.~Poria, and E.~Cambria.
\newblock {Recent Trends in Deep Learning Based Natural Language Processing}.
\newblock {\em IEEE Comput. Intell. Mag.}, 13(3):55--75, 2018.

\bibitem{Yu2017}
B.~Yu and C.~T. Silva.
\newblock {VisFlow - Web-based Visualization Framework for Tabular Data with a
  Subset Flow Model}.
\newblock {\em IEEE Trans. Vis. Comput. Graph.}, 23(1):251--260, 2017.

\bibitem{Yu2020b}
B.~Yu and C.~T. Silva.
\newblock {FlowSense: A Natural Language Interface for Visual Data Exploration
  within a Dataflow System}.
\newblock {\em IEEE Trans. Vis. Comput. Graph.}, 26(1):1--11, 2020.

\bibitem{Yu2020}
T.~Yu, R.~Zhang, K.~Yang, and et~al.
\newblock {Spider: A large-scale human-labeled dataset for complex and
  cross-domain semantic parsing and text-to-SQL task}.
\newblock In {\em Proc. EMNLP'18}. ACL, 2018.

\bibitem{Yuan2021}
L.~Yuan, Z.~Zhou, J.~Zhao, and et~al.
\newblock {InfoColorizer: Interactive Recommendation of Color Palettes for
  Infographics}.
\newblock {\em IEEE Trans. Vis. Comput. Graph.}, 2626:1--15, 2021.

\bibitem{Zehrung2021a}
R.~Zehrung, A.~Singhal, M.~Correll, and L.~Battle.
\newblock {Vis Ex Machina: An Analysis of Trust in Human versus Algorithmically
  Generated Visualization Recommendations}.
\newblock In {\em Proc. CHI'21}. ACM, 2021.

\bibitem{Zeng2021}
Z.~Zeng, P.~Moh, F.~Du, and et~al.
\newblock {An Evaluation-Focused Framework for Visualization Recommendation
  Algorithms}.
\newblock {\em IEEE Trans. Vis. Comput. Graph.}, pages 1--11, 2021.

\bibitem{Zhang2019a}
D.~Zhang, Y.~Suhara, J.~Li, and et~al.
\newblock {Sato: Contextual semantic type detection in tables}.
\newblock {\em Proc. VLDB Endow.}, 13(11):1835--1848, 2020.

\bibitem{zhang2019knowledgeaware}
H.~Zhang, Y.~Song, Y.~Song, and D.~Yu.
\newblock {Knowledge-aware Pronoun Coreference Resolution}.
\newblock In {\em Proc. ACL'19}. ACL, 2019.

\bibitem{zhang2020brief}
H.~Zhang, X.~Zhao, and Y.~Song.
\newblock {A Brief Survey and Comparative Study of Recent Development of
  Pronoun Coreference Resolution}.
\newblock {\em arXiv}, 2020.

\bibitem{Zhang2019}
S.~Zhang, L.~Yao, A.~Sun, and Y.~Tay.
\newblock {Deep Learning Based Recommender System}.
\newblock {\em ACM Comput. Surv.}, 52(1):1--38, 2019.

\bibitem{Zhang2017d}
Y.~Zhang, P.~Pasupat, and P.~Liang.
\newblock {Macro Grammars and Holistic Triggering for Efficient Semantic
  Parsing}.
\newblock In {\em Proc. EMNLP'17}. ACL.

\bibitem{Zhang2021a}
Z.~Zhang, Y.~Gu, X.~Han, and et~al.
\newblock {CPM-2: Large-scale Cost-effective Pre-trained Language Models}.
\newblock {\em arXiv}, 2021.

\bibitem{Zhou}
{Zhen Wen}, M.~Zhou, and V.~Aggarwal.
\newblock {An optimization-based approach to dynamic visual context
  management}.
\newblock In {\em Proc. InfoVis'05}. IEEE, 2005.

\bibitem{Zheng2017}
W.~Zheng, H.~Cheng, L.~Zou, and et~al.
\newblock {Natural Language Question/Answering}.
\newblock In {\em Proc. CIKM'17}. ACM, 2017.

\bibitem{Zheng2019a}
X.~Zheng, X.~Qiao, Y.~Cao, and R.~W.~H. Lau.
\newblock {Content-aware generative modeling of graphic design layouts}.
\newblock {\em ACM Trans. Graph.}, 2019.

\bibitem{zhong2017seq2sql}
V.~Zhong and et~al.
\newblock {Seq2SQL: Generating Structured Queries from Natural Language using
  Reinforcement Learning}.
\newblock {\em arXiv}, 2017.

\bibitem{Zhou2020}
M.~Zhou, Q.~Li, X.~He, and et~al.
\newblock {Table2Charts: Recommending Charts by Learning Shared Table
  Representations}.
\newblock In {\em Proc. KDD'21}. ACM.

\bibitem{Zhu2020}
S.~Zhu, G.~Sun, Q.~Jiang, and et~al.
\newblock {A survey on automatic infographics and visualization
  recommendations}.
\newblock {\em Vis. Informatics}, 4(3):24--40, 2020.

\bibitem{Azcan2020}
F.~Őzcan, A.~Quamar, J.~Sen, and et~al.
\newblock {State of the Art and Open Challenges in Natural Language Interfaces
  to Data}.
\newblock In {\em Proc. SIGMOD'20}. ACM, 2020.

\end{thebibliography}
	\vspace{-2.7cm}
	\begin{IEEEbiography}[{\includegraphics[width=1in,height=1.2in,clip,keepaspectratio]{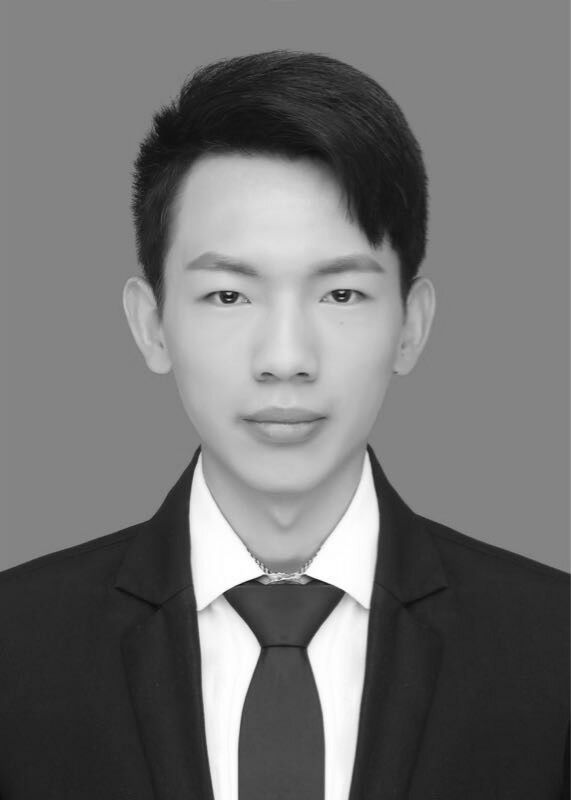}}]{Leixian Shen}
		received his bachelor's degree in Software Engineering from the Nanjing University of Posts and Telecommunications in 2020. He is currently a master student in the School of Software, Tsinghua University, Beijing, China. His research interests include data visualization and human computer interaction.	
	\end{IEEEbiography}
	\vspace{-1.3cm}
	\begin{IEEEbiography}[{\includegraphics[width=1in,height=1.2in,clip,keepaspectratio]{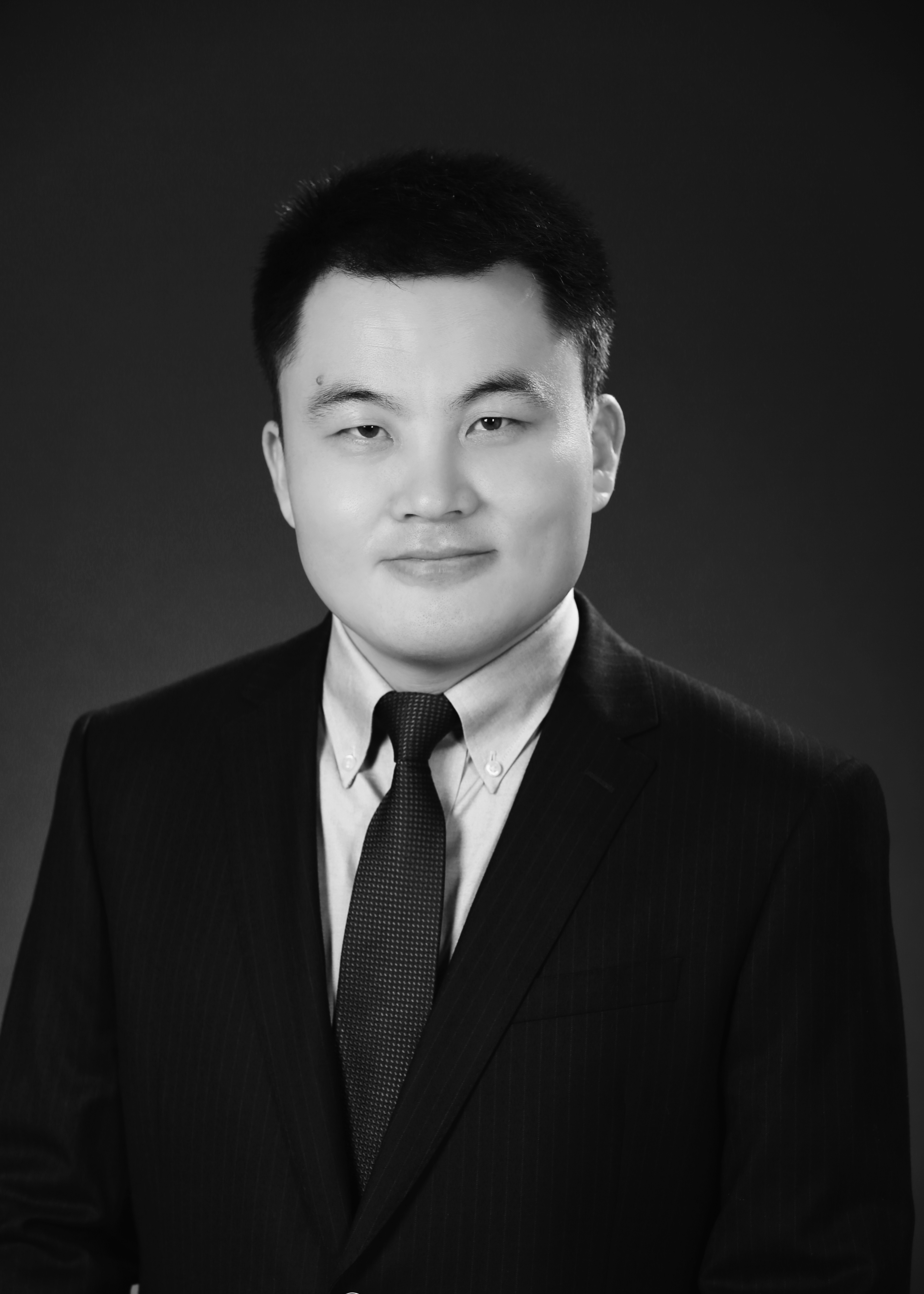}}]{Enya Shen}
		received the BSc degree from Nanjing University of Aeronautics and Astronautics, in 2008, and MSc and PhD degree from National University of Defense Technology, in 2010 and 2014, respectively. He works as a research assistant in Tsinghua University. His research interests include data visualization, human computer interaction, augmented reality, and geometry modeling.
	\end{IEEEbiography}
	\vspace{-1.3cm}
	\begin{IEEEbiography}[{\includegraphics[width=1in,height=1.2in,clip,keepaspectratio]{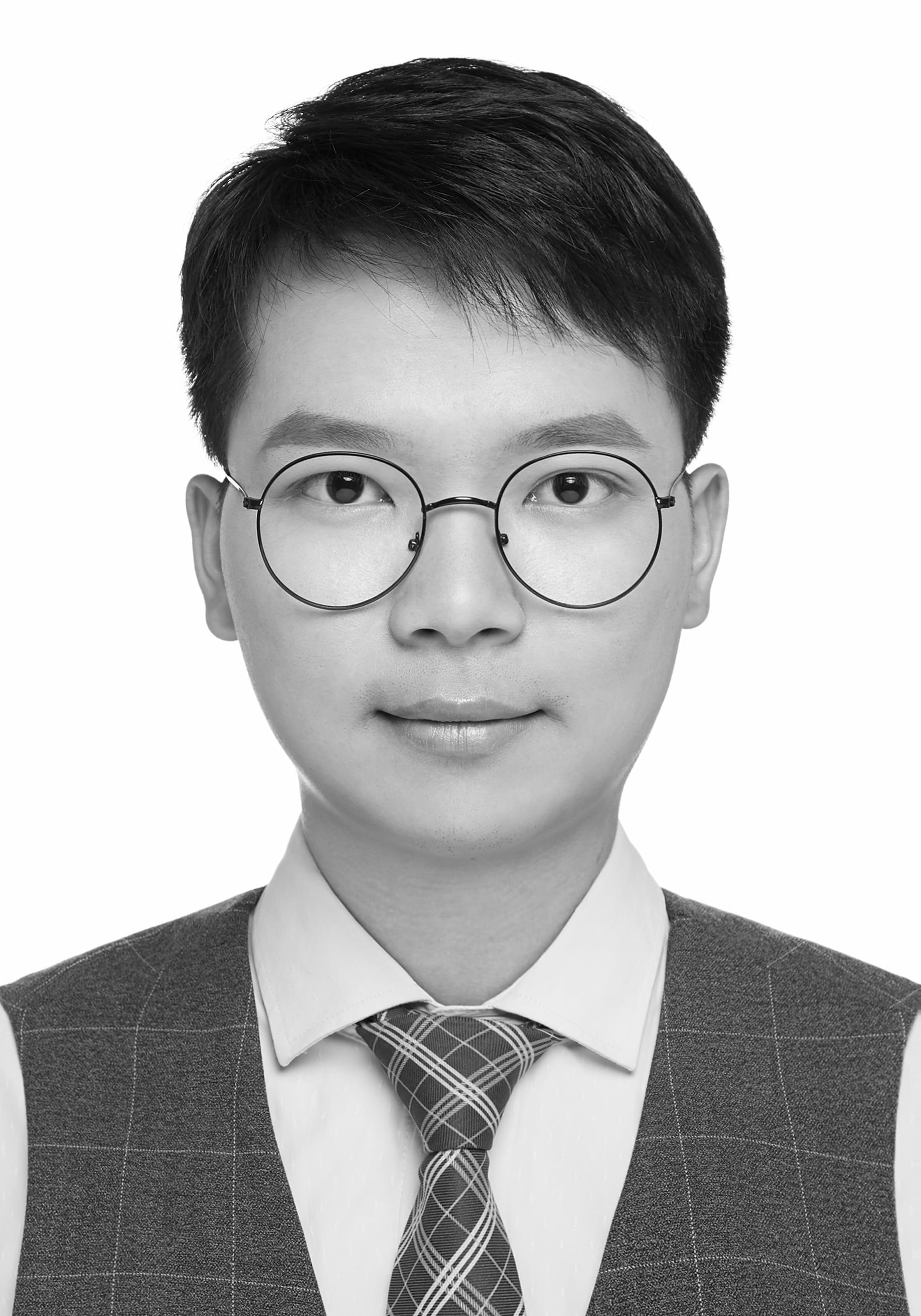}}]{Yuyu Luo}
		received his bachelor's degree in Software Engineering from the University of Electronic Science and Technology of China in 2018. He is currently a PhD student in the Department of Computer Science, Tsinghua University, Beijing, China. His research interests include data visualization and data cleaning.
	\end{IEEEbiography}
	\vspace{-1.3cm}
	\begin{IEEEbiography}[{\includegraphics[width=1in,height=1.2in,clip,keepaspectratio]{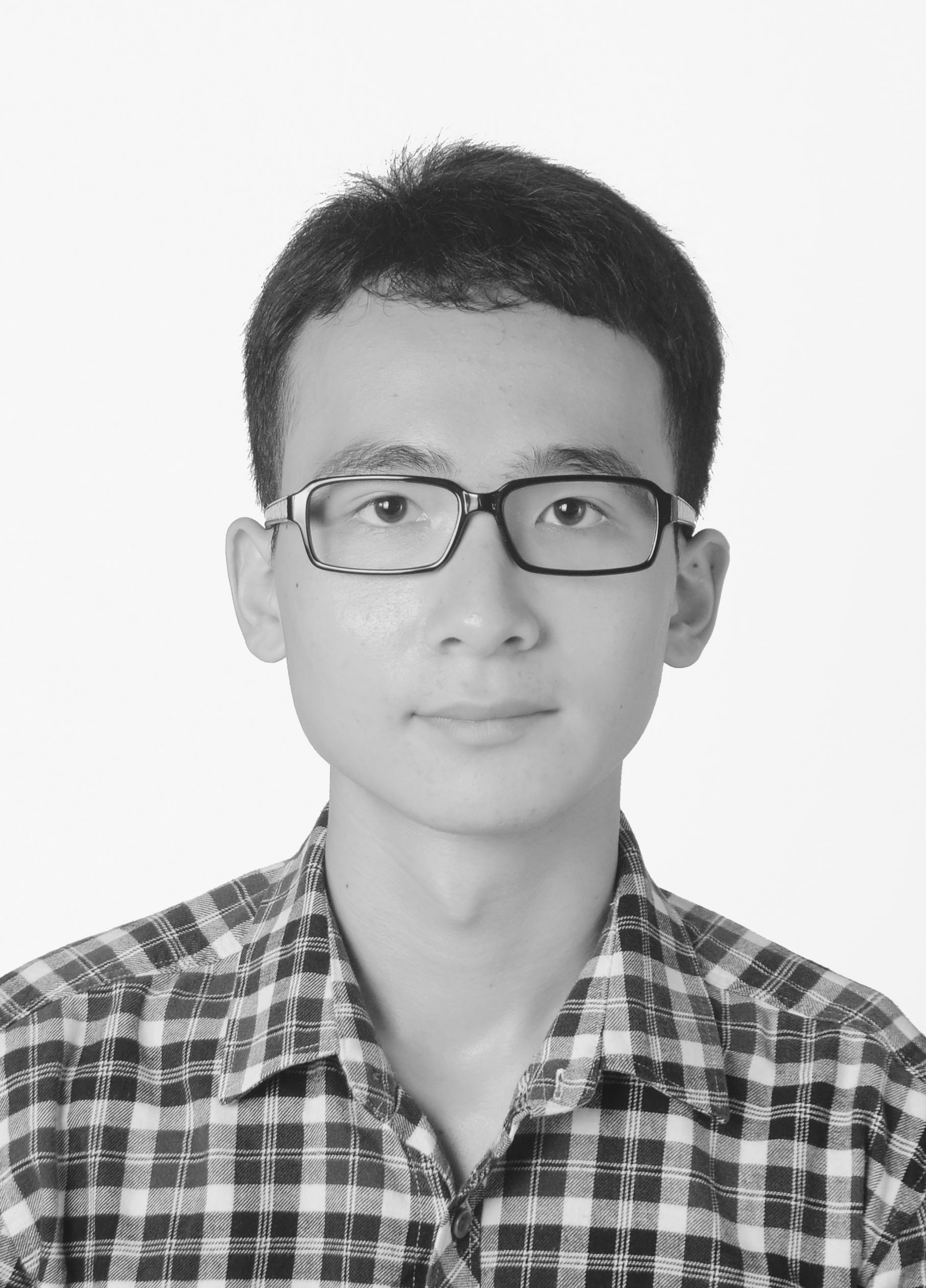}}]{Xiaocong Yang}
		is a fourth-year undergraduate working towards a BS degree at Tsinghua University, and plan to apply for a PhD position in Computer Science. His research interests include natural language processing and machine learning theory.
	\end{IEEEbiography}
	\vspace{-1.3cm}
	\begin{IEEEbiography}[{\includegraphics[width=1in,height=1.2in,clip,keepaspectratio]{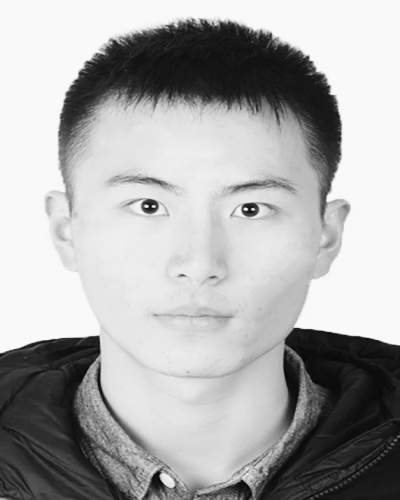}}]{Xuming Hu}
		received the BE degree in Computer Science and Technology, Dalian University of Technology. He is working toward the PhD degree at Tsinghua University. His research interests include natural language processing and information extraction.
	\end{IEEEbiography}
	\vspace{-1.3cm}
	\begin{IEEEbiography}[{\includegraphics[width=1in,height=1.2in,clip,keepaspectratio]{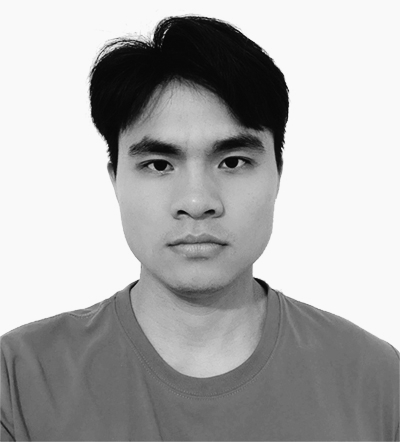}}]{Xiongshuai Zhang}
		received Bachelor’s Degree from South China University of Techinology, China, in 2019. He is now a master student with the School of Software, Tsinghua University, China. His research interests include data visualization and computer graphics.
	\end{IEEEbiography}
	\vspace{-1.3cm}
	\begin{IEEEbiography}[{\includegraphics[width=1.1in,height=1.2in,clip,keepaspectratio]{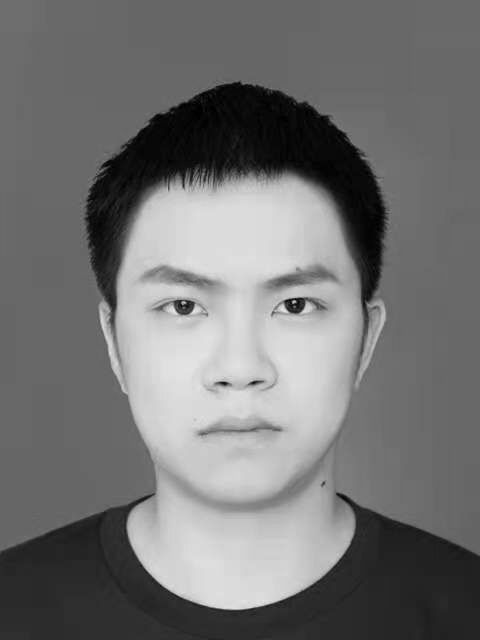}}]{Zhiwei Tai}
		received his BS degree from the School of Software, Tsinghua University in 2020. He is currently working toward the Master’s degree in Software Engineering at Tsinghua University. His research interests include data visualization, digital twin, and augmented reality.
	\end{IEEEbiography}
	\vspace{-1.3cm}
	\begin{IEEEbiography}[{\includegraphics[width=1.1in,height=1.2in,clip,keepaspectratio]{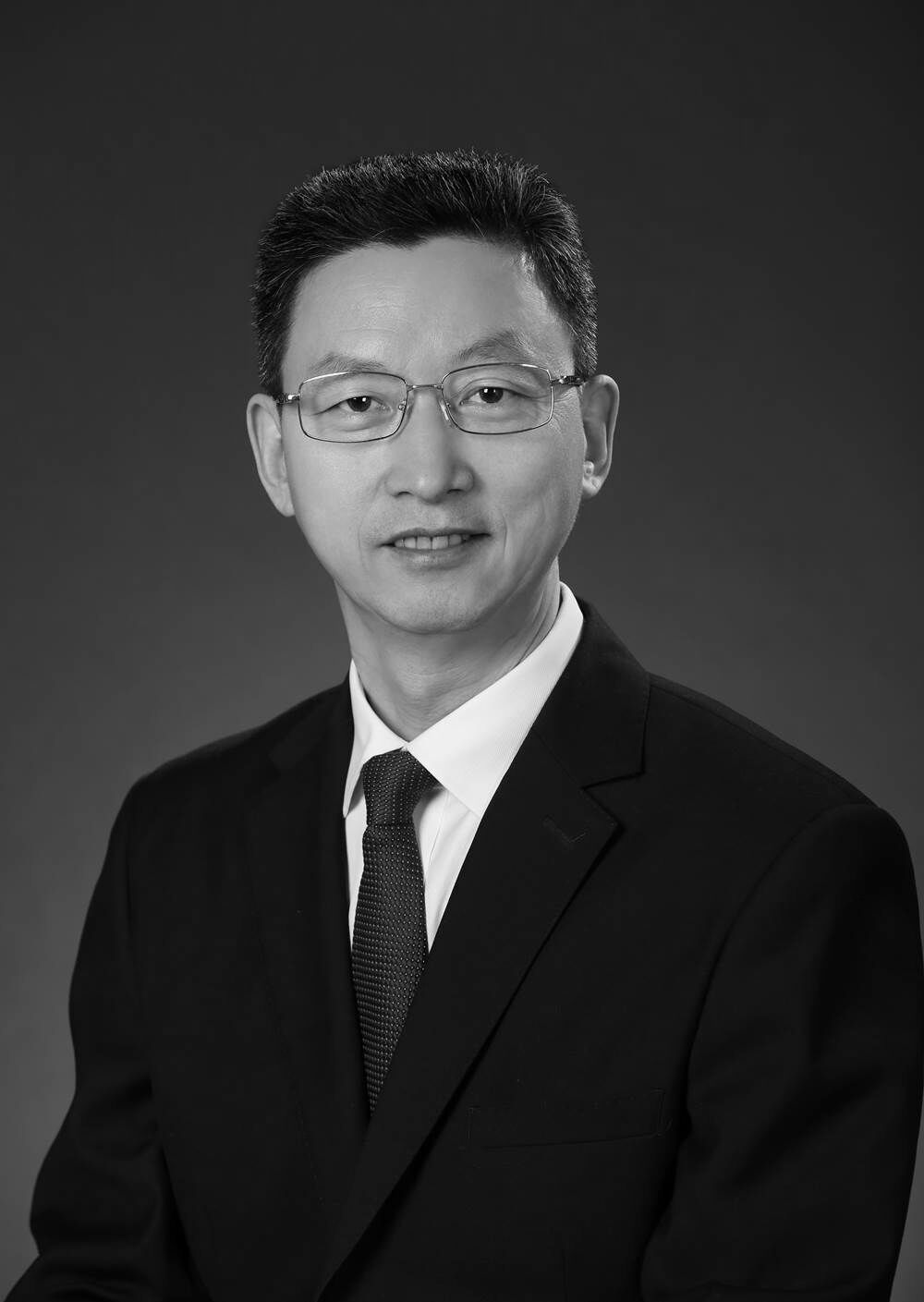}}]{Jianmin Wang}
		received his bachelor's degree from Peking University, China, in 1990, and the ME and PhD degrees in computer software from Tsinghua University, China, in 1992 and 1995, respectively. He is a full professor with the School of Software, Tsinghua University. His research interests include big data management and large-scale data analytics. 
		He is leading to develop a big data management system in the National Engineering Lab for Big Data Software.
	\end{IEEEbiography}

\end{document}